\documentclass[a4paper, 12pt]{article}
\usepackage{latexsym, amsmath, amssymb}
\usepackage{tikz}
\usetikzlibrary{calc, decorations.pathmorphing, decorations.markings, arrows.meta, ext.paths.arcto}
\usepackage[american]{babel}
\usepackage{graphicx}
\usepackage{bbm}
\usepackage{cite}
\usepackage{enumitem}
\usepackage{mathrsfs}
\usepackage[colorlinks=true, citecolor=blue!90!black, linkcolor=blue!90!black, linktocpage=true, urlcolor=red!70!black, bookmarksdepth=subsubsection]{hyperref}
\usepackage{braket}
\usepackage{caption}
\usepackage{cleveref}
\crefname{figure}{Fig.}{Figs.}
\crefname{equation}{eqn.}{eqns.}
\crefname{section}{Sec.}{Secs.}
\crefname{appendix}{App.}{Apps.}
\crefname{footnote}{fn.}{fns.}

\usepackage[top=2cm, bottom=3cm, left=2cm, right=2cm]{geometry}

\renewcommand{\baselinestretch}{1.2}
\providecommand{\abs}[1]{\lvert#1\rvert}

\renewcommand{\d}{\mathrm{d}}

\newcommand{\bdot}{{\boldsymbol{\cdot}}}
\newcommand{\ud}[2]{^{#1}_{\phantom{#1}#2}}

\newcommand{\ts}{\textstyle}
\newcommand{\ds}{\displaystyle}

\newcommand{\wb}{\overline}

\newcommand{\tps}[2]{\texorpdfstring{\ensuremath{#1}}{#2}}
\newcommand{\btps}[2]{\texorpdfstring{\ensuremath{\boldsymbol{#1}}}{#2}}

\newcommand{\eg}{\textit{e.g.}}

\newcommand{\ie}{\textit{i.e.}}

\newcommand{\aka}{\textit{a.k.a.}}
\newcommand{\rhs}{r.h.s.}
\newcommand{\lhs}{l.h.s.}

\numberwithin{equation}{section}

\newcommand{\nn}{\nonumber}
\newcommand{\mat}[1]{\begin{pmatrix} #1 \end{pmatrix}}

\newcommand{\be}{\begin{equation}} \newcommand{\ee}{\end{equation}}
\newcommand{\bea}{\begin{equation} \begin{aligned}} \newcommand{\eea}{\end{aligned} \end{equation}}

\newcommand{\cA}{\mathcal{A}}
\newcommand{\cB}{\mathcal{B}}

\newcommand{\scrC}{\mathscr{C}}

\newcommand{\cF}{\mathcal{F}}

\newcommand{\cH}{\mathcal{H}}

\newcommand{\cL}{\mathcal{L}}
\newcommand{\cM}{\mathcal{M}}
\newcommand{\scrM}{\mathscr{M}}
\newcommand{\cN}{\mathcal{N}}
\newcommand{\cO}{\mathcal{O}}

\newcommand{\cS}{\mathcal{S}}

\newcommand{\cV}{\mathcal{V}}

\newcommand{\cX}{\mathcal{X}}

\newcommand{\bC}{\mathbb{C}}

\newcommand{\bN}{\mathbb{N}}

\newcommand{\bP}{\mathbb{P}}

\newcommand{\bR}{\mathbb{R}}

\newcommand{\bZ}{\mathbb{Z}}
\newcommand{\fA}{\mathfrak{A}}

\newcommand{\fR}{\mathfrak{R}}

\newcommand{\sS}{\mathsf{S}}
\newcommand{\sT}{\mathsf{T}}
\newcommand{\sX}{\mathsf{X}}

\newcommand{\unit}{\mathbbm{1}}
\newcommand{\id}{\mathrm{id}}
\newcommand{\Id}{\mathbbm{1}}

\DeclareMathOperator{\Tr}{Tr}

\DeclareMathOperator{\diag}{diag}

\DeclareMathOperator{\End}{End}
\DeclareMathOperator{\Hom}{Hom}
\DeclareMathOperator{\Irr}{Irr}

\newcommand{\tikzmark}[1]{\tikz[overlay, remember picture] \coordinate (#1);}
\pgfmathsetmacro\mathaxis{-height("$\vcenter{}$")}

\def\colortopoline{blue!90!black}
\def\colortopojun{red!60!black}
\def\colorbound{blue!50!black}
\def\colorshade{gray!20}
\def\colorstate{black!60!white}

\def\stripw{0.55}
\def\striph{0.7}
\def\stripshadew{0.5}

\def\densmatradx{0.3}
\def\densmatrady{0.3}
\def\densmatgapw{1.3}
\def\densmatgaph{0.12}

\newcommand{\tube}[4]{%
\begin{tikzpicture}[line width=0.6, baseline={(0,-0.1)}]
        \draw (0, -0.5) node [shift={(0.2, 0.05)}] {\footnotesize$#1$} -- (0, 0.5) node [shift={(0.2, -0.05)}] {\footnotesize$#4$};
        \draw (-0.5, 0) node[shift={(0, 0.2)}] {\footnotesize$#2$} -- (0.5, 0);
        \node at (-0.15, 0.15) {\scriptsize$#3$};
\end{tikzpicture}}

\newcommand{\strip}[7]{%
  \begin{tikzpicture}[baseline=\mathaxis, line width=0.6]
      \filldraw [\colorshade] (\stripw, -\striph) rectangle (\stripw + \stripshadew, \striph);
      \filldraw [\colorshade] (-\stripw, -\striph) rectangle (-\stripw - \stripshadew, \striph);
      \draw[\colortopoline] (-\stripw, -0.2) node[\colortopojun, shift={(-0.20, 0.05)}] {\scriptsize$#6$} -- (\stripw, 0.2) node[\colortopojun, shift={(0.20, -0.05)}] {\scriptsize$#7$} node[pos=0.5, above] {\small$#3$};
      \draw[\colorbound] (\stripw, -\striph) node[right,shift={(0,0.15)}] {\small$#2$} -- (\stripw, \striph) node[right,shift={(0,-0.15)}] {\small$#5$};
      \draw[\colorbound, -stealth] (\stripw, -0.6) -- ++(90: 0.45); \draw [\colorbound,-stealth] (\stripw, -0.6) -- ++(90: 1.10);
      \draw[\colorbound] (-\stripw, \striph) node[left,shift={(0,-0.15)}] {\small$#4$} -- (-\stripw, -\striph) node[left,shift={(0,0.15)}] {\small$#1$};
      \draw[\colorbound, -stealth] (-\stripw, 0.6) -- ++(-90: 0.45); \draw [\colorbound, -stealth] (-\stripw, 0.6) -- ++(-90: 1.10);
      \fill[\colortopojun] (-\stripw, -0.2) circle (1.pt);
      \fill[\colortopojun] (\stripw, 0.2) circle (1.pt);
  \end{tikzpicture}%
}

\newcommand{\densmatsetup}[1]{
  \begin{tikzpicture}[
      baseline=\mathaxis,
      arrBound/.style = { decoration = { markings, mark = at position 0.55 with {\arrow[rotate=15,scale=0.8]{Stealth}}  }, postaction = decorate },
      arrBoundRev/.style = { decoration = { markings, mark = at position 0.45 with {\arrowreversed[rotate=15,scale=0.8]{Stealth}}  }, postaction = decorate },
      node/.style = {inner sep = 0, outer sep = 0},
      line width=0.6
      ]
      #1
  \end{tikzpicture}
}

\newcommand{\densmatvac}[2]{
  \coordinate (a) at (-\densmatgapw/2, \densmatgaph/2);
  \coordinate (b) at (\densmatgapw/2, \densmatgaph/2);
  \coordinate (c) at (\densmatgapw/2, -\densmatgaph/2);
  \coordinate (d) at (-\densmatgapw/2, -\densmatgaph/2);

  \draw[densely dashed] (a) -- (b);
  \draw[densely dashed] (c) -- (d);

  \draw[\colorbound, arrBound] (d) arc to[x radius=\densmatradx, y radius=\densmatrady, large, clockwise] (a);
  \draw[\colorbound, arrBound] (b) arc to[x radius=\densmatradx, y radius=\densmatrady, large, clockwise] (c);
  \node[\colorbound, below] at (-\densmatgapw/2 - \densmatradx, -\densmatrady) {\small $#1$};
  \node[\colorbound, below] at (\densmatgapw/2 + \densmatradx, -\densmatrady) {\small $#2$};
}

\newcommand{\densmatnorm}[2]{
  \coordinate (a) at (-\densmatgapw/2,0);
  \coordinate (b) at (\densmatgapw/2,0);
  \draw [\colorbound, arrBoundRev] (-\densmatgapw/2-\densmatradx,0) circle [x radius=\densmatradx, y radius=\densmatrady];
  \draw [\colorbound, arrBoundRev, rotate=180] (-\densmatgapw/2-\densmatradx,0) circle [x radius=\densmatradx, y radius=\densmatrady];
  \node[\colorbound, below] at (-\densmatgapw/2-\densmatradx,-\densmatrady) {\small $#1$};
  \node[\colorbound, below] at (\densmatgapw/2+\densmatradx,-\densmatrady) {\small $#2$};
}

\newcommand{\mysquare}{%
  \tikz[baseline=\mathaxis, scale=0.15]{
    \fill[gray] (-1,-1) rectangle (1,1);
    \draw[black] (-1,-1) rectangle (1,1);
  }%
}

\newcommand{\mydiagonalsquare}{%
  \tikz[baseline=\mathaxis, scale=0.15]{
    \draw[] (-1,-1) rectangle (1,1); 
    \draw[] (-1,1) -- (1,-1);        
  }%
}

\begin{document}
\pagenumbering{roman}
\thispagestyle{empty}
\fontsize{12pt}{16pt}

\begin{flushright}
SISSA 10/2025/FISI
\end{flushright}

\vspace{13mm}

\begin{center}
    {\huge Entanglement Asymmetry \\[.6em] for Higher and Noninvertible Symmetries}
    \\[13mm]
    {\large Francesco Benini, \, Pasquale Calabrese, \\[0.3em] Michele Fossati, \, Amartya H. Singh, \, Marco Venuti}

\bigskip
    {\it
      SISSA, Via Bonomea 265, 34136 Trieste, Italy \\[-0.0em]
      INFN, Sezione di Trieste, Via Valerio 2, 34127 Trieste, Italy
    }
\end{center}

\bigskip

\begin{abstract}
\noindent
Entanglement asymmetry is an observable in quantum systems, constructed using quantum-information methods, suited to detecting symmetry breaking in states --- possibly out of equilibrium --- relative to a subsystem. In this paper we define the asymmetry for generalized finite symmetries, including higher-form and noninvertible ones. To this end, we introduce a ``symmetrizer'' of (reduced) density matrices with respect to the $C^*$-algebra of symmetry operators acting on the subsystem Hilbert space. We study in detail applications to $(1{+}1)$-dimensional quantum field theories: First, we analyze spontaneous symmetry breaking of noninvertible symmetries, confirming that distinct vacua can exhibit different physical properties. Second, we compute the asymmetry of certain excited states in conformal field theories (including the Ising CFT), when the subsystem is either the full circle or an interval therein. The relevant symmetry algebras to consider are the fusion, tube, and strip algebras. Finally, we comment on the case that the symmetry algebra is a (weak) Hopf algebra.
\end{abstract}

\newpage

\pagenumbering{arabic}

{%
  \renewcommand{\baselinestretch}{.88}
  \parskip=0pt
  \setcounter{tocdepth}{2}
  \tableofcontents
}


\section{Introduction}
\label{sec: intro}

Symmetries are a powerful and universal tool in quantum field theory (QFT), applicable beyond the perturbative regime. In particular, it is useful to be able to detect whether symmetries are broken in a given quantum state. For instance, in the Landau paradigm one distinguishes different phases of a theory according to how symmetries are realized and possibly spontaneously broken in ground states. As another example, excited states produced by quenches can exhibit interesting physics revealed by symmetry breaking and its restoration (or lack thereof) during time evolution due to relaxation. The standard approach to detect symmetry breaking uses order parameters, \ie, expectation values of (usually local) operators. This is a valid approach, but there are alternatives that might have advantages.

One alternative approach, proposed in \cite{Ares:2022koq} to study the dynamics of out-of-equilibrium states, is named ``entanglement asymmetry.'' It builds on quantum-information methods, and it is equivalent to previously introduced quantities in different contexts, namely the ``asymmetry'' of \cite{Vaccaro:2007drw, Gour:2009hng, Marvian:2014awa} and the ``entropic order parameter'' of \cite{Casini:2019kex, Magan:2020ake, Casini:2020rgj, Magan:2021myk}. Given a state and a subregion of space $A$, the entanglement asymmetry $\Delta S$ with respect to a symmetry group $G$ is defined as the relative entropy between the reduced density matrix on $A$ and its average over the action of $G$.%
\footnote{There exist also R\'enyi versions of asymmetry, denoted $\Delta S^{(n)}$ where $n$ is a number. Entanglement asymmetry is recovered in the limit $n \to 1$.}
This definition applies to both continuous and discrete, Abelian and non-Abelian groups. A few advantages of asymmetry include the following: it does not require one to identify a suitable order parameter for the symmetry; it has an extra dependence on a subsystem, or subregion, $A$; it is faithful, meaning that $\Delta S > 0$ if and only if the state breaks the symmetry.
Notably, entanglement asymmetry is a measure of symmetry breaking rather than of entanglement itself, as it can be nonzero even for states with vanishing entanglement.

By now, entanglement asymmetry has been studied from the perspectives of both condensed-matter and high-energy physics. On the condensed-matter front, it was applied to a variety of systems, including ground states with explicit \cite{Murciano:2023qrv, Ferro:2023sbn, Fossati:2024xtn, Lastres:2024ohf, Kusuki:2024gss, Fossati:2024ekt} and spontaneous \cite{Capizzi:2023yka} symmetry breaking, matrix product states \cite{Capizzi:2023xaf}, integrable systems \cite{Rylands:2023yzx, Murciano:2023qrv, Klobas:2024mlb, Ares:2023kcz}, and quantum circuits \cite{Turkeshi:2024juo, Liu:2024kzv}, to name a few. It shed light on out-of-equilibrium dynamics, in particular revealing a ``quantum Mpemba effect'' taking place under time evolution \cite{Ares:2022koq} (see \cite{Ares:2025onj} for a review), later experimentally confirmed \cite{Joshi:2024sup}. On the high-energy front, asymmetry was studied in certain excited states in conformal field theories (CFTs) \cite{Chen:2023gql, Benini:2024xjv, Fujimura:2025rnm}. Besides, a holographic dual was proposed for a class of perturbative states \cite{Benini:2024xjv}. In \cite{Ares:2023ggj, Chen:2024lxe, Russotto:2024pqg, Russotto:2025cpn} the asymmetry with respect to $U(1)$ and $SU(N)$ symmetries was computed in the Page model of radiating black holes, yielding a cousin of the Page curve.

In this paper we aim to define asymmetry for \emph{generalized symmetries}, which include the cases of both higher-form \cite{Gaiotto:2014kfa} and noninvertible \cite{Bhardwaj:2017xup} symmetries.%
\footnote{The asymmetry for higher-form symmetries is studied in greater detail, from a different perspective, in \cite{Benini:2025hbj}. On the other hand, a generalized Landau paradigm for noninvertible symmetries was advocated in \cite{Bhardwaj:2023fca}.}
In the modern formulation, the (finite internal) symmetries of a local quantum field theory are described by topological defects of various dimensions \cite{Gaiotto:2014kfa}. These objects fit naturally into mathematical structures known as (higher) fusion categories \cite{Bhardwaj:2017xup, Chang:2018iay, Douglas:2018qfz, Gaiotto:2019xmp, JohnsonFreyd:2020, Bhardwaj:2022yxj}. The latter not only encode which defects are present in a given theory, but also how they can fuse and join together, as well as various manifestations of anomalies (\eg, braiding phases and transformations produced by manipulations of the topological defects). Such a generalized definition of ``symmetry'' goes well beyond ordinary symmetry groups. For the finite internal symmetries of unitary QFTs, we propose to define asymmetry through the following construction.

A QFT associates a Hilbert space of states to each possible spatial manifold. Consider thus a (possibly mixed) state, whose asymmetry one wishes to compute, described by a density matrix. One might be interested in studying the asymmetry of the full system, or of a subsystem. In the latter case, to each spatial subregion $A$ and choice of boundary conditions, one can associate another Hilbert space and a reduced density matrix \cite{Ohmori:2014eia}. Moreover, one may choose whether or not to include the twisted sectors in these Hilbert spaces. In any case, one ends up with a (sub)system Hilbert space $\cH$ and a (reduced) density matrix $\rho$. The symmetry defects, when laid parallel to the spatial manifold, give rise to an action on the (sub)system Hilbert space through a finite-dimensional $C^*$-algebra $\cA$. We then propose to construct a ``symmetrizer'' $\sS$, that produces a symmetrized density matrix $\rho_\sS$ from $\rho$. Concisely, $\sS$ is the projector to the space of operators on $\cH$ that commute with the symmetry algebra $\cA$ (which turns out to be a quantum channel). Lastly, one quantifies how much $\rho_\sS$ differs from $\rho$. Following \cite{Ares:2022koq}, we use relative entropy or its R\'enyi version.%
\footnote{First, one could define what a symmetric density matrix is in different ways, see for instance \cite{Schafer-Nameki:2025fiy, AliAhmad:2025bnd}. Second, other functions that quantify the difference between two density matrices could be used as well; see, for instance, \cite{Ares:2025onj} for a discussion.}
This defines, respectively, entanglement and R\'enyi asymmetries with respect to generalized symmetries.

In fact, this definition applies even beyond local quantum field theories --- for instance, to lattice models. Once the relevant Hilbert space and finite-dimensional $C^*$-algebra of symmetry operators have been identified, our definition of asymmetry can be used.

To be concrete, we analyze in detail the case of two-dimensional unitary quantum field theories, where symmetries are implemented by the topological lines of a unitary fusion category $\scrC$. First, we consider QFTs placed on a spatial circle $S^1$ and study the asymmetry of the full system. The simplest setup is to restrict to the untwisted Hilbert space, on which the symmetry acts through a fusion algebra. On the other hand, one can encompass the total Hilbert space of untwisted and twisted sectors, on which a noninvertible symmetry has a more sophisticated action that mixes the untwisted sector with the twisted ones, and that is described by the so-called tube algebra $\mathsf{Tube}(\scrC)$ \cite{Ocneanu:1994}. We compare the asymmetries computed with respect to the fusion and tube algebras, finding that the latter is a more refined detector of symmetry breaking, especially in the case of noninvertible symmetries. We investigate this issue in the context of a class of excited states in CFTs, created by the insertion of local operators.

Moreover, we compute entanglement asymmetry in two-dimensional gapped phases with spontaneous symmetry breaking (SSB), employing a topological quantum field theory (TQFT) description. We find that vacua related by noninvertible symmetries need not have the same asymmetry, whereas for group-like symmetries they always do. This is another manifestation of the fact that vacua arising from the spontaneous breaking of a noninvertible symmetry can exhibit different physics \cite{Huse:1984mn, Damia:2023ses}.

Next, we study the asymmetry of subsystems when the subregion $A$ is an interval. In order to properly define a subsystem Hilbert space and a symmetry action on it, one needs to specify boundary conditions, mathematically described by a $\scrC$-module category $\scrM$. The noninvertible symmetry then acts through the so-called strip algebra $\mathsf{Strip}_\scrC(\scrM)$ \cite{Ostrik:2001xnt, Kitaev:2011dxc, Cordova:2024iti, Choi:2024tri}.%
\footnote{See also \cite{Cordova:2024vsq, Copetti:2024dcz, Choi:2024wfm}.}
When the theory admits (strongly or weakly \cite{Choi:2023xjw}) symmetric boundary conditions, the definition of asymmetry is straightforward --- otherwise,%
\footnote{For instance, the presence of 't~Hooft anomalies in the theory prevents the existence of symmetric boundary conditions \cite{Jensen:2017eof, Thorngren:2020yht, Choi:2023xjw}, in a sense that we spell out more precisely in \cref{sec: interval}.}
we implement an averaging protocol over the boundary conditions related by symmetry. For excited states in CFTs (\eg, in the Ising CFT), we investigate how the asymmetry varies with the subsystem size, and compare it with the asymmetry of the full system. Interestingly, we observe that R\'enyi asymmetries are not always monotonically increasing with the subsystem size (whereas entanglement asymmetry is, see \eg\ \cite{Benini:2025hbj}). For group-like symmetries, we find that the choice of boundary conditions eventually does not affect the asymmetry, thus validating previous, less rigorous approaches in the literature in which the boundary conditions (and the possible influence of 't~Hooft anomalies) were not taken into account. For noninvertible symmetries, instead, a proper account of the boundary conditions seems inevitable in order to compute the asymmetry explicitly.

Finally, we note that when the symmetry algebra $\cA$ has special properties or additional structure, there might exist other equivalent presentations of the symmetrizer $\sS$ (and hence of the asymmetries). For instance, for those symmetry algebras that are Hopf algebras (including certain strip algebras) we find an alternative formula for $\sS$.

The paper is organized as follows. In \cref{sec: asymmetry} we put forward our definition of asymmetry for generalized symmetries, whose core is the symmetrizer $\sS$ with respect to a finite-dimensional semisimple algebra. In \cref{sec: circle}, focusing on two-dimensional theories, we study the asymmetry of states on a spatial circle, with respect to both the fusion and tube algebras. In \cref{sec: interval}, instead, we study the asymmetry of states on an interval with respect to the strip algebra. The possible definitions of the symmetrizer depend strongly on the available types of boundary conditions. We also present an alternative formula valid in the case of Hopf and weak Kac algebras. The asymmetry on the circle is applied in \cref{sec: SSB} to the detection of spontaneous symmetry breaking in 2d gapped phases. Finally, \cref{sec: examples} is devoted to concrete examples that we work out in detail: group-like symmetries, as well as a collection of fusion categories including the Ising, $\mathsf{Rep}(H_8)$, Fibonacci, and Haagerup ones. More technical aspects are collected in appendices.

\bigskip

\textbf{Note added.} While this work was under completion, we became aware of \cite{AliAhmad:2025bnd}. We thank its authors for coordinating the submission to arXiv.


\section{Entanglement asymmetry for generalized symmetries}
\label{sec: asymmetry}

In the modern formulation, the finite internal symmetries of a quantum field theory are described by a (higher) fusion category. Its objects, the morphisms between objects, the morphisms between morphisms, and so on, describe the topological operators of various codimensions in the quantum field theory and their intersections, and correspond to the symmetry elements. The full structure of the fusion category encodes phase factors (such as braiding phases and 't~Hooft anomalies) as well as more general transformations produced by moves of the topological operators, which constitute the symmetry structure.

Depending on the physical setup, the symmetry category gives rise to different algebras of symmetry operators, as we now describe. A quantum field theory associates a Hilbert space of states to each spatial manifold. If we are interested in a subsystem (\ie, a subregion) $A$, we can use the procedure of \cite{Ohmori:2014eia} (that we review in \cref{sec: interval}) to produce the Hilbert space of the QFT on a spatial manifold $A$ with boundaries. Besides, we may want to include or not to include twisted sectors, \ie, to allow the insertion of topological operators along the time direction so that they pierce the spatial manifold. In all cases, we end up with a Hilbert space $\cH$. We can construct symmetry operators acting on this space by taking topological defects laid along the spatial slice. Under the multiplication given by fusing the defects, they give rise to an algebra $\cA$. We call $\{\cX_a\}$ a basis of the algebra.

For technical reasons, in this paper we focus on unitary quantum field theories with finite internal symmetries. These are described by a unitary (higher) fusion category, which contains a finite number of simple objects. As a result, the algebra $\cA$ is a finite-dimensional $C^*$-algebra, which in particular is semisimple.

Entanglement asymmetry is a quantum-information-based measure of symmetry breaking in pure or mixed states. Given a density matrix $\rho$, in order to quantify how much this breaks the symmetry, we compare it with a symmetrized version of itself, that we call $\rho_\sS$. Entanglement asymmetry is then defined as the relative entropy
\be
\label{def asymmetry as relative entr}
\Delta S[\rho] \equiv S[\rho \| \rho_\sS] \equiv \Tr\bigl( \rho \, ( \log \rho - \log \rho_\sS ) \bigr) \;.
\ee
Because of the properties of $\rho_\sS$, entanglement asymmetry can also be written, as we show below around (\ref{equality of traces}), as a difference
\be
\label{def asymmetry as von Neumann}
\Delta S[\rho] = S[\rho_\sS] - S[\rho] \;,
\ee
where $S[\rho] \equiv - \Tr (\rho \log \rho)$ is the von~Neumann entropy.
From standard properties of relative entropy, it follows that entanglement asymmetry is nonnegative, $\Delta S[\rho] \geq 0$, and that it vanishes if and only if $\rho = \rho_\sS$. One can also define R\'enyi asymmetries, characterized by an index $n$:%
\footnote{If $\rho$ is a pure state, this reduces to the R\'enyi entropy, which is nonnegative (independently of the details of the symmetrization procedure).
For a generic state, it has been proven in \cite{Ma:2021zgf, Han:2022phu} that the R\'enyi asymmetries are nonnegative for every $n$ for the case of a $U(1)$ symmetry. The proof therein generalizes to any Abelian algebra, because the symmetrizer still takes the form \eqref{eq:symmetrizer-1d-irreps}. To the best of our knowledge, we are not aware of a proof of nonnegativity for a non-Abelian algebra or for a non-Abelian group.}
\be
\Delta S^{(n)}[\rho] = \frac1{1-n} \log \frac{ \Tr( \rho_\sS^n) }{ \Tr( \rho^n) } \;.
\ee
The entanglement asymmetry is reproduced in the limit $n \to 1$.

Given a Hilbert space $\cH$ and a density matrix $\rho \in \End(\cH)$, in order to obtain the symmetrized version of $\rho$ we need to construct a \emph{symmetrizer} map $\sS \in \End\bigl( \End(\cH) \bigr)$ such that
\be
\label{def of symmetrizer}
\mathsf{S}\colon \rho \,\mapsto\, \rho_\sS \;.
\ee
The symmetrizer is characterized by the following properties:
\begin{enumerate}
\item It is a linear projector, in particular $(\rho_\sS)_\sS = \rho_\sS$.
\item Its image is symmetry-invariant, namely $\cX_a \, \rho_\sS = \rho_\sS \, \cX_a$ for all $\cX_a \in \cA$.
\item If $\rho$ is already symmetry-invariant, then $\rho_\sS = \rho$.
\item It is trace-preserving, namely $\Tr(\rho_\sS) = \Tr(\rho)$.
\item It maps density matrices to density matrices: $\rho_\sS^\dag = \rho_\sS$ and $\rho_\sS \geq 0$.
\end{enumerate}
Notice that while properties 1.~--~4.\ only depend on the product structure on $\cA$, property 5.\ also depends on its $*$-structure.

Properties 1.~--~3.\ can be more compactly stated as follows: $\sS$ is a linear projector to the vector space of operators that commute with the algebra $\cA$. Indeed, property 2.\ states that the image of $\sS$ is contained in the space of commuting operators, while property 3.\ states that commuting operators are contained in the image of $\sS$. Properties 4.\ and 5., on the other hand, guarantee that $\rho_\sS$ is a normalized density matrix (as long as $\rho$ is) so that it makes sense to compute its von~Neumann or R\'enyi entropies.

An intuitive picture for the action of the symmetrizer is the following. Decomposing the Hilbert space $\cH$ into a direct sum of irreducible representations of $\cA$ (with multiplicities), we can decompose both $\rho$ and $\rho_\sS$ into blocks that map irreducible representations to irreducible representations. Property 2.\ states that $\rho_\sS$ commutes with every element of the algebra $\cA$. Then, Schur's lemma (see, \eg, \cite{Lang:2002}) implies that every block of $\rho_\sS$ that connects two different irreducible representations must be the zero map, and that every block of $\rho_\sS$ that connects two equal irreducible representations must be proportional to the identity. In \cref{sec:schur} we show that the action of the symmetrizer $\sS$ on $\rho$ is precisely that of setting to zero the blocks that connect different irreducible representations, and to replace the blocks that connect equal irreducible representations with blocks proportional to the identity matrix, with a prefactor such that the trace of each block is preserved. See \cref{fig: action on matrices} for a graphical representation.

\begin{figure}[t]
\centering$\ds
\setlength{\arrayrulewidth}{0.7pt}
\arraycolsep=1.6pt
\def\arraystretch{0.72}
\rho = \left( \begin{array}{cc|ccc|c}
        \mysquare & \mysquare & \mysquare & \mysquare & \mysquare & \mysquare \\
        \mysquare & \mysquare & \mysquare & \mysquare & \mysquare & \mysquare \\
        \hline
        \mysquare & \mysquare & \mysquare & \mysquare & \mysquare & \mysquare \\
        \mysquare & \mysquare & \mysquare & \mysquare & \mysquare & \mysquare \\
        \mysquare & \mysquare & \mysquare & \mysquare & \mysquare & \mysquare \\
        \hline
        \mysquare & \mysquare & \mysquare & \mysquare & \mysquare & \mysquare \\
\end{array} \right)
\qquad\Rightarrow\qquad
\rho_\sS = \left( \begin{array}{cc|ccc|c}
        \mydiagonalsquare & \mydiagonalsquare & 0 & 0 & 0 & 0 \\
        \mydiagonalsquare & \mydiagonalsquare & 0 & 0 & 0 & 0 \\
        \hline
        0 & 0 & \mydiagonalsquare & \mydiagonalsquare & \mydiagonalsquare & 0 \\
        0 & 0 & \mydiagonalsquare & \mydiagonalsquare & \mydiagonalsquare & 0 \\
        0 & 0 & \mydiagonalsquare & \mydiagonalsquare & \mydiagonalsquare & 0 \\
        \hline
        0 & 0 & 0 & 0 & 0 & \mydiagonalsquare \\
\end{array} \right)
\qquad
\text{where } \;\;\; \mydiagonalsquare = \frac{\Tr ( \mysquare )}{\dim(V_r)} \; \unit
$
\caption{\label{fig: action on matrices}%
The Hilbert space splits as $\cH = \bigoplus_r n_r V_r$ where $V_r$ is the $r$-th irreducible representation of $\cA$, while $n_r$ is its degeneracy. A similar block decomposition applies to density matrices. The symmetrizer acts by setting to zero the off-diagonal blocks, and keeping sub-blocks proportional to the identity on the diagonal.}
\end{figure}

It turns out that such a symmetrizer has a simple and universal construction%
\footnote{What we mean by ``universal'' is that the symmetrizer can be constructed using just elements of the symmetry algebra under consideration, without other specific knowledge of the particular theory.}
when the symmetry action on the Hilbert space gives rise to a \emph{strongly-separable algebra}. Consider a unital associative algebra $\cA$ over the complex field $\bC$ defined by the product
\be
\label{algebra product}
\cX_a \times \cX_b = \sum\nolimits_c T_{ab}^c \, \cX_c \;,
\ee
where $\{\cX_a\}$ is a basis while $T_{ab}^c \in \bC$ are the structure coefficients. We will often omit the product sign $\times$.
Associativity corresponds to the identity $\sum_m T_{ab}^m \, T_{mc}^d = \sum_m T_{am}^d \, T_{bc}^m$.
We indicate the unit element, which in general is a linear combination of the basis elements, as $\cX_\unit$.
The regular representation is defined by the left action of the algebra on itself, and is explicitly given by the matrices $(T_a)\ud{c}{b} = T_{ab}^c$.
Define a trace $\Tr\colon \cA \to \bC$ for every element $X = \sum_a x^a \cX_a \in \cA$ as
\be
\Tr(X) = \Tr \Bigl( {\ts \sum_a x^a \cX_a} \Bigr) \equiv \sum\nolimits_a x^a \Tr(T_a) = \sum\nolimits_{am} x^a \, T_{am}^m \;.
\ee
Then consider the symmetric bilinear form
\be
\label{bilinear form}
K_{ab} = \Tr(\cX_a \cX_b) = \Tr(T_a T_b) = \sum\nolimits_{mn} T_{an}^m \, T_{bm}^n \;.
\ee
As explained above, in this paper we will be dealing with finite-dimensional unital semisimple algebras over the complex field $\bC$. For such algebras, using the Wedderburn--Artin theorem (see \cref{app: strongly separable algebras}), one can verify that the bilinear form (\ref{bilinear form}) is non-degenerate. The inverse matrix $K^{-1} \equiv \tilde K$ allows one to construct a special element $e \in \cA \otimes \cA$ called the \emph{symmetric separability idempotent} (SSI):
\be
\label{def SSI}
e = \sum\nolimits_{ab} \tilde K^{ab} \, \cX_a \otimes \cX_b \;,
\ee
whose defining properties will be discussed shortly. An algebra for which such an object exists is called strongly separable. One can also prove (we give some details in \cref{app: strongly separable algebras}) that the structure coefficients satisfy the pivotal relation
\be
\label{pivotal relation}
T_{ca}^b = \sum\nolimits_{b'c'} \tilde K^{bb'} \, K_{cc'} \, T_{ab'}^{c'} \;.
\ee
Besides, one can explicitly construct the identity (or unit) element
\be
\label{identity element}
\cX_\unit = \sum\nolimits_{abc} T_{ab}^b \, \tilde K^{ac} \cX_c = \sum\nolimits_{abc} \tilde K^{ab} \, T_{ab}^c \, \cX_c
\ee
and verify that $\cX_\unit \cX_a = \cX_a \cX_\unit = \cX_a$ for all $\cX_a \in \cA$, as well as $\Tr( \cX_\unit) = \dim\cA$. The SSI has the following properties.
First, it is symmetric under the exchange of the two factors in the tensor product. Second, denoting the product as $XY = \mu(X\otimes Y)$ in terms of $\mu\colon \cA \otimes\cA \to \cA$, the contraction of the SSI gives the unit element: $\mu(e) = \cX_\unit$. Third, the SSI commutes with every element of the algebra, $\cX_a \, e = e \, \cX_a$, where multiplication from the left/right acts on the first/second factor. Fourth, the SSI is idempotent as an element of $\cA \otimes \cA^\text{op}$ (where $\cA^\text{op}$ is the same as $\cA$ but elements are multiplied in the opposite order): $e^2 = e$.
The one in (\ref{def SSI}) is the unique element with these properties.

The algebras we are interested in are also $C^*$-algebras and in particular have a $*$-structure, namely an anti-linear involution $\dag\colon \cA \to \cA$ such that $(XY)^\dag = Y^\dag X^\dag$. By a version of the Wedderburn--Artin theorem \cite{Murphy:1990}, one can show that the SSI is Hermitian, $e^\dag = e$, and that there exists a special basis $\{\cX_{\underline a}\}$ in which it takes the simple form $e = \sum_{\underline a} \cX_{\underline a}^\dag \otimes \cX_{\underline a}$. See \cref{app: strongly separable algebras} for more details.

The SSI provides then the general formula for the symmetrizer $\sS\colon \rho_\sS = e(\rho)$, where the first factor of $e$ acts on $\rho$ from the left while the second factor from the right. More explicitly:
\be
\label{symmetrizer}
\boxed{\;\rule[-0.8em]{0pt}{2em}
\rho_\sS = \sum\nolimits_{ab} \tilde K^{ab} \, \cX_a \, \rho \, \cX_b
\;} \;.
\ee
One can easily verify that the properties of the SSI guarantee the properties 1.~--~4.\ of the symmetrizer. Besides, assuming that $\rho^\dag = \rho$ and $\rho \geq 0$, one verifies that $\rho_\sS^\dag = \rho_\sS$ (using that $e^\dag = e$) and that $\rho_\sS \geq 0$ (using the simple form of $e$ in the special basis). In fact, it turns out that $\rho_\sS$ is a trace-preserving quantum channel. As we will show, one can also find equivalent alternative expressions for the SSI when the algebra $\cA$ has extra properties or extra structure. For instance, for commutative algebras one can rewrite (\ref{symmetrizer}) as a sum over projectors on one-dimensional representations, while for Hopf and weak Kac algebras one can rewrite (\ref{symmetrizer}) in terms of the coproduct and antipode operations.

In the special case that the symmetry is a standard 0-form invertible symmetry given by a group $G$,%
\footnote{In the modern language of \cite{Gaiotto:2014kfa}, a 0-form symmetry is a standard symmetry that acts on local operators and is implemented by spacetime-codimension-1 topological operators.}
previous literature has studied the simplest algebra $\cA$ that can describe the symmetry action, namely the group algebra $\bC[G]$, which is the fusion algebra associated to this symmetry. Its basis elements (and generators) $\cX_g$ correspond to the group elements $g \in G$ and to the simple objects $\cL_g$ of the fusion category $\mathsf{Vec}_{G}$. The structure coefficients are $T_{gh}^k = \delta_{(gh)}^k$, where $(gh)$ is the group multiplication. The bilinear form is then $K_{gh} = |G| \, \delta_{g, h^{-1}}$ and its inverse is $\tilde K^{gh} = |G|^{-1} \, \delta^{g, h^{-1}}$. The symmetrizer reads:
\be
\rho_\sS = \frac1{|G|} \, \sum\nolimits_{g \in G} \cL_g \, \rho \, \cL_{g^{-1}} \;.
\ee
This expression agrees with what already used in previous literature (\eg \cite{Vaccaro:2007drw, Ares:2022koq, Ferro:2023sbn, Capizzi:2023yka}).

For the symmetrizer (\ref{symmetrizer}) it is easy to prove that the two definitions (\ref{def asymmetry as relative entr}) and (\ref{def asymmetry as von Neumann}) coincide. It is enough to notice that
\be
\label{equality of traces}
\Tr\bigl( \rho_\sS \log \rho_\sS \bigr) = \sum\nolimits_{ab} \tilde K_{ab} \Tr \bigl( \cX_a \, \rho \, \cX_b \log \rho_\sS \bigr) = \sum\nolimits_{ab} \tilde K_{ab} \Tr \bigl( \rho \, \cX_b \log(\rho_\sS) \, \cX_a \bigr) = \Tr \bigl( \rho \, (\log \rho_\sS)_\sS \bigr) \;,
\ee
but $( \log \rho_\sS)_\sS = \log \rho_\sS$ because $\log \rho_\sS$ commutes with the algebra, as does $\rho_\sS$.

Finally, given the algebra $\cA$ corresponding to a chosen symmetry action, to what extent is the symmetrizer \eqref{symmetrizer} unique? If we require that the symmetrizer be expressible solely in terms of the symmetry operators $\cX_a$ --- \ie, that it be universal --- then the uniqueness of the SSI $e$ implies the uniqueness of $\sS$. Equivalently, as we show in \cref{sec:schur}, the form of the symmetrized density matrix is fixed as in \cref{fig: action on matrices}, and this completely specifies the action of the symmetrizer. One could relax the requirement of universality and look for more general maps (which would require theory-specific operators). Yet one can argue that $\sS$, being a self-adjoint projector onto the space of operators that commute with $\cA$, is completely determined.


\section{Asymmetry for the circle Hilbert space}
\label{sec: circle}

The construction in the previous section is general and it allows one to define the asymmetry for any type of finite internal symmetry, including higher-form \cite{Gaiotto:2014kfa} and noninvertible \cite{Bhardwaj:2017xup} generalized symmetries. In the following, however, we will present concrete constructions focusing on the case of two-dimensional unitary theories. In two dimensions, one can reduce to the case that only 0-form symmetries are present \cite{Hellerman:2006zs, Sharpe:2022ene}. Hence, the internal symmetries of a 2d unitary theory are described by a unitary fusion category $\scrC$ whose objects are the symmetry elements, namely the topological lines that implement the action of the symmetry in the theory.
We call $\{\cL_a\}$ the simple lines, that are finite in number.

In this work we study two natural Hilbert spaces: the circle Hilbert space and the interval Hilbert space for a choice of boundary conditions. Three algebras will play a central role: the fusion algebra of $\scrC$, the tube algebra $\mathsf{Tube}(\scrC)$, and the strip algebra $\mathsf{Strip}_\scrC(\scrM)$ for a chosen $\scrC$-module category $\scrM$. The first two act on the circle Hilbert space, while the latter acts on the interval Hilbert space.
We first consider the case that the two-dimensional theory is on a spatial circle $S^1$ and we are interested in the asymmetry of a (possibly mixed) state $\rho$ of the full system. We study the richer case of the interval Hilbert space (relevant for studying the asymmetry of subsystems, or the asymmetry of a theory on the interval) in \cref{sec: interval}. In the limit that the radius of $S^1$ goes to infinity, one can study the asymmetry of states at infinite volume. In particular, for CFTs or gapped systems (TQFTs) one can use conformal transformations or diffeomorphisms, respectively, to obtain the asymmetry of states on the infinite real line $\bR$.%
\footnote{For generic QFTs, on $\bR$ one usually chooses boundary conditions at infinity that fix a particular vacuum state that satisfies cluster decomposition. The infinite-volume limit of $S^1$, in contrast, keeps the Hilbert space constructed over the whole vector space of degenerate vacua.}

The simplest algebra $\cA$ to study is the one obtained by laying the topological lines on a spatial slice and letting them act on the untwisted Hilbert space $\cH_1$: this is the standard fusion algebra associated to the fusion category $\scrC$ (\cref{sec: fusion alg}). A more sophisticated algebra, that allows for a more accurate detection of symmetry breaking, is obtained by acting on the total Hilbert space $\cH_\text{Tube}$ of all untwisted and twisted sectors: this is the tube algebra of $\scrC$ (\cref{sec: tube alg}).

\subsection{Fusion algebra}
\label{sec: fusion alg}

Associated to every fusion category $\scrC$ is a fusion algebra $\cA$ spanned by the objects $\cL_a$. Such an algebra is unital and associative and takes the general form (\ref{algebra product}), however in the basis in which $\cX_a$ are the simple lines $\cL_a$, the coefficients $T_{ab}^c$ are constrained to be nonnegative integer numbers $N_{ab}^c \in \bN$. We thus write the fusion algebra as
\be
\label{fusion algebra}
\cL_a \times \cL_b = \sum\nolimits_c N_{ab}^c \, \cL_c \;.
\ee
Besides, associated to the rigid structure of $\scrC$ is a conjugation operation on $\cA$: there exists an element of the chosen basis, that we call $\cL_1$, such that the matrix $C_{ab} \equiv N_{ab}^1$ is involutive, namely $C^2 = \unit$. Since the coefficients are natural numbers, this implies that for every $a$ there exists a unique $\bar a$ such that $N_{ab}^1 = \delta_{b\bar a}$ and $\bar{\bar a} = a$. Conjugation is then the map
\be
\cL_a \;\mapsto\; \cL_a^\dag \equiv \cL_{\bar a} = \sum\nolimits_b C_{ab} \, \cL_b = \sum\nolimits_b N_{ab}^1 \, \cL_b \;.
\ee
This is also called duality. Conjugation is required to be an algebra anti-automorphism: $N_{\bar a \bar b}^{\,\bar c} = N_{ba}^c$.
This is then extended to the whole fusion algebra by anti-linearity. Conjugation and associativity allow one to derive that $N_{ab}^c = N_{b \bar c}^{\,\bar a}$, hence one obtains the pivotal relations
\be
N_{ab}^c = N_{\bar a c}^b = N_{c \bar b}^a = N_{b \bar c}^{\bar a} = N_{\bar c a}^{\bar b} = N_{\bar b \bar a}^{\bar c} \;.
\ee
When applied to the conjugation matrix, they imply that $N_{1a}^b = N_{a1}^b = \delta_a^b$ so that $\cL_1$ is in fact the unit element for the fusion algebra, it is unique and self-dual, and indeed it corresponds to the transparent identity line of $\scrC$.

\begin{figure}[t]
\centering
\begin{tikzpicture}[line width=0.6]
        \draw [
        \colortopoline,
        postaction=decorate,
        decoration={markings, mark=at position 0.5 with \arrow{>} }
        ] (-1.2, 0.9) arc [start angle = -180, end angle = 0, x radius = 1.2, y radius = 0.25] node[right] {\small$\cL_a$};
        \draw [\colortopoline, dashed] (-1.2, 0.9) arc [start angle = 180, end angle = 0, x radius = 1.2, y radius = 0.25];
        \draw [
        \colortopoline,
        postaction=decorate,
        decoration={markings, mark=at position 0.5 with \arrow{>} }
        ] (-1.2, 0) arc [start angle = -180, end angle = 0, x radius = 1.2, y radius = 0.25] node[right] {\small$\cL_b$};
        \draw[\colortopoline, dashed] (-1.2, 0) arc [start angle = 180, end angle = 0, x radius = 1.2, y radius = 0.25];
        \draw[\colorstate] (-1.2, -0.9) arc [start angle = -180, end angle = 0, x radius = 1.2, y radius = 0.25] node[right] {\small$|\psi\rangle$};
        \draw[\colorstate, dashed] (-1.2, -0.9) arc [start angle = 180, end angle = 0, x radius = 1.2, y radius = 0.25];
        \draw (-1.2, -1.2) -- (-1.2, 1.2);
        \draw (1.2, -1.2) -- (1.2, 1.2);
        \draw [->] (-0.5, 0.6) -- +(0, -0.2);
        \draw [->] (0.5, 0.6) -- +(0, -0.2);
        \draw [->] (-0.5, -0.3) -- +(0, -0.2);
        \draw [->] (0.5, -0.3) -- +(0, -0.2);
        \begin{scope}[shift={(8,0)}]
            \draw [fill] (0,0) node[shift={(0.3, 0)}] {\footnotesize$\cO$} circle [radius = 0.04];
            \draw [
            \colortopoline,
            postaction=decorate, decoration={markings, mark=at position 0.25 with \arrow{<}}
            ] circle [y radius = 0.6, x radius = 0.6] node[shift={(0:0.85)}] {\footnotesize$\cL_b$};
            \draw [
            \colortopoline,
            postaction=decorate, decoration={markings, mark=at position 0.25 with \arrow{<}}
            ] circle [y radius = 1.3, x radius = 1.3] node[shift={(0:1.55)}] {\footnotesize$\cL_a$};
            \foreach \i in {0,...,3} {
              \draw [->] (135+90*\i:0.55) -- +(-45+90*\i:0.2);
              \draw [->] (135+90*\i:1.25) -- +(-45+90*\i:0.2);
            }
        \end{scope}
\end{tikzpicture}
\caption{\label{fig: fusion algebra action}%
Left: action of the fusion algebra \eqref{fusion algebra} of the symmetry category $\scrC$ on the untwisted Hilbert space on $S^1$.
The product of two elements of the algebra is realized by stacking the corresponding lines on the cylinder and fusing them.
Right: the corresponding action of the fusion algebra on local operators.}
\end{figure}

The fusion algebra describes the action of the symmetry on the untwisted Hilbert space $\cH_1$ on $S^1$, as depicted in \cref{fig: fusion algebra action} on the left. It also describes the action of the symmetry on untwisted local operators, as depicted in \cref{fig: fusion algebra action} on the right.%
\footnote{In CFTs there is a one-to-one correspondence between untwisted states on $S^1$ and untwisted local operators. In generic QFTs there is no such a correspondence, but it is still true that both states and local operators transform in representations of the fusion algebra.}
One can prove (as we review in \cref{app: C* algebras}) that finite-dimensional fusion algebras are automatically $C^*$-algebras and thus semisimple and strongly separable. Therefore, if we are interested in defining and computing the asymmetry of (possibly mixed) states on $\cH_1$ with respect to the fusion algebra, we can define the symmetrizer $\sS$ --- and hence entanglement and R\'enyi asymmetries --- using the universal formula (\ref{symmetrizer}). We will describe some concrete results in \cref{sec: full system in CFT,sec: SSB}, and present detailed computations in various examples in \cref{sec: examples}.

\paragraph{Commutative case.}
In the case of an Abelian algebra, the symmetrizer $\sS$ can be described in an alternative but equivalent way, that we now present. The irreducible representations of a commutative algebra are one-dimensional. Therefore, the blocks shown in \cref{fig: action on matrices} are one-dimensional and the symmetrizer simply acts by setting to zero all blocks that connect different irreducible representations. Let $P_r\colon \cH \to \cH$ denote the projector onto the $r$-th irreducible representation. The symmetrizer then reads$\,$%
\footnote{The expression (\ref{eq:symmetrizer-1d-irreps}) agrees with the formula originally proposed in \cite{Ares:2022koq} for invertible Abelian symmetries.}
\be
\label{eq:symmetrizer-1d-irreps}
\rho_\sS = \sum\nolimits_r P_r \, \rho \, P_r \;.
\ee
If we further assume that the algebra is a commutative \emph{fusion algebra}, an explicit formula for $P_r$ in terms of simple lines is available. This was discussed in \cite{Lin:2022dhv} in the context of fusion algebras of modular fusion categories, but in fact it holds more generally for arbitrary fusion algebras. In the following we review the construction, partially based on \cite{Fuchs:1993et}.

The matrices $(N_a)\ud{b}{c} \equiv N_{ac}^b$ give the regular representation of the algebra. They satisfy $N_{\bar a} = N_a^{\,\sT} = N_a^{\,\dag}$ and in the Abelian case they all commute. It follows that they are normal matrices and can be simultaneously diagonalized with a unitary transformation. Let us call $w^{(r)}$ the common eigenvectors, that are orthogonal with respect to the standard Hermitian product, and $\nu_{a (r)}$ the eigenvalues:
\be
N_a \, w^{(r)} = \nu_{a(r)} \, w^{(r)} \;.
\ee
Note that $r = 1, \dots, \dim\cA$. For each $r$, the eigenvalues provide a one-dimensional representation of the fusion algebra, namely
\be
\label{1d reps of fusion algebra}
\nu_{a(r)} \, \nu_{b(r)} = \sum\nolimits_c N_{ab}^c \, \nu_{c(r)} \;.
\ee
This also shows that the eigenvectors can be taken to be equal to the eigenvalues: $w^{(r)}_a = \nu_{\bar a(r)}$, as one can check. Orthogonality of the vectors implies that $\sum_a \nu_{a(r)}^* \, \nu_{a(m)} = | s_r |^{-2} \, \delta_{r m}$ for some numbers $s_r$ that we can take in $\bR_{>0}$. One defines the matrix
\be
S_{ar} = \nu_{a(r)} \, s_r \;.
\ee
Note that $\nu_{1(r)} = 1$ in all representations $r$ since $\cL_1$ is the identity, thus $s_r = S_{1r}$. The action of the fusion algebra on its representations is usually written as
\be
\label{reps of Abelian fusion alg}
N_a \, w^{(r)} = \frac{S_{ar}}{S_{1r}} \, w^{(r)} \;,
\ee
and the matrix $S_{ar}$ is called the $S$-matrix. One checks that it is unitary, $S^\dag S = \unit$, and therefore also $S S^\dag = \unit$. Such identities correspond to the orthogonality and completeness of characters. The relations (\ref{1d reps of fusion algebra}) can then be written as
\be
\label{Verlinde formula}
N_{ab}^c = \sum\nolimits_r \frac{ S_{ar} \, S_{br} \, S_{cr}^* }{ S_{1r} } \;,
\ee
which is the Verlinde formula. It also shows that $S^\sT$ diagonalizes the matrices $N_a$, and that $\nu_{\bar a(r)} = \nu_{a(r)}^*$. Notice that when $\scrC$ is a modular fusion category, the $S$-matrix is also symmetric, but in general it is not.

The projector onto the $r$-th irreducible representation is then given by
\be
\label{eq:projector-commutative-fusion-algebra}
P_r = S_{1r} \sum\nolimits_a S_{ar}^* \, \cL_a \;.
\ee
One can verify, using the Verlinde formula and unitarity of the $S$-matrix, that the set $\{ P_r \}$ forms a family of orthogonal projectors, \ie\ they satisfy $P_r P_s = \delta_{rs} P_r$, and that if $| \psi \rangle$ belongs to the $r$-th irreducible representation then $P_{r'} | \psi \rangle = \delta_{rr'} | \psi \rangle$.
Finally, one can  verify that the expression \eqref{eq:symmetrizer-1d-irreps} with $P_r$ given by \eqref{eq:projector-commutative-fusion-algebra} agrees with \eqref{symmetrizer}.%
\footnote{The expression in \eqref{eq:symmetrizer-1d-irreps} is $\rho_\sS = \sum_{ab} \Bigl( \sum_r S_{1r}^2 S_{ar}^* S_{br}^* \Bigr) \, \cL_a \rho \cL_b$. We need to show that $\sum_r S_{1r}^2 S_{ar}^* S_{br}^* = \tilde K^{ab}$. This is easily verified by multiplying both sides by the matrix $K_{bc}$, using its definition (\ref{bilinear form}) and the Verlinde formula \eqref{Verlinde formula} twice.}

Note that for a noncommutative algebra the formula \eqref{eq:symmetrizer-1d-irreps} would fail. From \cref{fig: action on matrices} it is clear that this would only put the density matrix in a block-diagonal form, without further projection to the identity matrix inside each sub-block.
In \cref{sec: Haagerup} we study the simplest noninvertible and non-Abelian example: the Haagerup fusion algebra $H_{(3)}$.

\subsection{Tube algebra}
\label{sec: tube alg}

On a spatial $S^1$, a noninvertible symmetry can map (un)twisted sectors to other (un)twisted sectors. One can thus construct a more sophisticated and complete action of the symmetry on the total Hilbert space
\be
\label{eq:htube}
\cH_\text{Tube} = \bigoplus\nolimits_{a \,\in\, \Irr(\scrC)} \cH_a \;,
\ee
where $\cH_a$ are the Hilbert spaces on $S^1$ twisted by the lines $\cL_a$ (and thus $\cH_1$ is the untwisted sector), and $\Irr(\scrC)$ is the set of simple objects in $\scrC$. The finite-dimensional $C^*$-algebra $\cA$ that is obtained in this way is called the \emph{tube algebra} $\mathsf{Tube}(\scrC)$ \cite{Ocneanu:1994}. This algebra does not only depend on the fusion rules of $\scrC$, but rather also on its $F$-symbols (which are the components of the associator).

In order to describe the algebra, let us first fix our notation (that follows \cite{Barkeshli:2014cna}). The nonnegative integers $N_{ab}^c$ represent the dimensions of the vector spaces of morphisms between lines: $N_{ab}^c = \dim\bigl( \Hom(\cL_a \otimes \cL_b, \cL_c) \bigr) = \dim\bigl( \Hom(\cL_{\bar c}, \cL_{\bar b} \otimes \cL_{\bar a}) \bigr)$. The second equality is because conjugation corresponds to inverting the direction of a line:
\bea
\begin{tikzpicture}[line width=0.6]
        \draw [-{>[width=3mm, length=2mm]}] (-1.7, -0.6) -- (-1.7, 0.1) node[right, shift={(0.2, -0.08)}] {\small$\bar a$};
        \draw (-1.7, -0.6) -- (-1.7, 0.6);
        \node at (0,0) {$=$};
        \draw [-{>[width=3mm, length=2mm]}] (1.5, 0.6) -- (1.5, -0.1) node[right, shift={(0.2, 0.1)}] {\small$a$};
        \draw (1.5, -0.6) -- (1.5, 0.6);
\end{tikzpicture}
\eea
which physically is realized by a 180-degree rotation. The tensor product is associative, however it involves a change of basis in the vector spaces of morphisms described by the $F$-symbols:
\bea
\label{def F-symbols}
\begin{tikzpicture}[line width=0.6]
        \draw (-0.7, -1.2) node[left] {\small$a$} -- (0,0) node[shift={(-0.4, -0.2)}] {\small$e$} -- (0.7, -1.2) node[right] {\small$c$};
        \draw (0, -1.2) node[right] {\small$b$} -- (-0.35, -0.6) node[shift={(0.23, 0)}] {\scriptsize$\mu$};
        \draw (0,0) node[shift={(0.2, 0)}] {\scriptsize$\nu$} -- (0, 0.4) node[left] {\small$d$};
        \node at (3.3, -0.6) {$\ds{} = \; \sum_{f, \rho, \sigma} \; \bigl[ F_{abc}^d \bigr]_{(e; \mu\nu)(f; \rho\sigma)}$};
\begin{scope}[shift={(6.5,0)}]
        \draw (-0.7, -1.2) node[left] {\small$a$} -- (0,0) node[shift={(0.4, -0.2)}] {\small$f$} -- (0.7, -1.2) node[right] {\small$c$};
        \draw (0, -1.2) node[left] {\small$b$} -- (0.35, -0.6) node[shift={(-0.2, 0)}] {\scriptsize$\rho$};
        \draw (0,0) node[shift={(-0.15, 0.05)}] {\scriptsize$\sigma$} -- (0, 0.4) node[right] {\small$d$};
\end{scope}
        \node at (8, -0.65) {.};
\end{tikzpicture}
\eea
Here and in the following, unless stated otherwise, all lines have implicit arrows oriented from bottom to top. Roman letters label the lines, while Greek letters label the basis of morphisms.
The $F$-symbol $F_{abc}^d$ is an element of $\End\bigl( \Hom(\cL_a \otimes \cL_b \otimes \cL_c, \cL_d) \bigr)$ represented by an invertible square matrix of size $N_{abc}^d = \sum_e N_{ab}^e \, N_{ec}^d$. As in \cite{Barkeshli:2014cna}, we choose the bases of morphisms in such a way that the $F$-symbols are \emph{unitary} matrices%
\footnote{The existence of such bases is guaranteed by the fact that the fusion category is unitary. Without such an assumption, our formulas below would still be correct but in terms of inverse matrices, as opposed to Hermitian conjugate matrices.}
and $[F_{abc}^d] = \unit$ whenever $a$, $b$, or $c$ is $1$ (and fusion is possible). The $F$-symbols satisfy the pentagon equation. We spell it out, together with other details, in \cref{app: F-symbols}. The bases are normalized so that the following hold:
\bea
\label{fusion and q dim}
\begin{tikzpicture}[line width=0.6]
    \draw (-0.3, -0.7) node[left] {\small$a$} -- (-0.3, 0.7);
    \draw (0.3, -0.7) node[right] {\small$b$} -- (0.3, 0.7);
    \node at (2, 0) {$\ds{} = \sum_{c,\mu} \sqrt{ \dfrac{d_c}{d_ad_b}}$};
\begin{scope}[shift={(4.3,0)}]
    \draw (-0.3, -0.7) node[left] {\small$a$} -- (0, -0.3) node[shift={(-0.2,0)}] {\scriptsize$\mu$} -- (0, 0.3) node[shift={(-0.2,0)}] {\scriptsize$\mu$} -- (-0.3, 0.7) node[left] {\small$a$};
    \draw (0.3, -0.7) node[right] {\small$b$} -- (0, -0.3) -- (0, 0.3) node[pos=0.5, right] {\small$c$} -- (0.3, 0.7)  node[right] {\small$b$};
    \node at (1, -0.1) {,};
\end{scope}
\begin{scope}[shift={(8,0)}]
        \draw (0, -0.7) node[left] {\small$c$} -- (0, -0.4) node[shift={(0, 0.17)}] {\scriptsize$\mu$};
        \draw (0, 0.4) node[shift={(0, -0.17)}] {\scriptsize$\nu$} -- (0, 0.7) node[left] {\small$d$};
        \draw (0,0) circle [radius = 0.4];
        \node at (-0.6, 0) {\small$a$};
        \node at (0.6, 0) {\small$b$};
        \node at (2.4,0) {$\ds{} = \delta_{cd} \, \delta_{\mu\nu} \, \sqrt{ \frac{d_a d_b}{d_c}}$};
        \draw (4.5, -0.7) node[left] {\small$c$} -- (4.5, 0.7);
        \node at (5, -0.1) {.};
\end{scope}
\end{tikzpicture}
\eea
The positive real numbers $d_a$ are the quantum dimensions, which are associated to each line $\cL_a$ and furnish a 1d representation of the fusion algebra: $d_a d_b = \sum_c N_{ab}^c \, d_c$.

\begin{figure}[t]
\centering
\begin{tikzpicture}[line width=0.6]
        \draw [
        postaction=decorate,
        decoration={
          markings,
          mark=at position 0.15 with \arrow{>},
          mark=at position 0.4 with \arrow{>},
          mark=at position 0.85 with \arrow{>}
        },
        \colortopoline
        ] (0, -1.5) -- (0,1);
        \draw [
        \colortopoline,
        postaction=decorate,
        decoration={
          markings,
          mark=at position 0.25 with \arrow{>},
          mark=at position 0.75 with \arrow{>}
        },
        ] (-1.2, 0.1) node[left] {\small$\cL_a$} arc [start angle = -180, end angle = 0, x radius = 1.2, y radius = 0.25] node[right] {\small$\cL_a$};
        \draw [dashed, \colortopoline] (-1.2, 0.1) arc [start angle = 180, end angle = 0, x radius = 1.2, y radius = 0.25];
        \draw [\colorstate] (-1.2, -0.7) arc [start angle = -180, end angle = 0, x radius = 1.2, y radius = 0.25] node[right] {\small$|\psi_g\rangle$};
        \draw [\colorstate, dashed] (-1.2, -0.7) arc [start angle = 180, end angle = 0, x radius = 1.2, y radius = 0.25];
        \draw (-1.2, -1.2) -- (-1.2, 1.2);
        \draw (1.2, -1.2) -- (1.2, 1.2);
        \node[\colortopoline, right] at (0,-1.4) {$\cL_g$};
        \node[\colortopoline, right] at (0,-0.6) {$\cL_g$};
        \node[\colortopoline, right] at (0,0.9) {$\cL_h$};
        \filldraw [\colortopojun] (0, -0.15) node[shift={(130: 0.3)}] {\footnotesize$\Delta$} circle [radius=0.06];
\begin{scope}[shift={(8,-0.3)}]
            \draw[
            \colortopoline,
            postaction=decorate, decoration={markings, mark=at position 0.16 with \arrow{<}, mark=at position 0.35 with \arrow{<}}
            ] (0,0) circle [radius=0.9];
            \draw[
            \colortopoline,
            postaction=decorate, decoration={markings, mark=at position 0.33 with \arrow{>}, mark=at position 0.83 with \arrow{>}}
            ] (0,0) -- (0, 1.6);
            \draw [fill] (0,0) node[shift={(0.35, 0)}] {\footnotesize$\cO_g$} circle [radius=0.04];
            \filldraw [\colortopojun] (0, 0.9) node[shift={(40: 0.32)}] {\scriptsize$\Delta$} circle [radius=0.06];
            \node[\colortopoline, right] at (0.85, 0.5) {$\cL_a$};
            \node[\colortopoline, left = -0.5mm] at (0, 0.35) {$\cL_g$};
            \node[\colortopoline, left = -0mm] at (0, 1.5) {$\cL_h$};
\end{scope}
\end{tikzpicture}
\caption{\label{fig: tube algebra}%
Left: action of the tube algebra $\mathsf{Tube}(\scrC)$ on the total Hilbert space  on $S^1$ (left) and on (un)twisted point operators (right).
The product of two elements of the algebra is realized by stacking the corresponding elements on the cylinder and then resolving the configuration.}
\end{figure}

The tube algebra is generated by topological configurations in which a simple line $\cL_a$ --- stretched horizontally --- can interpolate between two (un)twisted sectors $\cH_g$ and $\cH_h$, as in \cref{fig: tube algebra} left. Equivalently, the generators map (un)twisted local operators $\cO_g$ (that live at the end of a line $\cL_g$) to (un)twisted operators $\cO_h$ via the so-called lasso action, as in \cref{fig: tube algebra} right. The configuration is described by a ``cross'', where the horizontal line $\cL_a$ is closed in a circle, while the vertical line $\cL_g$ lies below and $\cL_h$ lies above. The intersection $\Delta$ is an element of $\Hom( \cL_a \otimes \cL_g, \cL_h \otimes \cL_a)$, which has dimension $\sum_d N_{ag}^d \, N_{ha}^d$ (note that $N_{ha}^d = N_{\bar hd}^a$). It is convenient to decompose the intersection into trivalent junctions:
\be
\label{tube algebra basis}
\begin{tikzpicture}[line width=0.6, baseline=\mathaxis]
        \draw [line width = 0.4] (-0.9, -0.7) -- (-0.9, 0.9);
        \draw [line width = 0.4] (0.9, -0.7) -- (0.9, 0.9);
        \draw [\colortopoline] (0, -0.7) node[right] {\small$g$} -- (0, 0.9) node[right] {\small$h$};
        \draw [\colortopoline,->] (0, -0.7) -- (0, -0.3);
        \draw [\colortopoline,->] (0,0) -- (0, 0.7);
        \draw [\colortopoline] (-0.9, 0.2) node[left] {\small$a$} arc [start angle = -180, end angle = 0, x radius = 0.9, y radius = 0.2] node[right] {\small$a$};
        \draw [\colortopoline,->] (-0.9, 0.2) arc [\colortopoline, start angle = -180, end angle = -120, x radius = 0.9, y radius = 0.2];
        \draw [\colortopoline,->] (0,0) arc [\colortopoline, start angle = -90, end angle = -55, x radius = 0.9, y radius = 0.2];
        \draw [\colortopoline, dashed] (-0.9, 0.2) arc [start angle = 180, end angle = 0, x radius = 0.9, y radius = 0.2];
        \filldraw[\colortopojun] (0,0) node[shift={(130: 0.28)}] {\scriptsize$\Delta$} circle [radius = 0.06];
\end{tikzpicture}
\quad\equiv\quad
\begin{tikzpicture}[line width=0.6, baseline=\mathaxis]
        \draw (-0.2, -0.7) node[right] {\small$g$} -- (-0.2, -0.2) -- (0.2, 0.2) node[pos=0.6, shift={(135: 0.25)}] {\small$d$} -- (0.2, 0.7) node[right] {\small$h$};
        \draw [|-] (-0.7, -0.3) node[left] {\small$a$} -- (-0.2, -0.2) node[shift={(0.17, -0.07)}] {\scriptsize$\mu$};
        \draw [-|] (0.2, 0.2) node[shift={(0.07, -0.17)}] {\scriptsize$\nu$} -- (0.7, 0.3) node[right] {\small$a$};
\end{tikzpicture}
\;.
\ee
We will use the graphical notation on the \rhs, and use the indicated basis for the tube algebra. All lines have an implicit arrow from bottom to top, and the tick at the beginning or end of a line indicates the ``trace'' described in \cite{Barkeshli:2014cna}. The dimension of the tube algebra is
\be
\dim \bigl( \mathsf{Tube}(\scrC) \bigr) = \sum\nolimits_{g,a,d,h} \, N_{ag}^d \, N_{ha}^d \;.
\ee
In the tube algebra we multiply elements by placing one cylinder on top of the other (the product is such that bottom acts first, top acts second). The product is zero if there is no possible junction between adjacent vertical lines. In the case of the basis elements (\ref{tube algebra basis}), because only simple lines are involved, this forces the vertical lines to be the same.
We define a product matrix:
\be
\label{def product tube algebra}
\begin{tikzpicture}[line width=0.6, baseline=\mathaxis]
    \draw (-0.4, -1.3) node[right] {\small$g$} -- (-0.4, -0.8) -- (0, -0.4) node[pos=0.6, shift={(135: 0.25)}] {\small$d$} -- (0, 0.4) node[pos=0.45, shift={(0.15, 0)}] {\small$h$} -- (0.4, 0.8) node[pos=0.6, shift={(135: 0.22)}] {\small$e$} -- (0.4, 1.3) node[right] {\small$k$};
    \draw [|-] (-0.9, -0.9) node[left] {\small$a$} -- (-0.4, -0.8) node[shift={(0.17, -0.07)}] {\scriptsize$\mu$};
    \draw [-|] (0, -0.4) node[shift={(0.07, -0.17)}] {\scriptsize$\nu$} -- (0.9, -0.22) node[right] {\small$a$};
    \draw [|-] (-0.9, 0.22) node[left] {\small$b$} -- (0, 0.4) node[shift={(0.17, -0.05)}] {\scriptsize$\rho$};
    \draw [-|] (0.4, 0.8) node[shift={(0.07, -0.17)}] {\scriptsize$\sigma$} -- (0.9, 0.9) node[right] {\small$b$};
\end{tikzpicture}
\quad = \sum_{c,s,\alpha,\beta} \; M_{g, a, (d;\mu\nu), h, b, (e; \rho\sigma), k}^{c, (s; \alpha\beta)} \quad
\begin{tikzpicture}[line width=0.6, baseline=\mathaxis]
    \draw (-0.2, -0.7) node[right] {\small$g$} -- (-0.2, -0.2) -- (0.2, 0.2) node[pos=0.6, shift={(135: 0.25)}] {\small$s$} -- (0.2, 0.7) node[right] {\small$k$};
    \draw [|-] (-0.7, -0.3) node[left] {\small$c$} -- (-0.2, -0.2) node[shift={(0.17, -0.1)}] {\scriptsize$\alpha$};
    \draw [-|] (0.2, 0.2) node[shift={(0.07, -0.2)}] {\scriptsize$\beta$} -- (0.7, 0.3) node[right] {\small$c$};
\end{tikzpicture} \;.
\ee
The computation of $M$ is performed in \cref{app: F-symbols}, and the result is:
\be
\label{product tube algebra}
M_{g, a, (d;\mu\nu), h, b, (e; \rho\sigma), k}^{c, (s; \alpha\beta)} = \sqrt{ \frac{ d_a d_b d_h d_s}{ d_d d_e d_c}} \, \sum_{\gamma, \eta, \lambda} \, \bigl[ F_{kba}^s \bigr]^*_{(e; \sigma \eta)(c; \gamma\beta)} \, \bigl[ F_{bha}^s \bigr]_{(e; \rho\eta)(d; \nu\lambda)} \, \bigl[ F_{bag}^s \bigr]^*_{(c; \gamma\alpha)(d; \mu\lambda)} \,.
\ee
The sums are over a basis of morphisms $\cL_b \otimes \cL_a \xrightarrow{\gamma} \cL_c$, $\cL_e \otimes \cL_a \xrightarrow{\eta} \cL_s$, and $\cL_b \otimes \cL_d \xrightarrow{\lambda} \cL_s$. In the special case that all $N_{ab}^c \in \{0,1\}$ and thus junctions are multiplicity free, we have:
\be
\label{product tube algebra reduced}
M_{g,a,d,h,b,e,k}^{c,s} = \sqrt{ \frac{ d_a d_b d_h d_s}{ d_d d_e d_c} } \; \bigl[ F_{kba}^s \bigr]^*_{ec} \, \bigl[ F_{bha}^s \bigr]_{ed} \, \bigl[ F_{bag}^s \bigr]^*_{cd} \;.
\ee
In \cref{app: F-symbols} we verify that the product is associative. One can also check that the generators of the tube algebra that map the untwisted sector to itself, \ie, the elements with $g=h=1$ in (\ref{tube algebra basis}), form a subalgebra%
\footnote{In this paper, by subalgebra we simply mean a subspace of the algebra that is closed under product. We do not require that the identity of the subalgebra be equal to that of the original algebra, which indeed does not happen in this case.} that is identical to the fusion algebra.

Tube algebras are finite-dimensional $C^*$-algebras (see, \eg, \cite{Bartsch:2025drc} for a proof) and so in particular they are semisimple. Given the product in (\ref{def product tube algebra}), one constructs the bilinear form $K$ and then the symmetrizer $\sS$ with respect to the tube algebra using the universal formula (\ref{symmetrizer}). Since the fusion algebra is a subspace of the tube algebra closed under multiplication, we expect the asymmetry with respect to the tube algebra to be bigger or equal to the asymmetry with respect to the fusion algebra.%
\footnote{It would be interesting to prove this expectation, or to find a counterexample.}
In other words, we expect $\mathsf{Tube}(\scrC)$ to be a more accurate detector of symmetry breaking. We discuss explicit examples of this fact in the next section.

\subsection{Excited states in CFTs}
\label{sec: full system in CFT}

An interesting class of theories where certain asymmetries can be computed analytically is given by two-dimensional CFTs. The vacuum $|0\rangle$ of a 2d unitary CFT does not break any internal symmetry, therefore its asymmetries vanish. What is more interesting is to compute the asymmetry of \emph{excited states}. An accessible class of states is obtained by performing the Euclidean path-integral on a half-space with the insertion of local operators:%
\footnote{We can think of these states as arising from a local quench at Lorentzian time $t=0$.}
\be
\label{excited states}
|\psi\rangle = \Bigl( {\ts \sum\nolimits_i \cO_i (z_i) } \Bigr) \, |0\rangle \;.
\ee
Here $\cO_i$ are local operators, while $z_i \in \bC$ denote points in the Euclidean lower half-plane.

In the case of invertible symmetries, the asymmetries of such states have been computed, \eg, in \cite{Chen:2023gql, Benini:2024xjv, Fujimura:2025rnm}. Here we compare the asymmetries (of the whole system on $\bR$ or $S^1$) with respect to the fusion algebras of invertible and noninvertible symmetries, and with respect to the tube algebra. We show, in a specific example, that the larger the algebra is, the more powerful the asymmetry is in detecting symmetry breaking.

Consider the example of the Ising CFT, which has three untwisted primary scalar local operators: the identity $1$, the energy operator $\varepsilon$, and the spin operator $\sigma$. We study this example in \cref{sec: Ising}. First, consider the $\bZ_2$ spin-flip symmetry that maps $\sigma \mapsto - \sigma$. Since $1$ and $\varepsilon$ are neutral, any state constructed using them has vanishing asymmetry. The spin operator is odd, but the states $\sigma(z) |0\rangle$ have vanishing asymmetry because, as rays in the Hilbert space, they are mapped to themselves (equivalently, the density matrices $\rho \propto \sigma(z) |0\rangle \langle 0 | \sigma(\bar z)$ commute with the fusion or group algebra). Therefore, in order to detect any asymmetry, one should consider states such as $\bigl( 1 + \sigma(z) \bigr) |0\rangle$.

Next, consider the full noninvertible symmetry, namely the Ising fusion category $\scrC$ (see \cref{sec: Ising}), which includes Kramers--Wannier self-duality. This symmetry is more powerful because $\varepsilon$ is not invariant. We can thus detect symmetry breaking in states such as $\bigl( 1 + \sigma(z) \bigr) |0\rangle$ but also $\bigl( 1 + \varepsilon(z) \bigr) |0\rangle$. We compute the asymmetries of the latter on $\bR$ with respect to the fusion algebra of Ising in (\ref{eq:renyi-asymm-1+eps-full}), and plot the second R\'enyi asymmetry in \cref{fig: plot asym fusion Ising}.

Finally, consider the tube algebra of $\scrC$. Since Kramers--Wannier duality maps the spin operator $\sigma$ to the disorder (or twist) operator $\mu$, under the action of the tube algebra the operators $(\sigma,\mu)$ form a two-dimensional representation. This time even the states $\sigma(z) |0\rangle$ exhibit nonvanishing asymmetry, that we compute around (\ref{eq:symm-tube-ising-2d}), because they are mapped to different states in $\cH_\text{Tube}$ (even when considered as rays).

\paragraph{Commutative algebras.}
For CFTs, whenever the symmetry algebra $\cA$ is commutative, the asymmetries of the full system in the excited states (\ref{excited states}) can be easily computed in terms of two-point functions.

Consider a pure state prepared by a Euclidean path-integral with the insertion of a sum of local operators in the Euclidean past. The Hilbert space could be the one on the circle or on the real line. The operators transform in irreducible representations of the symmetry algebra $\cA$, which in the Abelian case are all one-dimensional. The state is
\be
|\psi\rangle = \sum\nolimits_a |\psi_a \rangle = Z^{-\frac12} \, \sum\nolimits_a \;\;
\begin{tikzpicture}[baseline=-0.5cm, line width=0.6]
	\node (origin) at (0,0) {};
	\draw[dotted] (-1,0) -- (-1,-0.8) -- (1,-0.8) -- (1,0);
	\draw[densely dashed] (-1,0) -- (1,0);
	\fill (0,-0.4) circle (1.5pt) node[right] {\( \phi_{a} \)};
\end{tikzpicture}
\;\;.
\ee
The sum is over irreducible representations of $\cA$, and each component is the Euclidean path-integral on a half-plane or a half-cylinder (in order to prepare a state on $\bR$ or $S^1$, respectively) with the insertion of $\phi_a(z)$, as depicted, while $Z$ is the path-integral on the plane or the cylinder with no insertions. The normalization of the states is provided by the 2-point functions:
\be
\label{norms as 2-pt functions}
\langle \psi_a | \psi_a \rangle = \bigl\langle \phi_a^\dag(\bar z) \, \phi_a(z) \bigr\rangle \;,
\ee
where $\bar z$ is the point specular to $z$ with respect to the spatial cut. The density matrix for $|\psi\rangle$ is
\be
\rho = \frac1R \, |\psi \rangle \langle \psi| = \frac1R \sum\nolimits_{ab} |\psi_a \rangle \langle \psi_b| \qquad\text{with}\qquad R = \sum\nolimits_a \langle \psi_a | \psi_a \rangle \;.
\ee
When the algebra is commutative we can use (\ref{eq:symmetrizer-1d-irreps}), hence the symmetrized matrix is
\be
\rho_\sS = \frac1R \sum\nolimits_a |\psi_a \rangle \langle \psi_a |
\ee
and the asymmetries are
\be
\label{asym Abelian on circle}
\Delta S^{(n)} = \frac1{1-n} \log \biggl[ \sum\nolimits_a \frac{\langle \psi_a | \psi_a \rangle^n}{R\rule{0pt}{0.7em}^n} \biggr] \;,\qquad
\Delta S = - \sum\nolimits_a \frac{\langle \psi_a | \psi_a \rangle}{R} \log \frac{\langle \psi_a | \psi_a \rangle}{R} \;,
\ee
where we used orthogonality between the states $|\psi_a\rangle$.
Because of (\ref{norms as 2-pt functions}), the asymmetries are expressed in terms of two-point functions.


\section{Asymmetry for the interval Hilbert space}
\label{sec: interval}

The entanglement entropy, and likewise the entanglement asymmetry, of a subsystem is normally defined assuming that the Hilbert space factorizes: $\cH = \cH_A \otimes \cH_B$. Here $A$ is the subsystem of interest, while $B$ is its complement. This allows one to define the reduced density matrix $\rho_A = \Tr_B(\rho)$ from the complete density matrix $\rho$ (that could be pure or mixed), and then compute the entropy or the asymmetry of $\rho_A$. In quantum field theory one usually takes $A$ to be a spatial subregion, and here we will consider the case that $A$ is an interval. However, in quantum field theory the Hilbert space does not actually factorize. A consequence is that entanglement entropy is divergent and it requires a regularization.

\begin{figure}[t]
\centering$\ds
        \psi_{\cB_L, \cB_R} \colon \
        \begin{tikzpicture}[line width=0.6, baseline=\mathaxis]
            \draw [dotted] (-3,1) -- (-3,-0.8);
            \draw [dotted] (3,1) -- (3,-0.8);
            \draw [densely dashed] (-3, -0.8) -- (3, -0.8) node [above, pos=0.5] {\small$\cH^{*}$};
            \draw [densely dashed] (-3, 1) -- (-1.8, 1) node [below, pos=0.5] {\small$\cH_B$};
            \draw [densely dashed] (-0.8, 1) -- (0.8, 1) node [below, pos=0.5] {\small$\cH_A$};
            \draw [densely dashed] (1.8, 1) -- (3, 1) node [below, pos=0.5] {\small$\cH_B$};
            \draw [\colorbound] (-1.8, 1) arc [start angle = -180, end angle = 0, radius = 0.5] node [below, pos=0.5] {\footnotesize$\cB_L$};
            \draw [->] (-1.3, 1) -- +(-120: 0.5) node[shift={(0.28, 0.17)}] {\scriptsize$\varepsilon$};
            \draw [\colorbound] (1.8, 1) arc [start angle = 0, end angle = -180, radius = 0.5] node [below, pos=0.5] {\footnotesize$\cB_R$};
            \draw [->] (1.3, 1) -- +(-120: 0.5) node[shift={(0.28, 0.17)}] {\scriptsize$\varepsilon$};
        \end{tikzpicture}
        \quad\quad
        \Psi_{\cB_L, \cB_R} \colon \
        \begin{tikzpicture}[line width=0.6, baseline=\mathaxis]
            \draw [dotted] (-3,1.2) -- (-3,-1.2);
            \draw [dotted] (3,1.2) -- (3,-1.2);
            \draw [densely dashed] (-3, 1.2) -- (3, 1.2) node[below, pos=0.6] {\small$\cH$};
            \draw [densely dashed] (-1, 0.05) -- (1, 0.05) node[above, pos=0.35] {\small$\cH_A^*$};
            \draw [densely dashed] (-1, -0.05) -- (1, -0.05) node[below, pos=0.35] {\small$\cH_A$};
            \draw [densely dashed] (-3, -1.2) -- (3, -1.2) node[above, pos=0.6] {\small$\cH^{*}$};
            \draw [\colorbound] (-1.4, 0) ++(5.5: 0.4) arc [start angle = 5.5, end angle = 354.5, radius = 0.4] node[left, pos=0.5] {\footnotesize$\cB_L$};
            \draw [->] (-1.4, 0) -- +(-120: 0.4) node[shift={(0.25, 0.15)}] {\scriptsize$\varepsilon$};
            \draw [\colorbound] (1.4, 0) ++(174.5: 0.4) arc [start angle = 174.5, end angle = -174.5, radius = 0.4] node[right, pos=0.5] {\footnotesize$\cB_R$};
            \draw [->] (1.4, 0) -- +(-120: 0.4) node[shift={(0.25, 0.15)}] {\scriptsize$\varepsilon$};
        \end{tikzpicture}
$
\caption{\label{fig: factorizing map}%
Left: the map $\psi_{\cB_L, \cB_R}\colon \cH \to \cH_A \otimes \cH_B$ constructed with the path-integral on a strip, where $\cB_{L,R}$ are boundary conditions at two removed half-discs of radius $\varepsilon$.
Right: the map $\Psi_{\cB_L, \cB_R}\colon \End(\mathcal{H}) \to \End(\mathcal{H}_{A})$ constructed via the path-integral on a strip with a finite-size cut, with the same boundary conditions imposed.}
\end{figure}

A convenient regularization was proposed in \cite{Ohmori:2014eia}. One constructs a linear map
\be
\label{map psi for states}
\psi_\cB \colon \cH \to \cH_A \otimes \cH_B
\ee
and computes the entanglement entropy in the image of $\psi_\cB$. Such a map can be constructed as the path-integral on a strip, in which the lower (or incoming) boundary has the geometry underlying $\cH$, while the upper (or outgoing) boundary is the union of two disjoint parts with geometries underlying $\cH_A$ and $\cH_B$, respectively. In the case that $\cH_A$ is an interval, one removes two small half-discs of radius $\varepsilon$ at the two entangling points that bound $A$, and specifies some boundary conditions $\cB_{L,R}$ there.%
\footnote{In \eqref{map psi for states} we used the compact notation $\cB$ to indicate the collection of boundary conditions used.}
Eventually, the limit $\varepsilon \to 0$ is taken. We depict the setup in \cref{fig: factorizing map} left. Given the map $\psi_\cB$ between Hilbert spaces, one also constructs the linear map
\be
\label{map Psi for matrices}
\Psi_\cB \colon \End(\cH) \to \End(\cH_A)
\ee
that acts on density matrices and already includes the partial trace on $\cH_B$. It can be constructed as the path-integral on a strip obtained from a copy of $\psi_\cB$ and a copy of its dual, glued along $B$. We depict the setup in \cref{fig: factorizing map} right. The path-integral constructions of those maps are particularly useful when also the states we want to analyze have a path-integral preparation, as we will see.

Once we have a (generically mixed) state in $\cH_A$, we can define asymmetry in the general way described in \cref{sec: asymmetry}. We need to identify how the symmetry acts on $\cH_A$, which is the Hilbert space on an interval with boundary conditions $\cB_{L,R}$. Such an action is given by the strip algebra, that we describe next. Note that $\cH_A$ could also describe the asymmetry of the full system on an interval with boundary conditions; in this case one would not take the limit $\varepsilon \to 0$. For invertible symmetries, asymmetry with boundaries have been studied in \cite{Fossati:2024ekt, Kusuki:2024gss}.

\subsection{Strip algebra}

The action of the symmetry on the interval Hilbert space is described by the strip algebra $\mathsf{Strip}_\scrC(\scrM)$ \cite{Kitaev:2011dxc, Cordova:2024iti, Choi:2024tri}. Let us first recall how the symmetry properties of boundary conditions are described in terms of \emph{module categories}. In a left-module category $\scrM$  the simple objects $\cB_m$ correspond to right boundary conditions. The fusion of simple bulk lines $\cL_a$ (from the left) with simple boundaries $\cB_m$ is described by a fusion algebra:
\be
\label{module fusion algebra}
    \cL_a \times \cB_m = \sum\nolimits_n \tilde N_{am}^n \, \cB_n \;,
    \qquad\qquad\text{depicted as}\qquad
    \begin{tikzpicture}[line width=0.6, baseline=\mathaxis]
        \filldraw [\colorshade] (0,-0.5) rectangle (0.5,0.5);
        \draw[\colortopoline] (-0.5, -0.5) node[left] {\small$a$} -- (0, 0) node[\colortopojun, shift={(0.15,0)}] {\scriptsize$\rho$};
        \draw[\colorbound] (0, -0.5) node[shift={(0.25,0.1)}] {\small$m$} -- (0, 0.5) node[shift={(0.25,-0.1)}] {\small$n$};
        \draw[\colorbound] [-stealth] (0, -0.5) -- +(90: 0.30);
        \draw[\colorbound] [-stealth] (0, -0.5) -- +(90: 0.80);
        \fill[\colortopojun] (0,0) circle (1.pt);
    \end{tikzpicture}
\ee
and where the fusion coefficients $\tilde N_{am}^n \in \bN$. The number $\tilde N_{am}^n$ measures the dimension of the vector space of morphisms $\Hom(\cL_a \otimes \cB_m, \cB_n)$, and the index $\rho$ labels a basis of vectors when such a dimension is larger than 1. Simplicity of the objects $\cB_m$ and the pivotal relation state that
\be
\tilde N_{1m}^n = \delta_m^n \;,\qquad\qquad \tilde N_{am}^n = \tilde N_{\bar a n}^m \;.
\ee
Associativity of the product $\cL_a \times \cL_b \times \cB_m$ requires
\be
\sum\nolimits_p \tilde N_{ap}^n \, \tilde N_{bm}^p = \sum\nolimits_c N_{ab}^c \, \tilde N_{cm}^n \,\equiv\, \tilde N_{abm}^n \;.
\ee
This means that the matrices $(\tilde N_a)\ud{n}{m}$ furnish a nonnegative-integer-valued matrix representation (NIM-rep) of the bulk fusion algebra. There is a change of basis involved in the associativity of the tensor product, described by the $\tilde F$-symbols:
\be
\label{def F tilde symbols}
    \begin{tikzpicture}[line width=0.6, baseline=\mathaxis]
        \filldraw [\colorshade] (0,-0.8) rectangle (0.5,0.8);
        \draw[\colortopoline] (0, 0.4) node[\colortopojun, shift={(0.20, -0.05)}] {\scriptsize$\mu$} -- (-1.2, -0.8) node[left] {\small$a$};
                \draw[\colortopoline] (0, 0.4) -- ++(-135: 0.5) node[shift={(100: 0.3)}] {\small$e$};
                \draw[\colortopoline] (-0.6, -0.2) node[\colortopojun,shift={(-0.15, 0.15)}] {\scriptsize$\alpha$} -- (-0.6, -0.8) node[right] {\small$b$};
                \draw[\colorbound] (0, -0.8) node[shift={(0.25,0.1)}] {\small$m$} -- (0, 0.8) node[shift={(0.25,-0.1)}] {\small$n$};
                \draw[\colorbound, -stealth] (0, -0.8) -- (0,-0.2);
        \draw[\colorbound, -stealth] (0, -0.8) -- (0, 0.7);
        \fill[\colortopojun] (0,0.4) circle (1.pt);
        \fill[\colortopojun] (-0.6, -0.2) circle (1.pt);
    \end{tikzpicture}
    = \;\; \sum_{p,\nu,\rho} \; \bigl[ \tilde F_{abm}^n \bigr]_{(e; \alpha\mu)(p; \nu\rho)}
    \begin{tikzpicture}[line width=0.6, baseline=\mathaxis]
        \filldraw [\colorshade] (0,-0.8) rectangle (0.5,0.8);
        \draw[\colortopoline] (0, 0.4) node[\colortopojun, shift={(0.18, 0.02)}] {\scriptsize$\rho$} -- (-1.2, -0.8) node[left] {\small$a$};
        \draw[\colortopoline] (0, -0.2) node[\colortopojun, shift={(0.18, -0.02)}] {\scriptsize$\nu$} -- (-0.6, -0.8) node[left] {\small$b$};
        \draw[\colorbound] (0, -0.8) node[shift={(0.25,0.1)}] {\small$m$} -- (0, 0.8) node[shift={(0.25,-0.1)}] {\small$n$} node[pos=0.57, right] {\small$p$};
        \draw[\colorbound, -stealth] (0, -0.8) -- (0,-0.45);
        \draw[\colorbound, -stealth] (0, -0.8) -- (0, 0.15);
        \draw[\colorbound, -stealth] (0, -0.8) -- (0, 0.7);
                \fill[\colortopojun] (0,0.4) circle (1.pt);
        \fill[\colortopojun] (0,-0.2) circle (1.pt);
    \end{tikzpicture}
\;\;.
\ee
We indicate right boundaries with upward arrows, while all bulk lines run from bottom to top. The $\tilde F$-symbols are unitary square matrices of size $\tilde N_{abm}^n$ and hence relations similar to (\ref{def F tilde symbols}) can be obtained using unitarity or the associator diagram. We work in a basis in which $\tilde F_{abm}^n = \unit$ whenever $a$ or $b$ are equal to 1. The $\tilde F$-symbols satisfy a left-module pentagon equation, that also involves the $F$-symbols, reported in (\ref{left-module pentagon}). One can define the quantum dimension $\tilde d_m$ of a boundary condition as the disc partition function with that boundary condition. Those quantum dimensions satisfy the relations:
\be
\label{constraint on boundary quant dim}
d_a \, \tilde d_m = \sum\nolimits_n \tilde N_{am}^n \, \tilde d_n \;.
\ee
Thus the boundary quantum dimensions are the components of a common eigenvector of the matrices $\tilde N_a$ (with eigenvalues $d_a$).%
\footnote{Indeed, using that $d_a = d_{\bar a}$ and the pivotal relation,  (\ref{constraint on boundary quant dim}) can be rewritten as $d_{\bar a} \, \tilde d_m = \sum_n \tilde N_{\bar a n}^m \, \tilde d_n$.}
In a unitary theory $\tilde d_m > 0$. Notice that the quantum dimensions $\tilde d_m$ are ambiguous by a common rescaling: this is because the disc partition functions can be rescaled by adding an Euler counterterm to the action.

A right-module category is defined in a similar way, but the objects correspond to left boundary conditions and bulk lines act on them from the right. We describe the associativity of the tensor product with symbols ${}^R\!\tilde F$:
\be
\label{def R F tilde symbols}
    \begin{tikzpicture}[line width=0.6, baseline=\mathaxis]
        \filldraw [\colorshade] (0,-0.8) rectangle (-0.5,0.8);
        \draw[\colortopoline] (0, -0.4) node[\colortopojun, shift={(-0.20, 0.05)}] {\scriptsize$\mu$} -- (1.2, 0.8) node[right] {\small$b$};
        \draw[\colortopoline] (0, -0.4) -- ++(45: 0.55) node[shift={(-80: 0.3)}] {\small$e$};
        \draw[\colortopoline] (0.6, 0.2) node[\colortopojun, shift={(0.15, -0.15)}] {\scriptsize$\alpha$} -- (0.6, 0.8) node[left] {\small$a$};
        \draw[\colorbound] (0, 0.8) node[shift={(-0.25,-0.1)}] {\small$n$} -- (0, -0.8) node[shift={(-0.25,0.1)}] {\small$m$};
        \draw[\colorbound, -stealth] (0, 0.8) -- (0, 0.2);
        \draw[\colorbound, -stealth] (0, 0.8) -- (0, -0.7);
                \fill[\colortopojun] (0,-0.4) circle (1.pt);
                \fill[\colortopojun] (0.6, 0.2) circle (1.pt);
    \end{tikzpicture}
    = \;\; \sum_{p,\nu,\rho} \; \bigl[ {}^R\!\tilde F_{\bar n ab}^{\bar m} \bigr]_{(\bar p; \rho\nu)(e; \alpha\mu)} \quad
    \begin{tikzpicture}[line width=0.6, baseline=\mathaxis]
        \filldraw [\colorshade] (0,-0.8) rectangle (-0.5,0.8);
        \draw[\colortopoline] (0, -0.4) node[\colortopojun, shift={(-0.18, -0.02)}] {\scriptsize$\nu$} -- (1.2, 0.8) node[right] {\small$b$};
        \draw[\colortopoline] (0, 0.2) node[\colortopojun, shift={(-0.18, 0.02)}] {\scriptsize$\rho$} -- (0.6, 0.8) node[right] {\small$a$};
        \draw[\colorbound] (0, 0.8) node[shift={(-0.25,-0.1)}] {\small$n$} -- (0, -0.8) node[shift={(-0.25,0.1)}] {\small$m$} node[pos=0.57, left] {\small$p$};
        \draw[\colorbound, -stealth] (0, 0.8) -- (0,0.45);
        \draw[\colorbound, -stealth] (0, 0.8) -- (0, -0.15);
        \draw[\colorbound, -stealth] (0, 0.8) -- (0, -0.7);
        \fill[\colortopojun] (0,-0.4) circle (1.pt);
        \fill[\colortopojun] (0,0.2) circle (1.pt);
    \end{tikzpicture}
\;\;.
\ee
We use downward arrows to indicate left boundary conditions. The symbols ${}^R\!\tilde F$ satisfy their own right-module pentagon equation, that we report in (\ref{right-module pentagon}). For simplicity, we will restrict to the case that the same module is used on the right and on the left. If $[ \tilde F]$ solves the left-module pentagon equation (\ref{left-module pentagon}), then $[ {}^R\!\tilde F]$ defined as
\be
\label{right module from left}
\bigl[ {}^R\!\tilde F_{\bar n ab}^{\bar m} \bigr]_{(\bar p; \rho\nu)(e; \alpha \mu)} \,\equiv\, \bigl[ \tilde F_{abm}^n \bigr]_{(e; \alpha\mu)(p; \nu\rho)}^*
\qquad\text{or simply}\qquad
\bigl[ {}^R\!\tilde F_{\bar n ab}^{\bar m} \bigr]_{\bar p e} \,\equiv\, \bigl[ \tilde F_{abm}^n \bigr]_{ep}^*
\ee
solves the right-module pentagon equation (\ref{right-module pentagon}). Besides, this identification guarantees that if right boundary conditions are bent into left boundary conditions, their junctions with bulk lines can be moved from right to left. We thus regard this $[ {}^R\!\tilde F]$ symbols as the right-module version of the left-module $[\tilde F]$.

A module category that always exists is the \emph{regular module}, in which we identify the boundary conditions with the bulk lines, $\cB_m = \cL_m$, and thus also the boundary and bulk fusion algebras are identified: $\tilde N = N$. This is the module in which the fusion category $\scrC$ acts on itself. The $\tilde F$-symbols are simply identified with the $F$-symbols:
\be
\text{regular module:} \qquad \tilde F = F \;.
\ee
For the right module, one could either use the gauge (\ref{right module from left}) as we will do, or use a different gauge in which $\tilde F = {}^R\!\tilde F = F$. Both solve the right-module pentagon equation (\ref{right-module pentagon}).

The strip algebra is generated by the following basis elements:
\bea
\strip{r}{m}{a}{s}{n}{\rho}{\mu}
\eea
where the line $\cL_a$ acts both on the bulk as well as on the boundary by changing the boundary conditions. We will always make the same choice of module category for the left and right boundary condition. It is however easy to extend all formulas to the general case. The dimension of the algebra is
\be
\dim \bigl( \mathsf{Strip}_\scrC(\scrM) \bigr) = \sum\nolimits_{m,r,a,n,s} \tilde N_{am}^n \, \tilde N_{ar}^s \;.
\ee
The product in the algebra is easily computed using the two moves above and \eqref{fusion and q dim}:
\be
\label{eq:strip-algebra-product-general}
    \strip{s}{n}{b}{t}{p}{\sigma}{\nu}
    \;\; \times \;\;
    \strip{r}{m}{a}{s}{n}{\rho}{\mu}
    \;\; = \;\; \sum_{c,\tau,\pi} \; Q_{m,r,(a; \rho\mu), n, s, (b; \sigma\nu), p, t}^{(c; \tau\pi)} \quad
    \strip{r}{m}{c}{t}{p}{\tau}{\pi}
    \;,
\ee
where
\be
Q_{m,r,(a; \rho\mu), n, s, (b; \sigma\nu), p, t}^{(c; \tau\pi)} = \sum\nolimits_\alpha \sqrt{ \frac{d_ad_b}{d_c} } \; \bigl[ \tilde F_{bar}^t \bigr]_{(c; \alpha\tau)(s; \rho\sigma)} \, \bigl[ \tilde F_{bam}^p \bigr]_{(c; \alpha\pi)(n; \mu\nu)}^* \;.
\ee
The sum is over a basis of morphisms $\cL_b \otimes \cL_a \xrightarrow{\alpha} \cL_c$. With no multiplicities this simplifies to
\be
Q_{m,r,a,n,s,b,p,t}^c = \sqrt{ \frac{d_a d_b}{d_c} } \;\; \bigl[ \tilde F_{bar}^t \bigr]_{cs} \, \bigl[ \tilde F_{bam}^p \bigr]_{cn}^* \;.
\ee
We verify in \cref{app: strip algebra} that such a product is associative.

Since, in our context, the boundary conditions are introduced as a tool to regularize the entanglement entropy, it is desirable to use symmetry-invariant boundary conditions, in order not to contaminate the measurement of symmetry breaking coming from the state with an effect due to the boundaries.%
\footnote{On the other hand, as previously mentioned, it is an interesting question to study the asymmetry of boundary conditions. This has been analyzed in \cite{Fossati:2024ekt, Kusuki:2024gss}.}
The type of symmetry-invariant boundary conditions available in a given theory depends on the structure of the symmetry, and in particular on the presence of 't~Hooft anomalies \cite{Jensen:2017eof, Thorngren:2020yht}. For noninvertible symmetries, one can distinguish three situations \cite{Choi:2023xjw}: theories with strongly-symmetric simple boundaries, theories with only weakly-symmetric simple boundaries, and theories with no symmetric simple boundaries. We will discuss these three cases in turn.

\subsection{Strongly-symmetric boundary conditions}
\label{sec:strongly-symm-bc}

The most favorable case is when there exists a module category made of a single simple boundary. Such a boundary condition is automatically symmetry-invariant, meaning that it is invariant (up to rescaling) under parallel fusion with the lines $\cL_a \in \scrC$. Yet, there might be multiple junctions between a given bulk line $\cL_a$ and the boundary. Such a boundary condition was called ``strongly-symmetric'' in \cite{Choi:2023xjw}. It makes the definition and computation of entanglement asymmetry particularly simple. Such a boundary condition exists if and only if the symmetry is free from 't~Hooft anomalies.%
\footnote{\label{foo: modules}%
More precisely, when the 't~Hooft anomaly vanishes there always exists a module category with a single simple object, that can describe a strongly-symmetric boundary condition. Whether such a boundary condition actually exists in a given theory depends on the theory itself and on extra properties we might require to the boundary condition, for instance conformality. A similar comment applies to the regular module category that always exists, irrespective of 't~Hooft anomalies.}

A strongly-symmetric boundary condition is described by a module category in which there is a single simple object: $\cB$. Since the boundary index takes a single value, we henceforth indicate it with a dot. The fusion coefficients and the $\tilde F$-symbols simplify:
\be
\label{redef to algebra object}
\tilde N_{1\, \bdot}^\bdot = 1 \;,\qquad \tilde N_{a \,\bdot}^\bdot = \tilde N_{\bar a\, \bdot}^\bdot = d_a \geq 1 \;,\qquad \bigl[ \tilde F_{ab \, \bdot}^{\,\bdot} \bigr]_{(c; \delta\rho)( \bdot\, ; \nu\mu)}^* \equiv m_{a\mu, b\nu}^{c\rho; \delta} \;.
\ee
The fact that the boundary fusion coefficients are equal to the quantum dimensions immediately follows from (\ref{constraint on boundary quant dim}). Thus a necessary condition for the existence of strongly-symmetric boundary conditions in $\scrC$ is that all quantum dimensions are integer. Notice that when $\tilde N_{a \,\bdot}^\bdot > 1$ there are multiple junctions between the line $\cL_a$ and the boundary.
It turns out that there is a one-to-one correspondence between:
\begin{itemize}
\item left-module categories over $\scrC$ with only one simple object;
\item fiber functors of the fusion category $\scrC$;
\item maximal haploid algebra objects $\fA$ in $\scrC$.%
\footnote{More precisely, one is typically interested in equivalence classes of the objects listed above. The relevant equivalence relations are, respectively: equivalence of $\scrC$-module categories; equivalence of tensor functors; Morita equivalence of algebras. In particular, the last one turns out to be the correct equivalence that corresponds to equivalent module categories, as opposed to the stronger isomorphism of algebras \cite{Ostrik:2001xnt, etingof2016tensor}.}
\end{itemize}
The first entry corresponds to a strongly-symmetric boundary condition.
The second entry corresponds to a trivially-gapped phase (\aka{} SPT phase, or invertible TQFT) for the symmetry, and thus the symmetry is non-anomalous. We see that strongly-symmetric boundary conditions exist if and only if the symmetry is non-anomalous. The third entry corresponds to a maximal gauging of the whole symmetry (possible only in the absence of anomalies).

The redefinitions in (\ref{redef to algebra object}) make the correspondence with the maximal algebra object $\fA$ explicit. This is
\be
\label{decomposition algebra object}
\fA = \bigoplus\nolimits_a \, d_a \, \cL_a \;,
\ee
where $d_a \in \bN$ counts the dimension of $\Hom(\cL_a , \fA)$.%
\footnote{Haploid means that $\dim\bigl( \Hom(\cL_1, \fA) \bigr) = 1$.}
The symbols $m$ are the components of the multiplication morphism $m\colon \fA \otimes \fA \to \fA$:
\be
\begin{tikzpicture}[baseline=\mathaxis,line width=0.6]
    \draw (0,0) -- (-140: 1) node[left] {\small$a$};
    \draw (0,0) -- (-40: 1) node[right] {\small$b$};
    \draw (0,0) -- (90: 1) node[left] {\small$c$};
    \draw [\colortopoline, line width=1.5] (0,0) node[above right] {\small$\fA$} -- (90: 0.6);
    \draw [\colortopoline, line width=1.5] (0,0) -- (-40: 0.6);
    \draw [\colortopoline, line width=1.5] (0,0) -- (-140: 0.6);
    \draw [fill=white] (90: 0.6) circle [radius = 0.05] node[left] {\scriptsize$\rho$};
    \draw [fill=white] (-40: 0.6) circle [radius = 0.05] node[shift={(0.15, 0.15)}] {\scriptsize$\nu$};
    \draw [fill=white] (-140: 0.6) circle [radius = 0.05] node[shift={(-0.15, 0.15)}] {\scriptsize$\mu$};
\end{tikzpicture}
= \;\; \sum_\delta \; m_{a\mu, b\nu}^{c\rho; \delta}
\begin{tikzpicture}[baseline=\mathaxis,line width=0.6]
    \draw (0,0) node[above left] {\footnotesize$\delta$} -- (90: 1) node[left] {\small$c$};
    \draw (0,0) -- (-40: 1) node[right] {\small$b$};
    \draw (0,0) -- (-140: 1) node[left] {\small$a$};
\end{tikzpicture}
\;.
\ee
The requirement that $m$ be associative, namely that $m \bigl( m (x\otimes y) \otimes z \bigr) = m \bigl( x \otimes m(y \otimes z) \bigr)$ for all $x,y,z \in \fA$, when written in components, boils down to the constraint \cite{Fuchs:2002cm}:
\be
\label{associativity of m}
\sum\nolimits_{e, \epsilon, \alpha, \beta} \, m_{a\mu, b\nu}^{e\epsilon; \alpha} \; m_{e\epsilon, c\rho}^{d\sigma; \beta} \; \bigl[ F_{abc}^d \bigr]_{(e; \alpha\beta)(f; \gamma\delta)} = \sum\nolimits_\phi \, m_{a\mu, f\phi}^{d\sigma; \delta} \; m_{b\nu, c\rho}^{f\phi; \gamma} \;.
\ee
Using the correspondence in (\ref{redef to algebra object}) and unitarity, this equation is equivalent to the left-module pentagon equation (\ref{left-module pentagon}) (for module categories with one simple object). Notice that a fusion category $\scrC$ can admit multiple inequivalent maximal algebra objects $\fA$: they will have the same decomposition (\ref{decomposition algebra object}), but would differ in the product $m$.%
\footnote{\label{foo: internal Hom}%
The construction is more general, see App.~A of \cite{Choi:2023xjw} as well as \cite{Ostrik:2001xnt, Bhardwaj:2017xup, Huang:2021zvu}. For any module, choose a simple object (a boundary), then the so-called internal Hom construction produces a (in general not maximal) haploid algebra object in $\scrC$. The algebra objects one obtains from the various simple objects (boundaries) of a given module can be different, but are all Morita equivalent. All haploid, semisimple, indecomposable algebra objects are obtained in this way, by considering all possible indecomposable module categories. Indeed, from the algebra object one can construct a module category such that the internal Hom construction gives back that algebra.}

The corresponding strip algebra is generated by the basis elements
\bea
\label{def strongly sym basis}
H_{(a; \rho\mu)} \;\equiv\;\;\; \strip{}{}{a}{}{}{\rho}{\mu}
\eea
and the product is given by
\bea
\label{strip alg product strongly-sym}
Q_{(a; \rho\mu)(b; \sigma\nu)}^{(c; \tau\pi)} &= \sum\nolimits_\alpha \sqrt{ \frac{d_a d_b}{d_c}} \; \bigl[ \tilde F_{ba\, \bdot}^{\,\bdot} \bigr]_{(c; \alpha\tau)(\bdot\, ; \rho\sigma)} \, \bigl[ \tilde F_{ba\, \bdot}^{\,\bdot} \bigr]_{(c; \alpha\pi)(\bdot\, ; \mu\nu)}^* \\
&= \sum\nolimits_\alpha \sqrt{ \frac{d_a d_b}{d_c}} \; \bigl( m_{b\sigma, a \rho}^{c\tau; \alpha} \bigr)^* \, \bigl( m_{b\nu, a \mu}^{c\pi; \alpha} \bigr) \;.
\eea
The sum is over a basis of morphisms $\cL_b \otimes \cL_a \xrightarrow{\alpha} \cL_c$. This defines a finite-dimensional semisimple algebra $\cA$, whose symmetrizer can be constructed using the universal formula (\ref{symmetrizer}).

\subsection{Weakly-symmetric boundary conditions}

When strongly-symmetric boundary conditions do not exist, we are forced to consider modules with multiple simple objects. This means that we have to deal with a set of boundary conditions that form a ``multiplet'' under fusion with the bulk symmetry lines. It might happen, however, that the module contains one particular simple boundary $\hat\cB$ (equal to $\cB_m$ for a particular value $m=\hat m$) such that the fusion of any bulk line $\cL_a$ with $\hat\cB$ contains $\hat\cB$ itself: $\cL_a \otimes \hat\cB \supseteq \hat\cB$. This happens when
\be
\tilde N_{a\hat m}^{\hat m} \geq 1 \qquad\text{for all } a \;.
\ee
This means that every bulk like $\cL_a$ has at least one topological junction with the boundary $\hat\cB$. Such a boundary condition was called ``weakly-symmetric'' in \cite{Choi:2023xjw}. It allows us to restrict to a subspace of the strip algebra in which we have all bulk lines but only the boundary $\hat\cB$, and that is closed under strip-algebra product. When multiple junctions exist, $N_{a\hat m}^{\hat m}>1$, in general one has to keep all of them in order to obtain a closed algebra.%
\footnote{In some cases it might be possible to restrict to a subspace of the junction space.}
We call the resulting algebra a \emph{reduced strip algebra}.

Note that the reduced strip algebra always admits a symmetrizer. Indeed, its $C^*$-structure is simply the restriction of the $C^*$-structure of the full strip algebra.
In particular, the $C^*$ condition \eqref{eq:c-star-condition} is still satisfied and the reduced strip algebra is thus semisimple and strongly separable. The symmetrizer can be constructed using the universal formula (\ref{symmetrizer}).

It turns out \cite{Choi:2023xjw} that a fusion category $\scrC$ admits a weakly-symmetric boundary condition if and only if there exists a (not necessarily maximal) haploid algebra object
\be
\fA = \bigoplus\nolimits_a \, N_a^{(\fA)} \, \cL_a \qquad\qquad \text{such that $N_a^{(\fA)} \geq 1$ for all $a$} \,.
\ee
We introduced the nonnegative integer numbers $N_a^{(\fA)} \equiv \dim\bigl( \Hom (\cL_a, \fA) \bigr)$, that for haploid algebra objects always satisfy $N_a^{(\fA)} \leq d_a$ \cite{Fuchs:2004dz}. The connection between boundaries and algebra objects goes through the internal Hom construction (see \cref{foo: internal Hom}). Haploid algebra objects are the noninvertible generalization of anomaly-free subgroups of an invertible symmetry group with a choice of discrete torsion. Thus, a weakly-symmetric boundary condition exists if and only if there exists a gauging of $\scrC$ that involves all simple lines in $\scrC$. When the gauging is maximal, $N_a^{(\fA)} = d_a$, which is possible only in the absence of 't~Hooft anomalies, the boundary condition is actually strongly symmetric.

\subsection{Non-symmetric boundary conditions}
\label{non-symmetric bc}

It might happen that a given theory does not admit neither strongly- nor weakly-symmetric boundary conditions, either because the symmetry has a severe enough 't~Hooft anomaly to prevent them, or because we insist on some extra property, for instance conformality, of the boundary conditions. In such cases, the boundary condition at the entangling surface necessarily breaks the symmetry, implying a nonzero asymmetry on subsystems even when the original state is invariant. To overcome this, and to detect symmetry breaking of states independently of the boundary conditions at the entangling surface, we consider a map $\Psi_\cB$ (see \eqref{map Psi for matrices}) that produces a statistical mixture of density matrices with different boundary conditions. We show that there exists a choice of coefficients such that the invariant state prepared by the empty path-integral exhibits indeed vanishing asymmetry. The procedure can be carried out for any choice of module category $\mathscr{M}$, as we show here. However, in the examples in \cref{sec: examples} we specialize to the regular module category.

At a formal level, we could define a symmetric ``composite'' boundary condition $\ket{\cB_\sS}$ if we allow for arbitrary linear combinations of boundaries (which we denote here as kets because we regard them as element of a vector space):
\be
\ket{\cB_\sS} = \sum\nolimits_m \tilde d_m \ket{\cB_m} \qquad\text{so that}\qquad \cL_a \ket{\cB_\sS} = d_a \ket{\cB_\sS} \;.
\ee
Here $d_a$ and $\tilde d_m$ are the bulk and boundary quantum dimensions, and we used \eqref{constraint on boundary quant dim} to prove the second equality.%
\footnote{One also uses the pivotal relation: $\cL_a \ket{\cB_\sS} = \sum_{mn} \tilde d_m \tilde N_{am}^n \ket{\cB_n} = \sum_{mn} \tilde N_{\bar a n}^m \tilde d_m \ket{\cB_n}$, as well as $d_a = d_{\bar a}$.}

We exploit this idea --- in the case of unitary and normalizable CFTs and insisting on conformal boundary conditions --- to construct a symmetric density matrix $\rho$ on the interval Hilbert space starting from an invariant state, even in the absence of invariant simple boundary conditions.
Consider a module category $\scrM$, and let $\rho_{mn}$ be the reduced density matrix of the vacuum (or of another state invariant under the symmetry) constructed with simple boundary conditions $\cB_m, \cB_n \in \scrM$. We construct an ``average'' over boundary conditions as follows:
\be
\label{eq:averaged-rho}
\rho = \frac1D \sum_{m,n \,\in\, \Irr(\scrM)} \!\!\! \tilde d_m \, \tilde d_n \, \rho_{mn}
= \frac1D \sum_{m,n \,\in\, \Irr(\scrM)} \! \frac{\tilde d_m \, \tilde d_n}{Z_{mn}} \,\, \tilde \rho_{mn} \;,\quad\qquad
D = \Bigl( {\ts \sum_{m \in \Irr(\scrM)} \, \tilde d_m} \Bigr)^2 .
\ee
Here $\tilde \rho_{mn}$ is the unnormalized density matrix produced by the path-integral realization of $\Psi_\cB$ in (\ref{map Psi for matrices}) with boundary conditions $\cB_m$ and $\cB_n$, as depicted in \cref{fig:ising-regular}, while $Z_{mn} = \Tr(\tilde \rho_{mn})$. We show that in the limit $\varepsilon \to 0$, $\rho$ commutes with the strip algebra and thus it has vanishing asymmetry. The interval Hilbert space $\cH$ decomposes in the various sectors with simple boundary conditions as $\cH = \bigoplus_{m,n} \cH_{m,n}$. We want to show that, for every $O \in \End(\cH)$:
\be
\lim_{\varepsilon \to 0} \, \Tr \bigl( \rho \, H_{m,n}^{r,s} \, O \bigr) = \lim_{\varepsilon \to 0} \, \Tr \bigl( H_{m,n}^{r,s} \, \rho \, O \bigr) \;.
\ee
Here $H_{m,n}^{r,s}$ is a basis element of the strip algebra in $\Hom(\cH_{m,n}, \cH_{r,s})$. By matching boundary conditions and using \eqref{eq:averaged-rho}, the previous equation is equivalent to
\be
\lim_{\varepsilon \to 0} \, \frac{\tilde d_r \, \tilde d_s}{Z_{rs}} \, \Tr\bigl( \tilde\rho_{rs} \, H_{m,n}^{r,s} \, O_{r,s}^{m,n} \bigr) = \lim_{\varepsilon \to 0} \, \frac{\tilde d_m \, \tilde d_n}{Z_{mn}} \, \Tr\bigl( H_{m,n}^{r,s} \, \tilde\rho_{mn} \, O_{r,s}^{m,n} \bigr) \;,
\ee
where $O_{r,s}^{m,n}$ is the component of $O$ in $\Hom(\cH_{r,s}, \cH_{m,n})$. Notice that $\tilde\rho_{rs} \, H_{m,n}^{r,s} = H_{m,n}^{r,s} \, \tilde\rho_{mn}$ since, by assumption, lines can be slid past the matrices $\tilde\rho_{mn}$ (even for finite $\varepsilon$). Thus, because the operators $O_{r,s}^{m,n}$ are completely arbitrary, we only need to check that
\be
\label{ratio with Z_mn}
\lim_{\varepsilon \to 0} \, \frac{\tilde d_r \, \tilde d_s}{Z_{rs}} = \lim_{\varepsilon \to 0} \, \frac{\tilde d_m \, \tilde d_n}{Z_{mn}} \;.
\ee
This equation is indeed satisfied because
\be
\label{eq:shrinking-discs-Z}
\lim_{\varepsilon \to 0} \, Z_{rs} = \tilde d_r \, \tilde d_s \, Z(\bC) \;,
\ee
where $Z(\bC)$ is the Euclidean partition function on the plane. This can be argued as follows. Recall that $Z_{rs}$ is the partition function on the plane with two discs of radius $\varepsilon$ removed (see \cref{fig:ising-regular}). The discs are centered, say, at $z=0$ and $z = \ell$. Following \cite{Ohmori:2014eia}, this geometry can be mapped to a cylinder of length $2 \log (\ell / \varepsilon)$ with the transformation $w = \log \bigl( z / (\ell - z) \bigr)$.
The partition function can then be rewritten as a closed-channel amplitude:
\be
Z_{rs} = W \, \bigl\langle s \big| \, (\varepsilon / \ell)^{L_0 + \bar L_0 - c/12)} \, \big| r \bigr\rangle \;,
\ee
where $W$ is a Weyl factor, while $|r\rangle, |s\rangle$ are Cardy states. Inserting a resolution of the identity with eigenstates of the cylinder Hamiltonian, in the limit $\varepsilon \to 0$ (which corresponds to a cylinder of infinite length) only the ground state $\ket{0}$ remains.%
\footnote{For unitary CFTs with discrete spectrum, $|0\rangle$ is the ground state. More generally, $|0\rangle$ is the lowest-dimension state that overlaps with the Cardy states $|r\rangle, |s\rangle$. See the discussion in \cite{Ohmori:2014eia}.}
Thus one gets
\be
\label{eq:Z-as-amplitude-cardy}
\lim_{\varepsilon \to 0} \, Z_{rs} = \lim_{\varepsilon \to 0} \,  g_r \, g_s \, W \, \bigl\langle 0 \big| \, (\varepsilon / \ell)^{L_0 + \bar L_0 -c/12} \, \big| 0 \bigr\rangle \;,
\ee
where $g_r \equiv \braket{r|0}$ is the $g$-factor \cite{Affleck:1991tk}. Because of the state/operator correspondence, the term $\lim_{\varepsilon \to 0} \bigl\langle 0 \big| \, (\varepsilon / \ell)^{L_0 + \bar L_0 -c/12} \, \big| 0 \bigr\rangle$ becomes the partition function of an infinite cylinder, with no insertions at infinity.
We can then undo the Weyl transformation, reabsorbing the factor $W$, to obtain $\lim_{\varepsilon \to 0} \, Z_{rs} = g_r \, g_s \, Z(\bC)$. The $g$-factors can be interpreted as disc partition functions \cite{Recknagel:2013uja} for a specific choice of Euler counterterm, and thus they satisfy the equation of boundary quantum dimensions \eqref{constraint on boundary quant dim} (see also \cite{Fuchs:2001qc}). Hence we can choose a normalization such that $\tilde d_r = g_r$, and obtain \eqref{eq:shrinking-discs-Z}.%
\footnote{With a generic normalization (\ie, choice of Euler counterterm), the $g$-factors $g_m$ are proportional to the boundary quantum dimensions $\tilde d_m$ and thus they satisfy \eqref{constraint on boundary quant dim}. This implies that $\lim_{\varepsilon\to0} \, \tilde d_m \tilde d_n / Z_{mn}$ is independent from $m,n$. This is enough to prove (\ref{ratio with Z_mn}), even with generic normalization.}

\paragraph{Large subsystem limits.}
Given the asymmetry of some state, computed on the interval subsystem using one of the types of boundary conditions discussed above, what is the relation between its large subsystem limit and the asymmetry of the full system, computed using the fusion or tube algebras?

In \cref{sec: group-like symm} we analyze invertible group-like symmetries. In particular, we consider excited CFT states created by the insertion of local operators in Euclidean time. We find that the entanglement and R\'enyi asymmetries on an interval, as long as our ``averaging'' protocol is used, are independent of the choice of boundary conditions. Moreover, for untwisted states, the large subsystem limits of the asymmetries coincide with the asymmetries of the full system, computed with respect to either the fusion or the tube algebra (since they coincide for untwisted states).

In \cref{sec: Z2xZ2 TY} we study certain excited states in two copies of the Ising CFT, utilizing the $\mathsf{Rep}(H_8)$ (\aka\ $\bZ_2 \times \bZ_2$ Tambara--Yamagami) noninvertible symmetry that admits a strongly-symmetric boundary condition. The result is similar: the second R\'enyi entropy on the interval subsystem asymptotes in the large subsystem limit to the second R\'enyi entropy of the full system, computed with respect to the tube algebra (not the fusion algebra, which in general is different).

On the other hand, in \cref{sec: Ising,sec:Ising-symm-interval-Hilbert-space} we study certain excited states in the Ising CFT, utilizing its Ising fusion category (\aka\ $\bZ_2$ Tambara--Yamagami) symmetry, which does not admit symmetric boundary conditions and which then forces us to resort to the averaging protocol. This time we find that the large subsystem limit of the second R\'enyi asymmetry on an interval is qualitatively similar to, but slightly smaller than, the asymmetry of the full system with respect to the tube algebra. A similar behavior is found in \cref{sec: Fibonacci} when studying the Fibonacci fusion category symmetry with its weakly-symmetric boundary condition. It would be interesting to understand these results more systematically.

\subsection{Symmetrizer from (weak) Hopf algebra structure}
\label{sec:hopf-symmetrization}

We can find an alternative, but equivalent, formula for the symmetrizer by leveraging the following reasoning that holds in the invertible case for a symmetry group $G$ (or equivalently, for the group algebra $\mathbb{C}[G]$).
Consider a Hilbert space $\cH$ that carries a representation of $G$. A density matrix $\rho \in \End(\cH)$ is an element of $\cH \otimes \cH^\vee$ (the second factor being the dual space). This tensor product itself carries a representation of $G$, which decomposes into a sum of irreducible representations.
Then, the symmetrization operation is reproduced by applying to this space the projector onto the trivial representation.
The obstacle in generalizing this reasoning to the noninvertible case is that for a generic algebra $\cA$ the concepts of tensor product of representations, of dual (or conjugate) representation, and of trivial representation, are not defined.
In the special case that the algebra is \emph{Hopf}, however, these concepts are available and the symmetrizer can be written as the projector onto the trivial representation. Moreover, we will see that if the algebra is only weak Hopf but its antipode map is involutive (such an algebra is called \emph{weak Kac} \cite{Bohm:1999}), a formula similar to that of the Hopf case applies, even though it lacks that interpretation. It was shown in \cite{Kitaev:2011dxc, Cordova:2024iti} that the strip algebras are always weak Hopf, therefore when they satisfy the weak Kac condition we can apply the strategy above to the case of the interval Hilbert space.%
\footnote{On the contrary, the tube algebra is in general not a weak Hopf algebra. One can still define a tensor product of representations using the so-called Day convolution product \cite{Day:1970}. We do not explore such a possibility here.}

Let us first consider the case that a strip algebra $\cA$ is constructed using a strongly-symmetric boundary condition. We can show that such a strip algebra is a Hopf algebra. We provide definitions and basic properties in \cref{app: Hopf algebras} and prove the claim in \cref{app: symm strip Hopf}.

Concisely, a Hopf algebra comes equipped with (besides its unit and its associative product) the following algebra homomorphisms: a coassociative coproduct $\Delta \colon \cA \to \cA \otimes \cA$ such that $\Delta(\unit) = \unit \otimes \unit$ and a counit $\epsilon \colon \cA \to \bC$.
In addition, it comes with an antipode $S \colon \cA \to \cA$, which is a linear algebra anti-homomorphism, \ie, $S(xy) = S(y) \, S(x)$. In finite-dimensional semisimple Hopf algebras one proves that $S \circ S = \id$. The one-dimensional ``trivial'' representation $r_\epsilon \colon \cA \to \End(\bC) \cong \bC$, which behaves as the identity under the tensor product of representations, is provided by the counit:
\be
r_\epsilon(x) = \epsilon(x) \qquad\qquad\forall \, x \in \cA \;.
\ee
The tensor product of representations is constructed using the coproduct $\Delta$. Given two representations $r_1\colon \cA \to \End(V_1)$ and $r_2\colon \cA \to \End(V_2)$ on vector spaces $V_{1,2}$, their tensor product representation $r_{12}\colon \cA \to \End(V_1 \otimes V_2)$ is defined as
\be
r_{12} = (r_1 \otimes r_2) \circ \Delta \qquad\text{\ie}\qquad
r_{12}(x) \; v_1 \otimes v_2 = \sum\nolimits_i \Bigl[ r_1\bigl( x_{(1)}^i \bigr) \, v_1 \Bigr] \otimes \Bigl[ r_2 \bigl( x_{(2)}^i \bigr) \, v_2 \Bigr] \,,
\ee
where $v_{1,2} \in V_{1,2}$ and we used Sweedler's notation to write $\Delta(x) = \sum_i x_{(1)}^i \otimes x_{(2)}^i$. Using (\ref{eq:counit-property}), one sees that the representation $r_\epsilon$ multiplies trivially with any other representation. Finally, given a representation $r\colon \cA \to \End(V)$, its conjugate representation $r^*\colon \cA \to \End(V^\vee)$ is constructed using the antipode:
\be
\bigl\langle r^*(x) \, f ,\, v \bigr\rangle \,\equiv\, \bigl\langle f,\, r \bigl( S(x) \bigr) \, v \bigr\rangle \qquad\qquad \forall \, x \in \mathcal{A} ,\;\;  v \in V ,\;\; f \in V^\vee ,
\ee
where the angular brackets denote the evaluation of a functional on a vector. More compactly, this can be written as $r^* = (r \circ S)^\sT$. Using \eqref{props of S} one sees that the trivial representation is self-conjugate.

Associated to $r_\epsilon$ there is a minimal central idempotent (MCI) $P_\epsilon \in \cA$, \ie, the projector to the trivial representation. It satisfies%
\footnote{The first equality is the centrality of projectors. For the second equality, since $P_\epsilon^2 = P_\epsilon$ and the trivial representation is one-dimensional, then $P_\epsilon$ is a basis for the image of the projector in the regular representation (recall that a 1d representation occurs only once inside the regular representation), and thus $P_\epsilon \, x = \alpha(x) \, P_\epsilon$ for some $\alpha \in \bC$. Applying $\epsilon$ to both sides we find $\alpha(x) = \epsilon(x)$.}
\be
P_\epsilon \, x = x \, P_\epsilon = \epsilon(x) \, P_\epsilon \qquad\qquad \forall\, x \in \cA \;.
\ee
Taking $x$ equal to one of the MCIs $P_i$ (\ie, a projector to an irreducible representation of $\cA$), one obtains $\epsilon(P_i) = \delta_{i \epsilon}$, confirming that $P_\epsilon$ projects to the trivial representation.

We claim that the symmetric separability idempotent can be written as
\be
\label{symmetrizer Hopf abstract}
e = ( \id \otimes S) \circ \Delta(P_\epsilon)
\ee
which was proven to be a separability idempotent for a Hopf algebra in \cite{Schweigert}.
When acting on the space $\cH \otimes \cH^\vee \cong \End(\cH)$, this operator is precisely the action of the projector to the trivial representation $P_\epsilon$ on density matrices. More explicitly:
\be
\label{symmetrizer Hopf explicit}
\rho_\sS = \sum\nolimits_i ( P_\epsilon)_{(1)}^i \; \rho \; S\bigl( (P_\epsilon)_{(2)}^i \bigr)
\ee
where $\Delta(P_\epsilon) = \sum_i \, (P_\epsilon)_{(1)}^i \otimes (P_\epsilon)_{(2)}^i$. The disadvantage of this formula is that one needs to explicitly construct the projector $P_\epsilon$ to the trivial representation. In \cref{app: symmetrizer in Hopf} we prove that (\ref{symmetrizer Hopf explicit}) satisfies properties 1.~--~4.\ listed in \cref{sec: asymmetry}. The expression in (\ref{symmetrizer Hopf explicit}) is also universal, in the sense that it is written solely in terms of the elements of $\cA$. As we commented at the end of \cref{sec: asymmetry}, this completely fixes the symmetrizer and thus (\ref{symmetrizer Hopf explicit}) agrees with (\ref{symmetrizer}). Notice that the universal formula (\ref{symmetrizer}) only depends on the product structure on $\cA$, therefore different Hopf algebras that share the same product structure will give rise, through (\ref{symmetrizer Hopf explicit}), to the same symmetrizer.

For weak Hopf algebras a similar recipe exists to compute the symmetric separability idempotent, provided they satisfy the additional condition \( S^{2} = \id \) (such algebras are called weak Kac).%
\footnote{This is not the case for a generic strip algebra (see \cite{AliAhmad:2025bnd}, eq.~(D.14)). We thank the authors of~\cite{AliAhmad:2025bnd} for bringing to our attention a few inaccuracies on this point in the first version of this manuscript.}
Referring to \cite{Bohm:1998iu}, one introduces the left and right projectors $\Pi^\mathrm{L}, \Pi^\mathrm{R}\colon \cA \to \cA$ as
\be
\Pi^\mathrm{L}(x) = \sum\nolimits_i \epsilon\bigl( \unit_{(1)}^i \, x \bigr) \, \unit_{(2)}^i \;,\qquad\qquad
\Pi^\mathrm{R}(x) = \sum\nolimits_i \unit_{(1)}^i \, \epsilon\bigl( x \, \unit_{(2)}^i \bigr) \;,
\ee
acting on $x \in \cA$. Then one defines the left and right normalized integrals as those elements $\ell,r \in \cA$ such that
\be
x\, \ell = \Pi^\mathrm{L}(x) \, \ell \;,\qquad \Pi^\mathrm{L}(\ell) = \unit \;, \qquad\quad\text{or}\quad\qquad r \, x = r \, \Pi^\mathrm{R}(x) \;,\qquad \Pi^\mathrm{R}(r) = \unit \;,
\ee
for all $x \in \cA$. In general, being a left integral does not imply being a right integral. When this happens, one talks about \emph{Haar integrals}. That is, a Haar integral $h$ is defined as a simultaneous left and right normalized integral. When it exists, it is unique. The necessary and sufficient conditions for its existence are spelled out in Th.~3.27 of \cite{Bohm:1998iu}.
Such conditions are indeed satisfied under our working hypotheses.%
\footnote{In the notation of Th.~3.27 in \cite{Bohm:1998iu}, since we are assuming $S^{2} = \id$, we have $g=\unit$. Then, the second condition therein is satisfied because the trace of $\unit$ in any representation is the dimension of that representation.}
The symmetric separability idempotent is then given by
\be\label{eq:wha-si}
e = (\id \otimes S) \circ \Delta(h) \;,
\ee
where $h \in \cA$ is the Haar integral. In \cref{app: symmetrizer in Hopf} we verify that such an expression satisfies the properties 1.~--~4.\ and thus, being also universal, it agrees with (\ref{symmetrizer}).%
\footnote{Note that, had we not required the property $S^{2} = \id$, the object \eqref{eq:wha-si} could still be defined. However, the resulting separability idempotent would not be symmetric. Hence, the symmetrized density matrix will fail to satisfy property 4. In this case, the existence of the Haar integral is guaranteed provided the weak Hopf algebra also possesses a $C^{*}$-structure compatible with the former, see Th.~4.5 of \cite{Bohm:1998iu}.}

Note that in the case of a Hopf algebra, both the left and right projectors reduce to
\be
\Pi^\mathrm{L}(x) = \Pi^\mathrm{R}(x) = \epsilon(x) \, \unit
\ee
because $\Delta(\unit) = \unit \otimes \unit$. Then the Haar integral $h$ is equal to the projector $P_\epsilon$.


\section{Spontaneous symmetry breaking}
\label{sec: SSB}

A natural application of entanglement asymmetry is to detect spontaneous symmetry breaking in gapped phases. In the far IR, a gapped phase is described by a topological field theory on which the symmetry acts. Therefore, for a given fusion category $\scrC$, the possible indecomposable and inequivalent (from the symmetry point of view) two-dimensional gapped phases are in one-to-one correspondence with indecomposable two-dimensional $\scrC$-symmetric TQFTs. In turn, the latter are in one-to-one correspondence with indecomposable module categories $\scrM$ for $\scrC$ \cite{Thorngren:2019iar}. We can then use entanglement asymmetry to try to distinguish the various phases.

Consider a 2d gapped phase described by a certain 2d $\scrC$-symmetric TQFT in the IR, and let $\scrM$ be the corresponding left-module category. On the circle $S^1$ the TQFT has a finite number of states.
If the symmetry is unbroken, the circle Hilbert space is one-dimensional and the entanglement asymmetry with respect to the fusion algebra vanishes.%
\footnote{Here we only consider indecomposable gapped phases.}
On the other hand, when a finite symmetry is spontaneously broken we expect a higher-dimensional circle Hilbert space on which the symmetry acts. We are not interested in computing the asymmetry of generic vectors in that Hilbert space, but only of very special vectors that we call ``vacua'': these are the states that satisfy the cluster decomposition property. Such states form a preferred basis of the Hilbert space. It turns out that they are in correspondence with the simple boundaries $\cB_m$ of the corresponding module category $\scrM$, and can be constructed using open/closed duality \cite{Huang:2021zvu}: one simply considers the cylinder with the boundary condition $\cB_m$ in the past, as in \cref{fig: clustering states} left.

\begin{figure}[t]
\centering$\ds
\begin{tikzpicture}[baseline=\mathaxis, line width=0.6]
    \draw [\colortopoline,
    postaction=decorate,
    decoration={markings,mark=at position 0.5 with \arrow{>}},
    ] (-1.2, 0.3) arc [start angle = -180, end angle = 0, x radius = 1.2, y radius = 0.25] node[right, shift={(0.3, 0)}] {\small$\cL_a$};
    \draw [\colortopoline, dashed] (-1.2, 0.3) arc [start angle = 180, end angle = 0, x radius = 1.2, y radius = 0.25];
    \draw [->] (-0.5, 0.0) -- +(0, -0.2); \draw [->] (0.5, 0.0) -- +(0, -0.2);
    \draw [\colorbound] (-1.2, -0.9) arc [start angle = -180, end angle = 0, x radius = 1.2, y radius = 0.25] node[right, shift={(0.3, 0)}] {\small$\cB_m$};
    \draw [\colorbound, -{Stealth[length=2.5mm]}] (-1.2, -0.9) arc [start angle = -180, end angle = -82, x radius = 1.2, y radius = 0.25];
    \draw [\colorbound, dashed] (-1.2, -0.9) arc [start angle = 180, end angle = 0, x radius = 1.2, y radius = 0.25];
    \draw (-1.2, -0.9) -- (-1.2, 1.2);
    \draw (1.2, -0.9) -- (1.2, 1.2);
\end{tikzpicture}
\qquad\qquad
\braket{\cB_m | \cO | \cB_m} \;=\;
\begin{tikzpicture}[baseline=\mathaxis, line width=0.6]
    \fill[\colorshade] (-1.4,-1.4) rectangle (1.4,1.4) (0,0) circle (1.2);
    \draw [\colorbound, fill=\colorshade] (0, 0) circle [radius = 0.3];
    \draw [\colorbound, -Stealth] (-0.292, 0.05) -- ++(0, 0.05) node[shift={(-0.2, 0.12)}] {\footnotesize$m$};
    \draw [\colorbound] (0, 0) circle [radius = 1.2];
    \draw [\colorbound, -Stealth] (-1.198, -0.04) -- ++(0, -0.05) node[shift={(0.25, -0.1)}] {\footnotesize$m$};
    \fill (0, 0.7) circle [radius = 0.05] node[right] {\small$\cO$};
\end{tikzpicture}
\;=\; \frac{1}{\tilde d_{m}} \;\;
\begin{tikzpicture}[baseline=\mathaxis, line width=0.6]
    \fill[\colorshade] (-1,-1) rectangle (1,1) (0,0) circle (0.8);
    \draw [\colorbound] (0, 0) circle [radius = 0.8];
    \draw [\colorbound, -Stealth] (0.797, 0.05) -- ++(0, 0.05);
    \draw [\colorbound, -Stealth] (-0.797, -0.05) -- ++(0, -0.05) node[shift={(0.3, -0.1)}] {\footnotesize$m$};
    \fill (0, 0) circle [radius = 0.05] node[right] {\small$\cO$};
\end{tikzpicture}
$
\caption{\label{fig: clustering states}%
Left: the clustering states, or vacua, $|\cB_m\rangle$ on the circle are created by the simple boundaries. Right: Correlation functions in those states are computed by disc partition functions.}
\end{figure}

To prove that the states created by simple boundaries satisfy cluster decomposition one uses a boundary crossing relation proven in \cite{Huang:2021zvu}:
\bea
\label{crossing relation}
\begin{tikzpicture}[baseline=\mathaxis,line width=0.6]
    \fill [\colorshade] (1, -1) arc [start angle = 225, end angle = 135, radius = 1.4142] -- cycle;
    \draw [\colorbound] (1, -1) arc [start angle = 225, end angle = 135, radius = 1.4142];
    \draw [\colorbound, -Stealth] (0.586, 0) -- ++(0, 0.1) node[shift={(-0.3, 0.15)}] {\small$m$};
    \fill [\colorshade] (-1, 1) arc [start angle = 45, end angle = -45, radius = 1.4142] -- cycle;
    \draw [\colorbound] (-1, 1) arc [start angle = 45, end angle = -45, radius = 1.4142];
    \draw [\colorbound, -Stealth] (-0.586, 0) -- ++(0, -0.1) node[shift={(0.25, -0.15)}] {\small$n$};
\end{tikzpicture}
\quad = \quad \frac{\delta_{m,n}}{\tilde d_m} \quad
\begin{tikzpicture}[baseline=\mathaxis,line width=0.6]
    \fill [\colorshade] (-1, 1) arc [start angle = -135, end angle = -45, radius = 1.4142] -- (1, -1) arc [start angle = 45, end angle = 135, radius = 1.4142] -- cycle;
    \draw [\colorbound] (-1, 1) arc [start angle = -135, end angle = -45, radius = 1.4142];
    \draw [\colorbound, -Stealth] (0, 0.586) -- ++(0.1, 0) node[shift={(-0.06, 0.3)}] {\small$m$};
    \draw [\colorbound] (1, -1) arc [start angle = 45, end angle = 135, radius = 1.4142];
    \draw [\colorbound, -Stealth] (0, -0.586) -- ++(-0.1, 0) node[shift={(0.06, -0.3)}] {\small$m$};
\end{tikzpicture}
\quad.
\eea
A correlator $\langle \cB_m | \cO | \cB_m \rangle$ is the annulus partition function with boundaries $\cB_m$ and with the insertion of a local operator $\cO$, as in \cref{fig: clustering states} right. Using the crossing relation, such a correlator is equal to $1/\tilde d_m$ times the disc one-point function of $\cO$ with boundary $\cB_m$, that we call $Z_{D_2}[\cO]$. If $\cO$ is the identity operator, the latter is equal to $\tilde d_m$ and thus the correlator is normalized. The crossing relation can then be used to break a disc into two discs, hence one obtains:
\be
\langle \cB_m| \cO_1 \cO_2 | \cB_m \rangle = \frac1{\tilde d_m} \, Z_{D_2}[\cO_1 \cO_2] = \frac1{{\tilde d_m}^{\,\raisebox{0.2em}{\scriptsize2}}} \, Z_{D_2}[\cO_1] \, Z_{D_2}[\cO_2] = \langle
\cB_m | \cO_1 | \cB_m \rangle \, \langle \cB_m | \cO_2 | \cB_m \rangle \;.
\ee
This shows that in the states $|\cB_m\rangle$ all correlation functions are equal to products of one-point functions, which is cluster decomposition for topological theories.

The fusion category $\scrC$ acts on the untwisted circle Hilbert space according to its fusion algebra, as described in \cref{sec: fusion alg}. As is clear from \cref{fig: clustering states} left, the action of the lines on the states is dictated by the fusion coefficients $\tilde N_{am}^n$ (\ref{module fusion algebra}) of the corresponding module category $\scrM$:
\be
\cL_a \, |\cB_m\rangle = \sum\nolimits_n \tilde N_{am}^n \, |\cB_n\rangle \;.
\ee
Thus the clustering states of a gapped phase furnish a NIM-rep of the fusion algebra. This allows one to easily compute the entanglement asymmetry of the various vacua.

In \cref{sec: examples} we analyze various examples of spontaneous symmetry breaking and compute the asymmetry of the different vacua. In \cref{sec: group-like symm} we consider an invertible finite symmetry group $G$ spontaneously broken to a subgroup $H$, and reproduce the result of \cite{Capizzi:2023xaf} that the entanglement and R\'enyi asymmetries are equal to $\Delta S = \log(N_\text{vac})$, where $N_\text{vac} = |G|/|H|$ is the number of vacua.%
\footnote{The result of \cite{Capizzi:2023xaf} used matrix product states. Our derivation, that uses the low-energy TQFT description, is valid in two or more spacetime dimensions.}
In the rest of \cref{sec: examples} we study broken noninvertible symmetries. One of the outcomes is that the asymmetries can take more general values, and are not directly related to the number of vacua. Another outcome is that, as opposed to the invertible case, different vacua can exhibit inequivalent physical properties (as already anticipated in \cite{Huse:1984mn, Damia:2023ses}), and in particular can have different asymmetries. For instance, in a theory that spontaneously breaks the Ising symmetry, one vacuum has $\Delta S = \log 2$ while two other vacua (which are related by the invertible part of the symmetry) have $\Delta S = \frac32 \log 2$ (see \cref{sec: fusion alg of Ising}).

\subsection{Detecting SPT phases}

In the case of an invertible anomaly-free finite symmetry $G$, two-dimensional gapped phases are classified by the conjugacy class of an unbroken subgroup $H \subseteq G$ as well as an SPT phase for $H$ (see \cref{app:group-modules} for the precise classification and construction). In particular, even when the full group $G$ is unbroken, one could have inequivalent trivially-gapped phases with a single vacuum but distinguished by the SPT phase. Since they have a single vacuum, ordinary order parameters cannot detect the different SPT phases, and the asymmetries necessarily vanish. Interestingly, it was noted in \cite{Bhardwaj:2023idu} that in some examples twisted order parameters can distinguish different SPT phases. Can, likewise, one of the symmetry algebras $\cA$ introduced so far be used to distinguish those phases? This question clearly generalizes to anomalous as well as noninvertible symmetries.

Below we show in one example that, indeed, the action of the tube algebra can often distinguish different gapped phases. For that, we need the tube algebra action on TQFT states, including the twisted states. By open/closed duality, this is computed from the data of the corresponding module category $\scrM$. Using the various moves already discussed in the case of fusion categories, in particular those in (\ref{fusion and q dim}) and in \cref{fig: F-moves,fig: F-moves bis}, adapted to module categories, we find the following relation:
\be
    \label{tube on TQFT general}
    \begin{tikzpicture}[line width=0.6,baseline=\mathaxis]
	\fill [\colorshade] (0, -1.2) rectangle (0.5, 1.2);
	\draw [\colorbound, -{|[sep=0.8pt].|}] (0, 0) -- (0, 1.2) node[shift={(0.34, -0.1)}] {\small$m$}; \draw [\colorbound, -stealth] (0, 0) -- +(0, 0.4);
	\draw [\colorbound, -{|[sep=0.8pt].|}] (0, 0) -- (0, -1.2) node[shift={(0.34, 0.1)}] {\small$m$}; \draw [\colorbound, -{stealth[reversed]}] (0, 0) -- +(0, -0.9);
	\draw [\colortopoline] (0, -0.6) node[right, \colortopojun] {\scriptsize$\mu$} -- (-2.4, 1.2) node[left] {\small$h$} node[pos=0.15, above] {\small$g$} node[pos=0.55, below] {\small$d$};
	\draw [\colortopoline, |-] (-0.8, -1.2) node[left] {\small$a$} -- (-0.8, 0) node[above, \colortopojun] {\scriptsize$\alpha$};
	\draw [\colortopoline, -|] (-1.6, 0.6) node[below, \colortopojun] {\scriptsize$\beta$} -- (-1.6, 1.2) node[right] {\small$a$};
	\fill [\colortopojun] (0, -0.6) circle [radius = 1pt];
	\fill [\colortopojun] (-0.8, 0) circle [radius = 1pt];
	\fill [\colortopojun] (-1.6, 0.6) circle [radius = 1pt];
    \end{tikzpicture}
    \, = \, \sum_{n,\nu,\rho,\sigma} \sqrt{ \frac{d_gd_a}{d_d}} \; \bigl[ \tilde F_{agm}^n \bigr]_{(d; \alpha\rho)(m; \mu\sigma)} \, \bigl[ \tilde F_{ham}^n \bigr]_{(d; \beta\rho)(n; \sigma\nu)}^*
    \hspace{-5mm}
    \begin{tikzpicture}[line width=0.6,baseline=\mathaxis]
        \fill [\colorshade] (0, -1.2) rectangle (0.5, 1.2);
        \draw [\colorbound, -{|[sep=0.8pt].|}] (0, 0) -- (0, 1.2) node[shift={(0.34, -0.1)}] {\small$n$}; \draw [\colorbound, -stealth] (0, 0) -- +(0, 0.68);
        \draw [\colorbound, -{|[sep=0.8pt].|}] (0, 0) -- (0, -1.2) node[shift={(0.34, 0.1)}] {\small$n$}; \draw [\colorbound, -{stealth[reversed]}] (0, 0) -- +(0, -0.62);
        \draw [\colortopoline] (0, 0) node[right, \colortopojun] {\scriptsize$\nu$} -- (-1.2, 1.2) node[left] {\small$h$};
        \fill [\colortopojun] (0, 0) circle [radius = 1pt];
    \end{tikzpicture}
    \;.
\ee
The sum is over a basis of morphisms $\cL_h \otimes \cB_n \xrightarrow{\nu} \cB_n$, $\cL_d \otimes \cB_m \xrightarrow{\rho} \cB_n$, $\cL_a \otimes \cB_m \xrightarrow{\sigma} \cB_n$, while the ticks indicate the ``trace'' of \cite{Barkeshli:2014cna}, \ie, that the lines are attached. The configuration on the \rhs, if rotated clockwise by 90 degrees, represents a twisted state $\bigl| \cB_{n, \nu}^{(h)} \bigr\rangle \in \cH_h$, where $\cH_h$ is an $h$-twisted sector of dimension
\be
\dim(\cH_h) = \sum\nolimits_n \tilde N_{hn}^n \;.
\ee
The configuration on the \lhs\ of \eqref{tube on TQFT general}, on the other hand, represents the action of the tube algebra basis element (\ref{tube algebra basis}) on a twisted state $\bigl| \cB_{m,\mu}^{(g)} \bigr\rangle$. In the special case that $g=h=1$ and $d=a$, \eqref{tube on TQFT general} reproduces the fusion algebra action on the untwisted circle Hilbert space $\cH_1$.

\paragraph{Example: SPT phases.} In order to exhibit how the tube algebra action distinguishes different phases, consider the example, proposed in \cite{Bhardwaj:2023idu}, of an invertible symmetry group $G = \bZ_2 \times \bZ_2$ with no anomaly, $\omega = 1$. The indecomposable gapped phases are labeled by conjugacy classes of subgroups $H \subseteq G$ and cocycles $\psi \in H^2\bigl( H, U(1) \bigr)$. In this case the possibilities are: for $H=\{1\}$ the completely broken phase; for the three inequivalent subgroups $H=\bZ_2$ a broken phase with unbroken $\bZ_2$; for $H = G$ two unbroken phases distinguished by the cocycle $\psi \in H^2 \bigl( \bZ_2 \times \bZ_2 , U(1) \bigr) = \bZ_2$.

Let us focus on the two unbroken phases, which are SPT phases. The corresponding module categories have a single simple object that we indicate with a dot, the modules are multiplicity-free since $N_{g \,\bdot}^{\,\bdot} = 1$ for all $g \in G$, and the $\tilde F$-symbols are
\be
\bigl[ \tilde F_{ab\, \bdot}^{\,\bdot} \bigr] = \psi(a,b)
\ee
where $a,b \in G$ and the missing indices are fixed. The trivial cocycle, corresponding to the trivial SPT phase, is $\psi=1$. The non-trivial cocycle, instead, corresponding to the nontrivial SPT phase, has a representative
\be
\psi\bigl( (a_1, a_2) \,,\, (b_1, b_2) \bigr) = (-1)^{a_1 b_2}
\ee
where $a_i, b_j \in \{0, 1\}$. Both phases have four (un)twisted sectors, each with a single state. Since the symmetry is invertible and Abelian, the tube algebra basis elements (\ref{tube algebra basis}) have $g=h$: we label them by $g,a \in G$. They do not change the sector of the state, but, in the nontrivial SPT phase, they act according to the following table:
\be
\begin{array}{|c|cccc|}
\hline
g \backslash a & (0,0) & (1,0) & (0,1) & (1,1) \\
\hline
(0,0) & 1 & 1 & 1 & 1 \\
(1,0) & 1 & 1 & -1 & -1 \\
(0,1) & 1 & -1 & 1 & -1 \\
(1,1) & 1 & -1 & -1 & 1 \\
\hline
\end{array}
\ee
This reproduces the results in Sec.~4.5.4 of \cite{Bhardwaj:2023idu}. In the untwisted sector $g=(0,0)$, all elements $a$ of $G$ act trivially. On the other hand, in each nontrivial twisted sector $\cH_g$, the element with $a=g$ acts trivially but the other two flip the sign of the state.

Extrapolating to the general case, one can often distinguish different SPT phases by the tube algebra representations under which their untwisted and twisted states transform. Unfortunately it may happen that two inequivalent SPT phases realize exactly the same tube algebra representation.%
\footnote{This happens if the Lagrangian algebras in the Drinfeld center of $\scrC$ that determine the gapped phases have exactly the same decomposition in simple lines, and only differ by the product morphism.}


\section{Examples}
\label{sec: examples}

In this section we collect a number of examples in two-dimensional theories with details about the computation of entanglement and R\'enyi asymmetries, in order to exhibit general features.

\subsection{Group-like symmetries}
\label{sec: group-like symm}

Consider an invertible symmetry given by a finite group $G$. The corresponding fusion category is $\mathsf{Vec}_G^\omega$, whose simple lines $\cL_g$ are in correspondence with the elements $g \in G$ and they all have dimension $d_g=1$. The fusion coefficients are $N_{gh}^k = \delta_{(gh)}^k$ where $(gh)$ denotes the group multiplication of $g$ with $h$. In particular, there are no multiplicities at the junctions. The symmetrizer with respect to the fusion algebra is
\be
\label{symmetrizer invertible group}
\rho_\sS = \frac1{|G|} \, \sum\nolimits_{g \in G} \, \cL_g \, \rho \, \cL_{g^{-1}} \;.
\ee
A possible 't~Hooft anomaly is described by a class $\omega \in H^3\bigl( G, U(1) \bigr)$, which fixes the nonvanishing $F$-symbols to be
\be
\bigl[ F_{g,h,k}^{ghk} \bigr]_{gh, hk} = \omega(g,h,k) \;.
\ee
We use normalized representatives $\omega$, meaning that $\omega(g,h,k)=1$ whenever one or more of $g,h,k$ are equal to the identity $1 \in G$.
Different representatives of the same class produce the $F$-symbols in different gauges.

\paragraph{Spontaneous symmetry breaking.} The possible indecomposable $G$-symmetric TQFTs, in correspondence with the possible indecomposable $\scrC$-module categories, are characterized by the conjugation class of a subgroup $H \subseteq G$ (besides a cochain $\psi$, which however will not play a role here), as reviewed in \cref{app:group-modules}. They describe the SSB pattern $G \to H$ at very low energies.
The clustering vacua $|m\rangle$, in correspondence with the simple boundaries $\cB_m$, are labeled by the cosets $m \in G/H$. The vacua furnish a simple NIM-rep of the fusion algebra:
\be
\cL_g \, |m\rangle = |gm\rangle \;.
\ee
Here $g m$ is the coset with representative $ga$, if $a \in G$ is a representative of $m$.

Consider the pure state $\rho = |m \rangle \langle m|$ of a vacuum on the circle. Using (\ref{symmetrizer invertible group}), the corresponding symmetrized density matrix with respect to the fusion algebra is
\be
\rho_\sS = \frac1{|G|} \sum\nolimits_{g\in G} \cL_g \, \rho \, \cL_{g^{-1}} = \frac1{|G|} \sum\nolimits_{g\in G} |gm \rangle \langle gm| = \frac1{|G/H|} \sum\nolimits_{p \in G/H} |p \rangle \langle p | \;.
\ee
In the second equality we used that $\cL_{g^{-1}} = \cL_g^\dag$. Because of the crossing relation \eqref{crossing relation}, the vacua are orthonormal and thus $\Tr(\rho_\sS^n) = \bigl\lvert G/H \bigr\rvert{}^{1-n}$. The asymmetries easily follow:
\be
\Delta S[\rho] = \Delta S^{(n)}[\rho] = \log{} \bigl\lvert G/H \bigr\rvert \;.
\ee
This computation reproduces the result of \cite{Capizzi:2023xaf}, obtained using matrix product states. Notice that the derivation here applies verbatim to spacetime dimensions higher than two as well.

\subsubsection{Tube algebra}

The tube algebra of $\mathsf{Vec}_G^\omega$ (which is given by the twisted quantum double of $G$) is generated by the basis elements:
\be
\tau_{g,a} \;\equiv\;
\begin{tikzpicture}[baseline=\mathaxis, line width=0.6]
    \draw (-0.2, -0.7) node[right] {\small$g$} -- (-0.2, -0.2) -- (0.2, 0.2) node[shift={(185:0.40)}] {\small$ag$} -- (0.2, 0.7) node[right, shift={(0, 0.1)}] {\small$aga^{-1}$};
    \draw [|-] (-0.7, -0.3) node[left] {\small$a$} -- (-0.2, -0.2);
    \draw [-|] (0.2, 0.2) -- (0.7, 0.3) node[right, shift={(0, -0.05)}] {\small$a$};
\end{tikzpicture}.
\ee
The product is given by:
\be
\tau_{h, b} \,\times\, \tau_{g,a} = \delta_{h, aga^{-1}} \; \frac{ \omega(b, aga^{-1}, a) }{ \omega(baga^{-1}b^{-1}, b, a)  \, \omega(b,a,g) } \; \tau_{g, ba} \;.
\ee
From here the structure constants $T_{h,b;\, g,a}^{k,c}$ are immediately read off. The phase factor appearing in the product can be interpreted as a twisted 2-cocycle. Indeed, define
\be
\beta_g(x,y) = \frac{ \omega(g,x,y) \, \omega(x,y, y^{-1} x^{-1} g x y) }{ \omega(x, x^{-1}gx, y) } \;.
\ee
One verifies that this is a normalized twisted 2-cocycle, where $x$ has an adjoint action on $g \in G$ (this also appeared in \cite{Hu:2012wx} Sec.~4.B).%
\footnote{The twisted cocycle condition is $d\beta_g(x,y,z) = \beta_{x^{-1}gx}(y,z) \, \beta_g(x,yz) / \beta_g(xy, z) \beta_g(x,y) = 1$. See also App.~A of \cite{Benini:2018reh} for a review on twisted cocycles.}
We can write the tube algebra product as
\be
\tau_{h, b} \,\times\, \tau_{g,a} = \frac{ \delta_{h, aga^{-1}} }{ \beta_{baga^{-1}b^{-1}}(b,a) } \; \tau_{g, ba} \;.
\ee
The identity element is $\cX_\unit = \sum_g \tau_{g,1}$. Associativity of the product coincides with the twisted cocycle condition for $\beta$. The non-degenerate bilinear form can be written as
\be
K_{h,b;\, g,a} = \frac{ |G| \; \delta_{h, aga^{-1}} \, \delta_{b, a^{-1}} }{ \beta_g(a^{-1}, a) } \;.
\ee
From its inverse matrix $\tilde K^{g,a;\, h,b}$ one obtains the symmetrizer:
\be
\rho_\sS = \frac1{|G|} \, \sum\nolimits_{g,a} \, \beta_g(a^{-1}, a) \; \tau_{g,a} \, \rho \, \tau_{aga^{-1}, a^{-1}} \;.
\ee
Using the twisted cocycle condition one proves the relation $\beta_{aga^{-1}}(a,a^{-1}) = \beta_g(a^{-1},a)$ which is useful to check directly some properties of the symmetrizer.

\subsubsection{Strip algebras}
\label{sec:group-regular-strip-algebras}

Let us now study the case of the interval Hilbert space using the strip algebras. The possible boundary conditions are given in terms of module categories, whose classification in the case of invertible finite symmetries is reviewed in \cref{app:group-modules}.

In the non-anomalous case, one can always choose strongly-symmetric boundary conditions (but see \cref{foo: modules}). From \cref{app:group-modules}, those are classified by $\psi \in H^2\bigl( G,U(1) \bigr)$. Explicitly, one has
\be
\tilde N_{a\, \bdot}^{\,\bdot} = 1 \;,\qquad\qquad \bigl[ \tilde F_{a,b, \,\bdot}^{\,\bdot} \bigr]{}_{c,\,\bdot}^{\phantom{\bdot}} = \bigl( m_{a,b}^c \bigr)^* = \delta^c_{ab} \; \psi(a,b) \;.
\ee
Note that there are no multiplicities at the boundary junctions. In terms of the basis elements $H_g$ defined in (\ref{def strongly sym basis}), we see that the cocycle cancels out from the product \eqref{strip alg product strongly-sym} so that $H_g \times H_h = H_{gh}$. The strip algebra $\mathsf{Strip}_{\mathsf{Vec}_{G}}(\mathsf{Vec})$ is thus isomorphic to the group algebra $\bC[G]$ (which is also the fusion algebra) regardless of the class of $\psi$.%
\footnote{Alternatively, the cocycle can be reabsorbed with a change of basis of the junction vector spaces \cite{Cordova:2024iti}.}
One finds the symmetrizer
\be
\rho_\sS = \frac1{|G|} \, \sum\nolimits_{g\in G} \,  H_g \, \rho \, H_{g^{-1}} \;.
\ee

On the other hand, a boundary condition that can always be chosen regardless of the value of the anomaly cocycle $\omega \in H^3 \bigl( G,U(1) \bigr)$ is the regular module (but see \cref{foo: modules}). In this case we obtain the algebra $\mathsf{Strip}_{\mathsf{Vec}_G^\omega}(\mathsf{Vec}_G^\omega)$, generated by the basis elements:
\be
\def\stripshadew{0.7}
H_{r,m}^g \;\equiv\;\; \strip{r}{m}{g}{gr}{gm}{}{}
\ee
with $g,m,r \in G$ and $\tilde N_{gm}^n = \delta_{(gm)}^n$. The product is given by:
\be
\label{eq:strip-group-anom-prod}
H_{s,n}^b \,\times\, H_{r,m}^a = \delta_{s,ar} \, \delta_{n,am} \; \frac{\omega(b,a,r)}{ \omega(b,a,m)} \; H_{r,m}^{ba} \;.
\ee
Also in this case the phase can be interpreted as a twisted cocycle (in which the action on $r,m$ is in the regular representation) and associativity as its closure. In this case, however, there exists an isomorphism with $\mathsf{Strip}_{\mathsf{Vec}_G}(\mathsf{Vec}_G)$, namely, with the strip algebra of the symmetry without anomaly \cite{Cordova:2024iti}. The isomorphism is given by$\,$%
\footnote{The two algebras share the same underlying vector space, hence we denote a basis of both with the same notation $H_{r,m}^{a}$.}
\bea
\varphi \colon \mathsf{Strip}_{\mathsf{Vec}_G}(\mathsf{Vec}_G) &\to \mathsf{Strip}_{\mathsf{Vec}_G^{\omega}}(\mathsf{Vec}_G^{\omega}) \;, \\
\varphi \bigl(H_{r,m}^a \bigr) &= \omega(a, r, r^{-1}m) \, H_{r,m}^a \;,
\eea
then extended by linearity. Indeed one verifies that
\be
\varphi \bigl( H_{ar, am}^b \bigr) \times_{\omega} \varphi \bigl( H_{r,m}^a \bigr) = \varphi \bigl( H_{ar, am}^b \times_{\omega=1} H_{r,m}^a \bigr)
\ee
using that $d\omega(b,a,r,r^{-1}m) =1$. Here we made explicit where we used the anomalous product \eqref{eq:strip-group-anom-prod} or the non-anomalous one.
The non-degenerate bilinear form can be written as
\be
K_{s,n,b;\, r,m,a} = |G| \; \frac{\omega(a^{-1}, a, r)}{\omega(a^{-1},a,m)} \; \delta_{s,ar} \, \delta_{n,am} \, \delta_{b,a^{-1}} \;.
\ee
From its inverse one obtains the symmetrizer
\be
\label{symmetrizer group G strip regular module}
\rho_\sS = \frac1{|G|} \sum_{r,m,a} \frac{\omega(a^{-1},a,m)}{\omega(a^{-1},a,r)} \; H^a_{r,m} \, \rho \, H^{a^{-1}}_{ar,am} =  \frac1{|G|} \sum_{r,m,a} \, \varphi(H^a_{r,m}) \, \rho \, \varphi(H^{a^{-1}}_{ar,am}) \;.
\ee
The rewriting in terms of $\varphi$ makes it manifest that the symmetrizers for the two strip algebras are the same up to a phase redefinition of the basis elements.

\begin{figure}
\centering$\ds
    \rho^\mathrm{full} = \frac{1}{Z} \times \,
    \begin{tikzpicture}[baseline=\mathaxis, line width = 0.6]
        \node (origin) at (0,0) {};
        \draw[dotted] (-1,-0.05) -- (-1,-0.8) -- (1,-0.8) -- (1,-0.05);
        \draw[dotted] (-1,0.05) -- (-1,0.8) -- (1,0.8) -- (1,0.05);
        \draw[densely dashed] (-1,0.05) -- (1,0.05);
        \draw[densely dashed] (-1,-0.05) -- (1,-0.05);
        \fill (0,0.5) circle (1.5pt) node[right] {$\mathcal O^\dagger$};
        \fill (0,-0.5) circle (1.5pt) node[right] {$\mathcal O$};
    \end{tikzpicture}
    \;\;, \qquad\quad
    \cL_g \, \rho^{\mathrm{full}} \, \cL_{g^{-1}} = \frac{1}{Z} \times \,
    \begin{tikzpicture}[baseline=\mathaxis, line width = 0.6]
        \draw[dotted] (-1,-0.05) -- (-1,-0.8) -- (1,-0.8) -- (1,-0.05);
        \draw[dotted] (-1,0.05) -- (-1,0.8) -- (1,0.8) -- (1,0.05);
        \draw[dashed] (-1,0.05) -- (1,0.05);
        \draw[dashed] (-1,-0.05) -- (1,-0.05);
        \draw[\colortopoline, decoration={markings, mark=at position 0.5 with {\arrow{>}}}, postaction={decorate}] (-1,0.2) -- (1,0.2) node[right, shift={(0, 0.1)}] {\small$g^{-1}$};
        \draw[\colortopoline, decoration={markings, mark=at position 0.5 with {\arrow{>}}}, postaction={decorate}] (-1,-0.2) -- (1,-0.2) node[right, shift={(0, -0.05)}] {\small$g$};
        \fill (0,0.5) circle (1.5pt) node[right] {$\cO^\dag$};
        \fill (0,-0.5) circle (1.5pt) node[right] {$\cO$};
    \end{tikzpicture}
    \!\! = \frac{1}{Z} \times \,
    \begin{tikzpicture}[baseline=\mathaxis, line width = 0.6]
        \draw[dotted] (-1,-0.05) -- (-1,-0.8) -- (1,-0.8) -- (1,-0.05);
        \draw[dotted] (-1,0.05) -- (-1,0.8) -- (1,0.8) -- (1,0.05);
        \draw[dashed] (-1,0.05) -- (1,0.05);
        \draw[dashed] (-1,-0.05) -- (1,-0.05);
        \fill (0,0.5) circle (1.5pt) node[right] {${}^{g\hspace{-0.05em}}\cO^\dag$};
        \fill (0,-0.5) circle (1.5pt) node[right] {${}^{g\hspace{-0.05em}}\cO$};
    \end{tikzpicture}
$
\caption{\label{fig:rho-full}%
Left: path-integral representation of the state on the real line. The operator $\cO$ is inserted at $z=x-i\tau$, while the operator $\cO^\dag$ is inserted at the specular image $\bar z = x + i \tau$. The normalization by the partition function $Z$ ensures that $\Tr \bigl( \rho^\mathrm{full} \bigr) = 1$. Right: path-integral representation of $\cL_g \, \rho^\mathrm{full} \, \cL_{g^{-1}}$.}
\end{figure}

\begin{figure}[t]
\centering$\ds
        \rho_{a,b} \equiv \frac{1}{Z_{a,b}} \times
        \densmatsetup{%
          \densmatvac{a}{b}
          \fill (0, -0.6) circle [radius = 0.04] node[right] {\small$\cO$};
          \fill (0, 0.6) circle [radius = 0.04] node[right] {\small$\cO^\dag$};
        }
        \;, \qquad
        \rho_{ga,gb}^{(g)} \equiv \frac{1}{Z_{ga,gb}^{(g)}} \times
        \densmatsetup{%
          \coordinate (a) at ($(-\densmatgapw/2-\densmatradx,0) + (25:\densmatradx)$);
          \coordinate (b) at ($(\densmatgapw/2+\densmatradx,0) + (135:\densmatradx)$);
          \draw [\colortopoline] (a) -- (b) node [midway,shift={(-0.28, 0.18)}] {\scriptsize$g^{-1}$};
          \draw [\colortopoline] ($(0,0)-(a)$) -- ($(0,0)-(b)$) node [midway,shift={(-0.2, -0.18)}] {\scriptsize$g$};
          \fill (0.1, -0.6) circle [radius = 0.04] node[right] {\small $\cO$};
          \fill (0.1, 0.6) circle [radius = 0.04] node[right] {\small $\cO^\dag$};
          \densmatvac{a}{b}
        }
        = \frac{1}{Z_{a,b}} \times
        \densmatsetup{%
          \densmatvac{ga}{gb}
          \fill (-0.2, -0.6) circle [radius = 0.04] node[right] {\small ${}^{g\hspace{-0.05em}}\cO$};
          \fill (-0.2, 0.6) circle [radius = 0.04] node[right] {\small ${}^{g\hspace{-0.05em}}\cO^\dag$};
        }
$
\caption{\label{fig: group symmetry}%
Left: path-integral construction of the reduced density matrix $\rho_{a,b}$. The dashed lines are the two open boundaries (cuts). The constant $Z_{a,b}$ is the partition function obtained by gluing the two cuts, so as to ensure that $\Tr(\rho_{a,b})=1$. Right: path-integral construction of the rotated density matrix $\rho_{ga,gb}^{(g)}$ where $Z_{ga,gb}^{(g)} \equiv Z_{a,b} \; \omega(g^{-1}, g, a) / \omega(g^{-1}, g, b)$.}
\end{figure}

\paragraph{Asymmetries in CFTs.}
Consider a CFT and a state on the real line $\bR$ created by the Euclidean path-integral on a half space with the insertion of local operators. The corresponding density matrix $\rho^\mathrm{full}$ is prepared by a Euclidean path-integral in which the future half-space is the specular image of the past half-space, and there is a cut along the real line. This is depicted in \cref{fig:rho-full} left. There, $Z$ is the partition function on the plane obtained by closing the cut, so as to ensure that $\Tr\bigl( \rho^\mathrm{full} \bigr) = 1$.

To construct the reduced density matrix for a single interval along the real line, we apply the procedure of \cref{sec: interval}. Consider first the case that we use boundary conditions $a,b$ chosen from the regular module. We depict the path-integral construction of $\rho_{a,b}$ in \cref{fig: group symmetry} left, where $Z_{a,b}$ is the partition function obtained by closing the cut so as to ensure that $\Tr(\rho_{a,b})=1$. For simplicity, in the figures we indicate only one operator insertion $\cO$ and its specular image $\cO^\dag$, but the argument applies to any number of insertions. To obtain the symmetrized density matrix we apply (\ref{symmetrizer group G strip regular module}). Since the symmetry elements $H^g_{r,m}$ project to zero when their boundary conditions do not match those of the state, we remain with a single sum:
\be
\rho_\sS = \frac1{|G|} \sum\nolimits_g \frac{\omega(g^{-1}, g , b) }{ \omega(g^{-1}, g, a)} \, H_{a,b}^g \, \rho_{a,b} \, H_{ga ,gb}^{g^{-1}} \,\equiv\, \frac1{|G|} \sum\nolimits_g \rho_{ga,gb}^{(g)} \;.
\ee
In the last equality we introduced the rotated density matrices $\rho_{ga,gb}^{(g)}$ whose path-integral construction is depicted in \cref{fig: group symmetry} right. They belong to $\End(\cH_{ga,gb})$. The constant prefactor is $Z_{ga,gb}^{(g)} = Z_{a,b} \; \omega(g^{-1},g,a) / \omega(g^{-1}, g, b)$. To obtain the second presentation one moves the line of $H_{a,b}^g$ through infinity and then annihilates it with $H_{ga,gb}^{g^{-1}}$. When the line crosses an operator $\cO$, the latter gets transformed to ${}^{g\hspace{-0.05em}}\cO$. Since Hilbert spaces constructed with different simple boundary conditions are orthogonal, the moments of $\rho_\sS$ simplify:
\be
\Tr \bigl( \rho_\sS^n \bigr) = \frac1{|G|^n} \sum\nolimits_g \Tr \bigl[ (\rho_{ga,gb}^{(g)})^n \bigr] = \frac1{|G|^{n-1}} \, \Tr \bigl( \rho_{a,b}^n \bigr) \;.
\ee
The R\'enyi and entanglement asymmetries are then
\be
\Delta S^{(n)}[ \rho_{a,b} ] = \Delta S[\rho_{a,b}] = \log |G| \qquad\qquad\text{for all $n$} \,.
\ee
This result does not depend on the state, and in particular it applies to the vacuum as well. Its interpretation is that the boundary conditions completely break the symmetry, therefore $\Delta S^{(n)}$ and $\Delta S$ reach their maximal value irrespective of the state. This observable is clearly not very interesting. To make it interesting, we should rather use a symmetric combination of boundary conditions, as described in \cref{non-symmetric bc}.

Consider then the ``averaged'' reduced density matrix
\be
\rho = \frac1{|G|^2} \sum\nolimits_{a,b} \rho_{a,b} \;.
\ee
The symmetrized matrix this time can be written as
\be
\rho_\sS = \frac1{|G|^3} \sum\nolimits_{a,b,g} \rho_{a, b}^{(g)} \;,
\ee
where $\rho_{a,b}^{(g)}$ is defined in \cref{fig: group symmetry} right. One computes
\be
\Tr\bigl( \rho^n \bigr) = \frac1{|G|^{2n}} \sum\nolimits_{a,b} \Tr \bigl( \rho_{a,b}^n \bigr) \;,\qquad
\Tr\bigl( \rho_\sS^n \bigr) = \frac1{|G|^{3n}} \sum_{a,b} \sum_{g_1, \dots, g_n} \Tr\Bigl( \rho_{a, b}^{(g_1)} \cdots \rho_{a, b}^{(g_n)} \Bigr) \;.
\ee
From the path-integral representation of $\rho_{a,b}$ (see \cref{fig: group symmetry}),  $\Tr \bigl( \rho_{a,b}^n \bigr)$ is equal to $Z_{a,b}^{-n}$ times the partition function on an $n$-sheeted Riemann surface with operators $\cO$ and $\cO^\dag$ inserted in each replica. In this manifold, the $n$ copies of the boundary with boundary condition $a$ are glued together to form a single boundary that spans all replicas, and the same holds for the boundaries with boundary condition $b$.
Similarly, $\Tr\bigl( \rho_{a, b}^{(g_1)} \cdots \rho_{a, b}^{(g_n)} \bigr)$ is equal to $\prod_i 1/Z_{g_i^{-1}a, g_i^{-1}b}$ times the partition function on an $n$-sheeted Riemann surface with operators ${}^{g_j\hspace{-0.05em}} \cO$ and ${}^{g_j\hspace{-0.05em}} \cO^\dag$ inserted in the $j$-th replica and with one $a$ and one $b$ boundary.

In the limit $\varepsilon \to 0$, the dependence of the traces on the boundary conditions becomes very simple. As described in (\ref{eq:shrinking-discs-Z}), in the limit each partition function produces a factor of the boundary quantum dimension from each shrinking disc. Thus, the partition function on the $n$-sheeted Riemann surface becomes $\tilde d_a \tilde d_b$ times the partition function on a similar Riemann surface, where the boundary discs are shrunk to a branch point and all insertions of fields remain at the same position. For a group-like symmetry all boundary quantum dimensions $\tilde d$ are equal. We thus obtain the same factor ${\tilde d}^{\, 2(1-n)}$ from both $\lim_{\varepsilon \to 0} \Tr(\rho^n)$ and $\lim_{\varepsilon \to 0} \Tr(\rho_\sS^n)$, which simplifies in the ratio.
Besides, once the $\varepsilon \to 0$ limit is taken, there is no dependence on $a$ and $b$ left and we can easily sum over them. The result can be expressed in terms of correlation functions on a Riemann surface with branch points:
\be
\label{eq:asymmetry-group-after-shrink}
\lim_{\varepsilon \to 0} \; \Delta S^{(n)}[\rho] = \frac1{1-n} \log \biggl[\, \frac{1}{|G|^n} \sum_{g_1, \dots, g_n} \frac{ \langle\, {}^{g_1\hspace{-0.05em}} \cO^\dag \, {}^{g_1\hspace{-0.05em}} \cO \;\cdots\; {}^{g_n\hspace{-0.05em}} \cO^\dag \, {}^{g_n\hspace{-0.05em}} \cO \,\rangle_{\cM_n} }{ \langle \, \cO^\dag \, \cO \,\cdots\, \cO^\dag \, \cO \, \rangle_{\cM_n} } \biggr] \;,
\ee
where the notation $\langle \, \cdot \, \rangle_{\cM_n}$ indicates a correlation function on the branched Riemann surface, as detailed in \cref{fig:correlator-riemann-surface}.

\begin{figure}[t]
\centering$\ds
    \def\tikzscale{0.8}
    \def\baseline{-10.6ex}
    \def\disp{0.2}
	\bigl\langle \, {}^{g_1\hspace{-0.05em}}\cO^\dag \, {}^{g_1\hspace{-0.05em}}\cO \;\cdots\; {}^{g_n\hspace{-0.05em}}\cO^\dag \, {}^{g_n\hspace{-0.05em}}\cO \, \bigr\rangle_{\cM_n} \;\equiv\;\;
        \left\langle \;
            \scalebox{\tikzscale}{
            \begin{tikzpicture}[scale=1/\tikzscale, baseline=\baseline, line width = 0.6]
                  \draw[line width=0.8, dotted] (0, 0.1) -- (2, 0.1) -- (1.5, -1.1) -- (-0.5, -1.1) -- cycle; 
                  \fill (0.3, -0.5) circle (1.pt); 
                  \fill (1.2, -0.5) circle (1.pt); 
                  \draw[densely dashed]  (0.3, -0.5) -- (1.2, -0.5); 
                  \coordinate (middle) at (0.75, -0.5); 
                  \fill (middle) +(\disp/2, \disp/1.1) circle (0.8pt) node[shift={(0.1, 0.27)}] {${}^{g_1\hspace{-0.05em}} \cO^\dag$}; 
                  \fill (middle) +(-\disp/2, -\disp/1.1) circle (0.8pt) node[shift={(0.0, -0.24)}] {${}^{g_1\hspace{-0.05em}} \cO$}; 
                  \node at (0.75, -1.26) {$\vdots$}; 
                  \begin{scope}[yshift = -1.7cm]
                      \draw[line width=0.8, dotted] (0, 0.1) -- (2, 0.1) -- (1.5, -1.1) -- (-0.5, -1.1) -- cycle; 
                      \fill (0.3, -0.5) circle (1.pt);
                      \fill (1.2, -0.5) circle (1.pt);
                      \draw[densely dashed]  (0.3, -0.5) -- (1.2, -0.5);
                      \coordinate (middle) at (0.75, -0.5);
                      \fill (middle) +(\disp/2, \disp/1.1) circle (0.8pt) node[shift={(0.1, 0.27)}] {${}^{g_n\hspace{-0.05em}} \cO^\dag$};
                      \fill (middle) +(-\disp/2, -\disp/1.1) circle (0.8pt) node[shift={(0.0, -0.24)}] {${}^{g_n\hspace{-0.05em}} \cO$};
                  \end{scope}
            \end{tikzpicture}}
\; \right\rangle
$
\caption{\label{fig:correlator-riemann-surface}%
Correlator on the $n$-sheeted Riemann surface $\cM_n$. It is understood that the lower rim of each cut (denoted by a dashed segment) is glued to the upper rim of the cut of the sheet above it (and the uppermost sheet is connected to the lowermost sheet). Two branch points sit at the ends of the branch cut. In the $j$-th replica the operator ${}^{g_j\hspace{-0.05em}}\cO$ is inserted at $z = x-i \tau$ while the operator ${}^{g_j\hspace{-0.05em}} \cO^\dag$ is inserted at $\bar z = x + i \tau$.}
\end{figure}

We make three observations. First, if there are no insertions, \ie, if we compute the asymmetry of the vacuum, then all correlators in \eqref{eq:asymmetry-group-after-shrink} are equal to $1$ and thus $\Delta S^{(n)} = \Delta S = 0$, as expected from the general discussion in \cref{non-symmetric bc}. Second, the asymmetries have no explicit dependence on the anomaly, as already anticipated in \cite{Benini:2024xjv}. Third, in the absence of an anomaly, one could compute the asymmetries using the strongly-symmetric boundary condition, and this would lead to exactly the same formula. In particular, earlier works on asymmetry for groups ignored the entangling surface and used the group algebra instead of the strip algebra; we see here that those less rigorous approaches are in fact validated by our more rigorous approach of using the strip algebra.

\paragraph{Large subsystem limit.}
In the limit of a large subsystem of size $\ell$, the branch cut extends along the whole real axis. Then effectively each sheet gets split into two half-planes, and each lower half-plane is glued to the upper half-plane of the next sheet (\textit{cfr.}\ with \cref{fig:correlator-riemann-surface}). One can perform a conformal block expansion of the $2n$-point functions appearing in \eqref{eq:asymmetry-group-after-shrink}, in the channel in which each operator is first contracted with the operator found across the cut after the gluing. When that channel is dominated by exchanges of the identity operator, the $2n$-point function factorizes into the product of $n$ two-point functions. This is always the case for the $2n$-point function at denominator in \eqref{eq:asymmetry-group-after-shrink}. When exchanges of the identity operator are not possible in the numerator, the contribution to the sum in \eqref{eq:asymmetry-group-after-shrink} is subdominant --- and on the other hand some of the two-point functions vanish. Therefore, in the large subsystem limit we find the formula:
\be
\label{eq:half-space-fact}
\lim_{\ell \to \infty} \; \lim_{\varepsilon \to 0} \; \Delta S^{(n)}[\rho] = \frac{1}{1-n} \log \Biggl[ \, \frac{1}{|G|^n} \sum_{g_1, \dots, g_n}
\frac{ \langle {}^{g_1\hspace{-0.05em}} \cO^\dag \, {}^{g_2\hspace{-0.05em}} \cO \rangle_\bC \,\cdots\, \langle {}^{g_n\hspace{-0.05em}} \cO^\dag \, {}^{g_1\hspace{-0.05em}} \cO \rangle_\bC }{ \langle  \cO^\dag \cO \rangle_\bC^n } \Biggr] \,.
\ee
We claim that this is the same as the R\'enyi asymmetry of the full system, \ie, of the state on the real line symmetrized with respect to the fusion algebra $\bC[G]$. Indeed, let $\rho^\mathrm{full}$ be the state on the real line, depicted in \cref{fig:rho-full} left. The symmetrized density matrix for the full system is
\be
\rho^\mathrm{full}_\sS = \frac{1}{|G|} \, \sum\nolimits_{g\in G} \, \cL_g \, \rho^\mathrm{full} \, \cL_{g^{-1}} \;.
\ee
The summand $\cL_g \, \rho^\mathrm{full} \, \cL_{g^{-1}}$ is depicted in \cref{fig:rho-full} right. The R\'enyi asymmetry is then
\be
\label{eq:full-system-renyies-group}
\Delta S^{(n)} \bigl[ \rho^\mathrm{full} \bigr] = \frac{1}{1-n} \log \Biggl\{ \, \frac{1}{\abs{G}^n} \, \Tr \Biggl[ \, \frac1{Z^n} \, \Biggl( \, \sum_{g \in G} \;
            \begin{tikzpicture}[baseline=\mathaxis]
                \draw[dotted] (-1, -0.05) -- (-1, -0.6) -- (1, -0.6) -- (1, -0.05);
                \draw[dotted] (-1, 0.05) -- (-1, 0.6) -- (1, 0.6) -- (1, 0.05);
                \draw[dashed] (-1, 0.05) -- (1, 0.05);
                \draw[dashed] (-1, -0.05) -- (1, -0.05);
                \fill (-0.1, 0.35) circle (1.5pt) node[right] {${}^g \mathcal O^\dag$};
                \fill (-0.1, -0.35) circle (1.5pt) node[right] {${}^g \mathcal O$};
            \end{tikzpicture}
        \; \Biggr)^{\!\!n} \; \Biggr]
\Biggr\} \,.
\ee
We used that $\rho^\mathrm{full}$ is pure and thus $\Tr\bigl[ (\rho^\mathrm{full})^n \bigr] = 1$.
The $n$-th power of the operator appearing inside the trace is obtained by taking $n$ copies of the cut plane and then connecting the lower rim of each cut (which is the ``output'' of the operator) to the upper rim of the next sheet (which is the ``input''). The trace connects them cyclically, in a fashion similar to  \cref{fig:correlator-riemann-surface}. The partition functions can then be replaced by expectation values, since both at numerator and denominator the required normalization is given by the same partition function with no operator insertions. We conclude that (\ref{eq:full-system-renyies-group}) can be written as in (\ref{eq:half-space-fact}) and therefore
\be
\lim_{\ell \to \infty} \; \lim_{\varepsilon \to 0} \; \Delta S^{(n)} [\rho] = \Delta S^{(n)} \bigl[ \rho^\mathrm{full} \bigr] \;.
\ee

\subsection{Ising symmetry --- circle Hilbert space}
\label{sec: Ising}

The internal symmetry of the two-dimensional Ising CFT is described by a fusion category $\scrC$ (that we call the Ising fusion category) with 3 simple lines, that we indicate as $\cL_1$, $\cL_\eta$, $\cL_\cN$, satisfying the commutative fusion ring:
\be
\cL_\eta \times \cL_\eta = \cL_1 \;,\qquad\qquad \cL_\eta \times \cL_\cN = \cL_\cN \;,\qquad\qquad \cL_\cN \times \cL_\cN = \cL_1 + \cL_\eta \;.
\ee
Here $\cL_1$ is the identity, $\cL_\eta$ generates the $\bZ_2$ spin-flip symmetry, and $\cL_\cN$ is the line of Kramers--Wannier duality. The junctions have no multiplicities. The nontrivial $F$-symbols are:
\be
\bigl[ F_{\eta\cN\eta}^\cN \bigr]_{\cN\cN} = \bigl[ F_{\cN \eta \cN}^\eta \bigr]_{\cN\cN} = -1 \;,\qquad \bigl[ F_{\cN \cN \cN}^\cN \bigr]_{ab} = \frac1{\sqrt2} \mat{ 1 & 1 \\ 1 & -1 } \;,
\ee
where $a,b \in \{ 1,\eta \}$, while all other $F$-symbols allowed by fusion are equal to 1. All lines are self-dual. One reads off the Frobenius--Schur (FS) indicators $\varkappa_\eta = \varkappa_\cN = 1$ and the quantum dimensions $d_\eta =1$, $d_\cN = \sqrt2$. The category $\scrC$ is the Tambara--Yamagami fusion category $\mathsf{TY}(\bZ_2, \chi, +)$ \cite{Tambara:1998vmj} where $\chi(a,b) = e^{i\pi ab}$ is a symmetric nondegenerate $\bZ_2$ bicharacter that equals $-1$ only if $a,b$ are both the nontrivial element $\eta$, otherwise it equals 1.%
\footnote{The $\mathsf{TY}(\bZ_2, \chi, +)$ symmetry appears, for instance, in a system of $\nu = 1,7,9, 15$ (mod 16) Dirac fermions. A slight generalization is $\mathsf{TY}(\bZ_2, \chi, -)$, whose only difference is that $\bigl[ F_{\cN\cN\cN}^\cN \bigr]_{ab} = - \chi(a,b)/\sqrt2$ and thus $\varkappa_\cN=-1$, which appears, for instance, in a system of $\nu=3,5,11,13$ (mod 16) Dirac fermions.}

\subsubsection{Fusion algebra}
\label{sec: fusion alg of Ising}

The fusion algebra has three 1-dimensional representations that we indicate as $\unit$, $\varepsilon$, $\sigma$.
In the Ising CFT these are realized in the Verma module of the identity, the energy, and the spin operator, respectively. The fusion algebra acts on them through (\ref{reps of Abelian fusion alg}), in terms of the unitary and symmetric modular $S$-matrix:
\be
\label{eq:ising-smatrix}
S = \frac12 \,\mat{ 1 & 1 & \sqrt2 \\ 1 & 1 & - \sqrt2 \\ \sqrt2 & -\sqrt2 & 0 } \;.
\ee
The symmetrizer for the fusion algebra is given by the inverse bilinear form:
\be
\tilde K = \mat{ \frac38 & -\frac18 & 0 \\ -\frac18 & \frac38 & 0 \\ 0 & 0 & \frac14} \;,
\ee
and in particular it takes the form:
\be
\label{symmetrizer fusion Ising}
\rho_\mathsf{S} = \frac18 \Bigl[ 3 \, \rho + 3 \, \cL_\eta \, \rho \, \cL_\eta - \cL_\eta \, \rho - \rho \, \cL_\eta + 2 \, \cL_\cN \, \rho \, \cL_\cN \Bigr] \;.
\ee

\paragraph{Asymmetry of excited states in CFTs.}
We consider a CFT whose symmetry includes the Ising fusion category. This could be the Ising CFT itself, or a more complicated one (for instance, the tricritical Ising model). The local operators fall into representations of the fusion algebra, that we indicate as $\unit$, $\varepsilon$, $\sigma$ as in \eqref{eq:ising-smatrix}.
We consider excited states on the line $\bR$ produced using a scalar operator $\varepsilon(z)$ of dimension $\Delta_\varepsilon$ in the Euclidean past, which by definition transforms in the representation $\varepsilon$:
\be
\label{eq:state-1-epsilon}
\ket{\psi} \;=\;\;
        \begin{tikzpicture}[baseline=-0.5cm, line width = 0.6]
        \draw[dotted] (-1,0) -- (-1,-0.8) -- (1,-0.8) -- (1,0);
        \draw[densely dashed] (-1,0) -- (1,0);
        \fill (-0.6, -0.4) circle (1.5pt) node[right, shift={(0.05,0)}] {$1\! +\! \lambda \, \varepsilon $};
    \end{tikzpicture}
\;\;.
\ee
Here $\lambda$ is a constant of mass dimension $-\Delta_\varepsilon$. Notice that this state is neutral under the $\bZ_2$ spin-flip symmetry. To compute its asymmetry with respect to the fusion algebra we can use (\ref{asym Abelian on circle}). Setting the operator insertion at $z = - i\tau$ with $\tau>0$ (thus $\bar z = i\tau$), the relevant two-point function is $\bigl\langle \varepsilon(\bar z) \, \varepsilon(z) \bigr\rangle \equiv \langle \varepsilon^2 \rangle = 1 / (2\tau)^{2\Delta_\varepsilon}$. The second R\'enyi and entanglement asymmetries are then
\be
\label{eq:renyi-asymm-1+eps-full}
\Delta S^{(2)} \bigl[ |\psi \rangle \langle \psi | \bigr] = - \log \Biggl[ \frac{1 + \lambda^4 \langle \varepsilon^2 \rangle^2}{ \bigl( 1 + \lambda^2 \langle \varepsilon^2 \rangle \bigr)^2} \Biggr] \;,\qquad
\Delta S \bigl[ |\psi \rangle \langle \psi | \bigr] = f \biggl[ \frac1{1 + \lambda^2 \langle \varepsilon^2 \rangle} \biggr] + f \biggl[ \frac{ \lambda^2 \langle \varepsilon^2 \rangle }{ 1 + \lambda^2 \langle \varepsilon^2 \rangle} \biggr] \;,
\ee
where we used the function $f(x) = - x \log x$.

\begin{figure}[t]
\centering
\includegraphics[width = 0.5\textwidth]{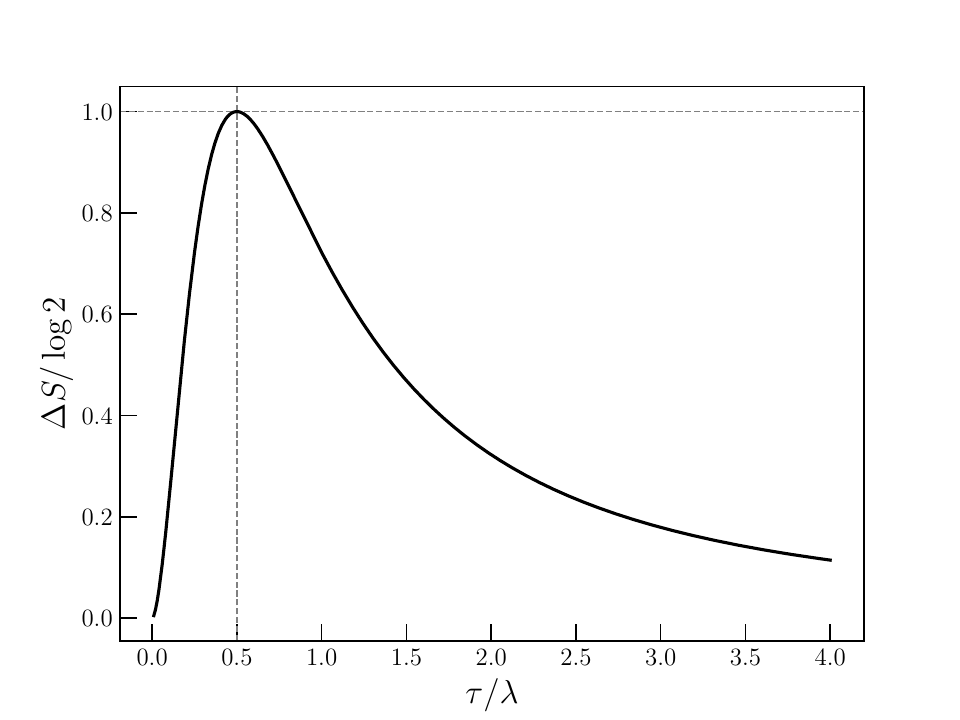}
\caption{\label{fig: plot asym fusion Ising}%
Entanglement asymmetry $\Delta S$ with respect to the Ising fusion algebra of a state on the real line prepared by inserting $(1 + \lambda\,\varepsilon)$ with $\Delta_\varepsilon = 1$ in the path-integral. The asymmetry is plotted as a function of $\tau / \lambda$, where $-\tau$ is the Euclidean time of the insertion. In the limit $\tau / \lambda \to \infty$ only the identity contributes, so the state becomes symmetric and the asymmetry vanishes. Similarly, in the limit $\tau / \lambda \to 0$ only $\varepsilon$ contributes, leading again to a symmetric state with vanishing asymmetry. The asymmetry reaches its maximum at $\tau / \lambda = 1/2$.}
\end{figure}

In the case of the Ising CFT we can take $\varepsilon(x)$ to be the energy conformal primary operator of dimension $\Delta_\varepsilon=1$. In \cref{fig: plot asym fusion Ising} we plot the entanglement asymmetry as a function of $\tau/\lambda$.

\paragraph{Spontaneous symmetry breaking.}
The only indecomposable $\scrC$-module category of the Ising fusion category is the regular one.%
\footnote{By Th.~6 of \cite{Ostrik:2001xnt} (where the Ising category is called $\scrC_2$ and the module is of type $A_3$), the Ising fusion category admits only one indecomposable module category. Hence, this must be the regular module, which always exists.}
This means that in a gapped phase the Ising symmetry is always completely spontaneously broken, and this is due to the presence of an 't~Hooft anomaly.
In the regular module one could indicate both lines and boundary conditions using the same set of labels. However, in order to adhere to a more commonly used notation, we relabel the boundary conditions as:%
\footnote{These boundary conditions have a counterpart in a lattice spin model: $\pm$ correspond to up/down fixed boundary conditions for the spins, while $f$ corresponds to the free boundary condition.}
\be
1 \;\to\; + \;,\qquad\qquad \eta \;\to\; - \;,\qquad\qquad \cN \;\to\; f \;.
\ee
Correspondingly, we indicate the three normalized vacua (\ie, the three clustering states) as $\bigl\{ |+\rangle \,,\, |-\rangle \,,\, |f\rangle \bigr\}$. They furnish a NIM-rep of the fusion algebra:
\bea
\cL_\eta \, |+\rangle &= |-\rangle \;,\qquad& \cL_\eta\, |-\rangle &= |+\rangle \;,\qquad& \cL_\eta\, |f\rangle &= |f\rangle \;, \\
\cL_\cN \, |+\rangle &= |f\rangle \;,\qquad& \cL_\cN \, |-\rangle &= |f\rangle \;,\qquad& \cL_\cN \, |f\rangle &= |+\rangle + |-\rangle \;.
\eea
The gapped phase (some curious properties thereof were already noticed in \cite{Huse:1984mn}) can be obtained, for instance, from the tricritical Ising model deforming it by the relevant operator $\varepsilon'$, as explained in Sec.~7.2.3 of \cite{Chang:2018iay}.

Consider the pure state $\rho = |f \rangle \langle f |$ for the vacuum $|f\rangle$. Using the symmetrizer (\ref{symmetrizer fusion Ising}) with respect to the fusion algebra, one computes $\rho_\sS$. Its diagonalization is $\rho_\sS \sim \text{diag}\bigl( \frac12, \frac12, 0 \bigr)$ and therefore $\Tr (\rho_\sS^n) = 2^{1-n}$. The R\'enyi and entanglement asymmetries are then
\be
\Delta S^{(n)} = \Delta S = \log 2 \;.
\ee
Notice that this is the same asymmetry as for a spontaneously broken $\bZ_2$ symmetry \cite{Capizzi:2023yka}.%
\footnote{This is explained by the fact that the ``orbit'' of $|f\rangle$ under the noninvertible symmetry is given by two rays, the ones of $|f\rangle$ and $\bigl( |+\rangle + |-\rangle \bigr)$, as it happens for a $\bZ_2$ symmetry.}
On the other hand, consider the pure state $\rho = |+\rangle \langle + |$ for the vacuum $|+\rangle$ (the vacuum $|-\rangle$ would give the same result). The symmetrized matrix $\rho_\sS$ has diagonalization $\rho_\sS \sim \text{diag}\bigl( \frac12, \frac14, \frac14 \bigr)$ and hence $\Tr( \rho_\sS^n ) = \bigl( 2^{n-1} + 1 \bigr) / 2^{2n-1}$, from which the R\'enyi and entanglement asymmetries are
\be
\Delta S^{(n)} = \frac1{1-n} \log \frac{ 2^{n-1} + 1 }{ 2^{2n-1}}  \;,\qquad\qquad \Delta S = \frac32 \log 2 \;.
\ee
This time the entanglement asymmetry is larger than for a broken $\bZ_2$, but smaller than for a broken $\bZ_3$. We conclude that, while for a completely broken invertible symmetry $G$ the entanglement asymmetry of a vacuum is $\Delta S = \log |G|$ \cite{Capizzi:2023xaf}, for noninvertible symmetries the asymmetry may take more general values. We also observe that different vacua in the same gapped phase can have inequivalent physical properties (as already stressed in \cite{Huse:1984mn, Damia:2023ses}) and in particular different asymmetries.

\subsubsection{Tube algebra}

The tube algebra is generated by the following 12 basis elements (see also \cite{Bartsch:2023wvv}):
\bea
\label{Ising tube elements}
& \tube{1}{1}{}{1} \;,\; \tube{1}{\eta}{}{1} \;,\; \tube{1}{\cN}{}{1} \;,\; \tube{1}{\cN}{}{\eta} \;,\; \tube{\eta}{1}{}{\eta} \;,\; \tube{\eta}{\eta}{}{\eta} \;,\; \tube{\eta}{\cN}{}{1} \;,\; \tube{\eta}{\cN}{}{\eta} \;, \\
& \tube{\cN}{1}{}{\cN} \;,\; \tube{\cN}{\eta}{}{\cN} \;,\; \tube{\cN}{\cN}{1}{\cN} \;,\; \tube{\cN}{\cN}{\eta}{\cN} \;.
\eea
By applying the formula (\ref{product tube algebra reduced}) we compute the algebra product. We use the bicharacter $\chi(a,b)$ defined on $\bZ_2 = \{1, \eta\}$ above. As in the general discussion, we adopt the convention that the element on the left is stacked on top of the one on the right. We find:
\bea
\tube{a}{c}{}{a} \,\times \tube{a}{b}{}{a} &\,= \tube{a}{cb}{}{a} \;,\qquad& \tube{\cN}{c}{}{\cN} \,\times \tube{\cN}{b}{}{\cN} &\,= \chi(b,c) \, \tube{\cN}{cb}{}{\cN} \;, \\
\tube{b}{c}{}{b} \,\times \tube{a}{\cN}{}{b} &\,= \chi(a,c) \, \tube{a}{\cN}{}{b} \;,\qquad& \tube{a}{\cN}{}{b} \,\times \tube{a}{c}{}{a} &\,= \chi(b,c) \, \tube{a}{\cN}{}{b} \;,
\eea
as well as
\bea
\tube{\cN}{b}{}{\cN} \,\times \tube{\cN}{\cN}{a}{\cN} &\,= \makebox[0pt][l]{$\ds \tube{\cN}{\cN}{a}{\cN} \,\times \tube{\cN}{b}{}{\cN} = \chi(b, ab) \, \tube{\cN}{\cN}{ab\,}{\cN}\;, $} \\
\tube{b}{\cN}{}{c} \,\times \tube{a}{\cN}{}{b} &\,= \delta_{ac} \sum_d \chi(b, da) \, \tube{a}{d}{}{a} \;, \\
\tube{\cN}{\cN}{b}{\cN} \,\times \tube{\cN}{\cN}{a}{\cN} &\,= \sum_d \frac{ \chi(a,b) \chi(ab,d) }{\sqrt2} \, \tube{\cN}{d}{}{\cN} \;.
\eea
The unit of the algebra is
\be
\cX_\unit = \tube{1}{1}{}{1} \;+\; \tube{\eta}{1}{}{\eta} \;+\; \tube{\cN}{1}{}{\cN} \;,
\ee
which agrees with \eqref{identity element}.

Untwisted and twisted point operators fall into representations of the tube algebra. Such representations are in one-to-one correspondence with simple objects in the Drinfeld center of $\scrC$. Because $\scrC$ is modular, its Drinfeld center is isomorphic to the product of two copies of the TY fusion category, therefore it has 9 simple lines, listed in \cite{Bartsch:2023wvv}.

The total circle Hilbert space \eqref{eq:htube} is decomposed as a sum of twisted sectors, each graded by a simple element $a \in \Irr(\scrC)$ of the symmetry category. Similarly, one can show that (see \eg \cite{Kawagoe:2024tgv}) any representation of the tube algebra has a grading by elements in $\Irr(\scrC)$: $V = \bigoplus_a V_a$. This grading is such that elements of the tube algebra that have $g$ as the lower vertical line act as zero on those $V_{g'}$ with $g' \neq g$. In other words, the twist of the tube algebra must be compatible with that of the state it is acting on. When listing the irreducible representations of the tube algebra, we thus explicitly report how each vector space decomposes in graded components.
In this case one finds:
\begin{itemize}
\item Two 1-dimensional irreducible representations $\cF_1^\pm$ with space $V = \bC_1$, and
\be
\cF_1^\pm \biggl(\!\! \tube{1}{a}{}{1} \biggr) = 1 \;,\qquad\qquad \cF_1^\pm \biggl(\!\! \tube{1}{\cN}{}{1} \biggr) = \pm \sqrt2 \;,
\ee
while all other ones are mapped to zero. They represent genuine local operators, not charged under $\bZ_2$, which could be charged under $\cN$ but that remain genuine local operators under its action.

\item Two 1-dimensional irreducible representations $\cF_\eta$ and $\cF_\eta^*$ with space $V = \bC_\eta$, and
\be
\cF_\eta \biggl(\!\! \tube{\eta}{a}{}{\eta} \biggr) = \chi(a,\eta) \;,\qquad\qquad \cF_\eta \biggl(\!\! \tube{\eta}{\cN}{}{\eta} \biggr) = i \, \sqrt2 \;,
\ee
while all other ones are mapped to zero. They represent $\eta$-twisted operators that are charged under $\bZ_2$, and that remain twisted-sector operators under the action of $\cN$.

\item Four 1-dimensional irreducible representations $\cF_\cN^\pm$ and $(\cF_\cN^\pm)^*$ with space $V = \bC_\cN$, and
\bea
\cF_\cN^\pm\biggl(\!\! \tube{\cN}{1}{}{\cN} \biggr) &= 1 \;,\qquad&
\cF_\cN^\pm\biggl(\!\! \tube{\cN}{\eta}{}{\cN} \biggr) &= i \;, \\
\cF_\cN^\pm\biggl(\!\! \tube{\cN}{\cN}{1}{\cN} \biggr) &= \pm e^{\frac{i\pi}8} \;,\qquad&
\cF_\cN^\pm\biggl(\!\! \tube{\cN}{\cN}{\eta}{\cN} \biggr) &= \pm e^{-\frac{3i\pi}8} \;,
\eea
while all other ones are mapped to zero. They represent $\cN$-twisted operators, charged under $\eta$ and $\cN$, that remain $\cN$-twisted operators under the action of $\cN$.

\item One 2-dimensional irreducible representation $\cF_2$ with space $V = \bC_1 \oplus \bC_\eta$, and
\bea
\label{eq:irrep-tube-ising-2d}
\cF_2\biggl(\!\! \tube{1}{1}{}{1} \biggr) &= \biggl( \begin{matrix} 1 & 0 \\ 0 & 0 \end{matrix} \biggr) ,\; &
\cF_2\biggl(\!\! \tube{1}{\eta}{}{1} \biggr) &= \biggl( \begin{matrix} -1 & 0 \\ 0 & 0 \end{matrix} \biggr) ,\; &
\cF_2\biggl(\!\! \tube{1}{\cN}{}{\eta} \biggr) &= \biggl( \begin{matrix} 0 & 0 \\ \sqrt2 & 0 \end{matrix} \biggr) \\
\cF_2\biggl(\!\! \tube{\eta}{1}{}{\eta} \biggr) &= \biggl( \begin{matrix} 0 & 0 \\ 0 & 1 \end{matrix} \biggr) ,\; &
\cF_2\biggl(\!\! \tube{\eta}{\eta}{}{\eta} \biggr) &= \biggl( \begin{matrix} 0 & 0 \\ 0 & 1 \end{matrix} \biggr) ,\; &
\cF_2\biggl(\!\! \tube{\eta}{\cN}{}{1} \biggr) &= \biggl( \begin{matrix} 0 & \sqrt2 \\ 0 & 0 \end{matrix} \biggr)
\eea
while all other ones are mapped to zero. It represents a doublet of a genuine local operator $\cO_1$ and an $\eta$-twisted operator $\cO_\eta$ that are exchanged by $\cN$. Note that $\cO_1$ is charged under $\bZ_2$ while $\cO_\eta$ is neutral.
\end{itemize}

\paragraph{Asymmetry of the tube algebra.} The inverse bilinear form, in the ordered basis (\ref{Ising tube elements}), reads: $\tilde K = \frac14 \diag\bigl( 1, 1, 1, 0, 1, 1, 0, -1, 1, -1, \frac1{\sqrt2}, -\frac1{\sqrt2} \bigr) + \frac14 \operatorname{offdiag}_{[4,7]} + \frac1{4\sqrt2} \operatorname{offdiag}_{[11,12]}$, where the first term is a diagonal matrix while $\operatorname{offdiag}_{[i,j]}$ is an off-diagonal matrix with a 1 at positions $(i,j)$ and $(j,i)$. The symmetrizer can be computed from $\tilde K$.

Let us specialize to a state belonging to the 2d irreducible representation \eqref{eq:irrep-tube-ising-2d} of the tube algebra. For example, we can choose a pure state $\cO_1(z) |0\rangle$ created by the untwisted local operator $\cO_1$, which in the Ising model is $\sigma(z)$. In this case the only terms that survive in the symmetrized density matrix are
\be
\label{eq:symm-tube-ising-2d}
\rho_\sS = \frac14 \biggl( \rho \; + \tube{1}{\eta}{}{1} \,\, \rho \, \tube{1}{\eta}{}{1} + \tube{1}{\cN}{}{\eta} \,\, \rho \, \tube{\eta}{\cN}{}{1} \biggr) \;.
\ee
One immediately obtains $\Delta S = \log(2)$.
More in general, the asymmetry of a pure state that belongs to a single $d$-dimensional irreducible representation of an algebra is $\Delta S = \log(d)$.
Indeed, following the discussion in \cref{fig: action on matrices}, the symmetrized density matrix is supported on a single block, where it takes the form $\frac{1}{\d} \Id_{d\times d}$. Because the entropy of the initial state is zero, one obtains that both the entanglement and R\'enyi asymmetries are equal to $\log(d)$.

\subsection[\tps{\bZ_2 \times \bZ_2}{Z₂×Z₂} Tambara--Yamagami fusion category]{\btps{\bZ_2 \times \bZ_2}{Z₂×Z₂} Tambara--Yamagami fusion category}
\label{sec: Z2xZ2 TY}

The Ising (or $\bZ_2$ TY) fusion category does not admit a fiber functor. This can be seen, for example, from the fact that the noninvertible line has non-integer quantum dimension. Therefore, the symmetry carries an 't~Hooft anomaly and does not admit strongly-symmetric boundary conditions (in the language of \cite{Choi:2023xjw}). Furthermore, it does not admit weakly-symmetric boundary conditions either, because the only indecomposable module is the regular one, which is not weakly symmetric.

However, one can take the stacking of two decoupled copies of the Ising symmetry and then look at a subsymmetry therein \cite{Choi:2023xjw, Perez-Lona:2023djo, Diatlyk:2023fwf}. The full set of simple lines of the stacking is made of objects of the form $a_1 \boxtimes a_2$ with $a_j \in \{ 1, \eta, \cN\}$. We consider the subcategory generated by
\be
\cL_1 = 1 \boxtimes 1 \;,\quad \cL_\eta = \eta \boxtimes 1 \;,\quad \cL_{\eta'} = 1 \boxtimes \eta \;,\quad \cL_{\eta\eta'} = \eta \boxtimes \eta \;,\quad \cL_\cV = \cN \boxtimes \cN \;.
\ee
Their fusion ring is Abelian and given by:
\bea
\label{tyz2z2 fusion ring}
\cL_\eta^2 &= \cL_{\eta'}^2 = \cL_{\eta\eta'}^2 = \cL_1 \;,\qquad\qquad& \cL_\eta \times \cL_\cV &= \cL_{\eta'} \times \cL_\cV = \cL_\cV \;, \\
\cL_\eta &\times \cL_{\eta'} = \cL_{\eta\eta'} \;, & \cL_\cV \times \cL_\cV &= \cL_1 + \cL_\eta + \cL_{\eta'} + \cL_{\eta\eta'} \;.
\eea
This is in fact the fusion ring of $\bZ_2 \times \bZ_2$ Tambara--Yamagami fusion categories \cite{Tambara:1998vmj}. In this case, the $\bZ_2 \times \bZ_2$ symmetric bilinear character is $\gamma\bigl( (s_1, s_2) , (s_3, s_4) \bigr) = e^{i\pi (s_1 s_3 + s_2 s_4)}$ where each $s_j \in \{0,1\}$, the Frobenius--Schur indicator is $\varkappa_\cV = 1$, therefore the nontrivial $F$-symbols are
\be
\bigl[ F_{i \,\cV\, j}^\cV \bigr]_{\cV\cV} = \bigl[ F_{\cV \, i \, \cV}^j \bigr]_{\cV \cV} = \gamma(i,j) \;,\qquad \bigl[ F_{\cV \cV \cV}^\cV \bigr]_{ij} = \frac12 \, \gamma(i,j) \;,
\ee
where $i,j \in \bZ_2^2$. The four invertible lines have quantum dimension 1, while $\cL_\cV$ has quantum dimension 2. This fusion category is also equivalent to $\mathsf{Rep}(H_8)$, where $H_8$ is the Kac--Paljutkin Hopf algebra \cite{kats1966finite}. As all categories of representations of a Hopf algebra, it admits at least one fiber functor given by the forgetful functor that maps a representation to its underlying vector space \cite{etingof2016tensor}. In this case one can show that this is the only fiber functor.%
\footnote{Indeed, there is a unique maximal haploid algebra object in $\mathsf{Rep}(H_8)$ as shown in \cite{Diatlyk:2023fwf}.}
We conclude that the $\mathsf{Rep}(H_8)$ symmetry admits one (and only one) strongly-symmetric boundary condition. This is our motivation to study this example.

\subsubsection{Fusion algebra}

The fusion algebra has five 1-dimensional representations that we call $\unit$, $\varepsilon$, $\sigma$, $\sigma'$, $\sigma''$.%
\footnote{In the case of two decoupled copies of the Ising CFT, the nine conformal primaries split as $1 \supset 1 \boxtimes 1, \varepsilon \boxtimes \varepsilon$; $\varepsilon \supset \varepsilon \boxtimes 1, 1 \boxtimes \varepsilon$; $\; \sigma \supset \sigma \boxtimes 1, \sigma \boxtimes \varepsilon$; $\; \sigma' \supset 1 \boxtimes \sigma, \varepsilon \boxtimes \sigma$; $\; \sigma'' \supset \sigma \boxtimes \sigma$.}
The unitary (but not symmetric) $S$-matrix is
\be
S = \frac1{2\sqrt2} \, \left( \begin{matrix} 1 & 1 & \sqrt2 & \sqrt2 & \sqrt2 \\ 1 & 1 & -\sqrt2 & \sqrt2 & -\sqrt2 \\ 1 & 1 & \sqrt2 & -\sqrt2 & -\sqrt2 \\ 1 & 1 & -\sqrt2 & -\sqrt2 & \sqrt2 \\ 2 & -2 & 0 & 0 & 0 \end{matrix} \right) \;.
\ee
The symmetrizer for the fusion algebra is
\be
\rho_\sS = \frac1{32} \biggl( 8 \!\! \sum_{a \,\in\, \bZ_2 \times \bZ_2} \!\! \cL_a \,\rho\, \cL_a - B \,\rho\, B + 4 \, \cL_\cV \,\rho\, \cL_\cV \biggr) \qquad\text{with}\qquad B = \! \sum_{a \,\in\, \bZ_2 \times \bZ_2} \!\! \cL_a \;.
\ee

\paragraph{Asymmetry of an excited state in CFT.}
Let us compute the asymmetry of excited states on the line $\bR$ produced by Euclidean operator insertions in CFTs. For concreteness, consider the states
\be
|\psi \rangle = \Bigl( 1 + \lambda\, \varepsilon(-i\tau) \Bigr) \, |0\rangle \;,
\ee
that are neutral under the invertible part of the symmetry. Using (\ref{asym Abelian on circle}), we obtain exactly the same formula (\ref{eq:renyi-asymm-1+eps-full}) as before. In the case of two decoupled copies of the Ising CFT, we can take $\varepsilon(z)$ to be the energy operator of one of the two copies of Ising, so that $\Delta_\varepsilon = 1$. We obtain exactly the same function as in \cref{fig: plot asym fusion Ising}.

\subsubsection{Strip algebra}

\paragraph{Strongly-symmetric boundary condition.}
The left-module category with a single object is described by the fusion coefficients $\tilde N_{a \,\bdot}^\bdot = \{1,1,1,1,2\}$. This means that the $\bZ_2^2$ lines $\cL_{\eta^m \eta^{\prime n}}$ have a single junction to the boundary, while the noninvertible line $\cL_\cV$ has two. Writing the $\tilde F$-symbols in terms of $m$-symbols,
\be
\bigl[ \tilde F_{ab \, \bdot}^{\,\bdot} \bigr]_{(c; - \rho)( \cdot\, ; \nu\mu)} = \bigl( m_{a\mu, b\nu}^{c\rho} \bigr)^* \;,
\ee
the nontrivial ones are given by \cite{Perez-Lona:2023djo, Diatlyk:2023fwf}:%
\footnote{Which we have adapted to our choice of unitary gauge.}
\begin{align}
m_{i,j}^{(ij)} &= K_{ij} \;,\qquad
m_{\eta ,\, \cV \mu}^{\cV \rho} = m_{\cV \rho ,\, \eta}^{\cV \mu} = \sqrt2 \, m_{\cV \mu ,\, \cV \rho}^\eta = m_{\eta' ,\, \cV \rho}^{\cV \mu} = m_{ \cV \mu ,\, \eta'}^{\cV \rho} = \sqrt2 \, m_{\cV \rho ,\, \cV \mu}^{\eta'} = M_{\mu\rho} \;, \nn \\[0.5em]
m_{1 ,\, \cV \mu}^{\cV \rho} &= m_{\cV \mu ,\, 1}^{\cV \rho} = \sqrt2 \, m_{\cV \mu ,\, \cV \rho}^1 = \delta_{\mu\rho} \;,\qquad
m_{\eta\eta' ,\, \cV \mu}^{\cV \rho} = m_{\cV \mu ,\, \eta\eta'}^{\cV \rho} = \sqrt2 \, m_{\cV \mu ,\, \cV \rho}^{\eta\eta'} = (\sigma_z)_{\mu\rho} \;,
\end{align}
in terms of matrices
\be
K_{ij} = \left( \begin{matrix} 1 & 1 & 1 & 1 \\ 1 & 1 & -i & i \\ 1 & i & 1 & -i \\ 1 & -i & i & 1 \end{matrix} \right) \,,\quad
M_{\mu\rho} = \left( \begin{matrix} 0 & \xi^{-1} \\ \xi & 0 \end{matrix} \right) \,,\quad \xi = e^{\frac{3\pi i}4} \,,\quad (\sigma_z)_{\mu\rho} = \left( \begin{matrix} 1 & 0 \\ 0 & -1 \end{matrix} \right) \;.
\ee
The $m$-symbols satisfy the associativity condition (\ref{associativity of m}), while the $\tilde F$-symbols satisfy the left-module pentagon equation (\ref{left-module pentagon}). We chose a gauge in which the $\tilde F$-symbols are unitary.

\paragraph{Strip algebra.} For the unique strongly-symmetric boundary condition described above, the strip algebra $\mathsf{Strip}_{\mathsf{Rep}(H_8)}(\mathsf{Vec})$ has dimension 8. The fusion coefficients are $\tilde N_{a \, \bdot}^{\bdot} = d_a$ and the basis elements are
\bea
\strip{}{}{a}{}{}{\rho}{\mu}
\eea
where $a \in \{1,\eta,\eta', \eta\eta', \cV\}$, and for $a=\cV$ the junctions are labeled by $\rho,\mu \in \{0,1\}$. We will use the notation $H_i$ with $i \in \bZ_2 \times \bZ_2$ and $H_\cV^{\rho,\mu}$ for the basis elements. Applying (\ref{strip alg product strongly-sym}) we find the following algebra product (see also \cite{Choi:2024wfm}):
\begin{align}
    H_\cV^{s,s'} \times H_{\eta^m \eta^{\prime n}} &= e^{- \frac{i\pi}2 (s - s')(m - n + 2mn)} \, H_\cV^{s+m+n, s' + m + n} \;, \qquad\qquad
    H_i \times H_j = H_{(ij)} \;, \nn\\
    H_{\eta^m \eta^{\prime n}} \times H_\cV^{s,s'} &= e^{\frac{i\pi}2 (s - s')(m - n + 2mn)} \, H_\cV^{s+m+n, s' + m + n} \;, \nn\\
    H_\cV^{s_1, s_2} \times H_\cV^{s_3, s_4} &= \delta_{s_1 s_3} \delta_{s_2 s_4} \Bigl( H_1 + (-1)^{s_1 + s_2} H_{\eta\eta'} \Bigr) \\
    &\quad + \delta_{s_1, 1-s_3} \delta_{s_2, 1-s_4} \Bigl( e^{\frac{i\pi}2 (s_1 - s_2)} H_\eta + e^{\frac{i\pi}2 (s_2 - s_1)} H_{\eta'} \Bigr) \;. \nn
\end{align}
The symmetrizer, computed with (\ref{symmetrizer}), reads:
\be
\label{symmetrizer of H8}
\rho_\sS = \frac18 \, \Biggl[ \; \sum_{j \,\in\, \bZ_2 \times \bZ_2} H_j \, \rho \, H_j + \sum_{s,\, s' \,\in\, \{0,1\}} H_\cV^{s,s'} \, \rho \, H_\cV^{s,s'} \,\Biggr] \;.
\ee

Strip algebras constructed using strongly-symmetric boundary conditions are Hopf algebras. We review the definition and some properties of Hopf algebras in \cref{app: Hopf algebras}. In this case one obtains an eight-dimensional Hopf algebra which is neither commutative nor cocommutative. It is known that there is only one such Hopf algebra, the Kac--Paljutkin $H_8$ Hopf algebra \cite{kats1966finite}. This algebra is not isomorphic to the group algebra of some group, nor to the dual of a group algebra, and it is the smallest-dimensional algebra with such properties \cite{Masuoka:1995, Sage:2012}. The coproduct is given by
\be
\Delta(H_i) = H_i \otimes H_i \;,\qquad\qquad \Delta\bigl( H_\cV^{s,s'} \bigr) = \frac1{\sqrt2} \, \sum\nolimits_{t \,\in\, \{0,1\}} \, H_\cV^{s,t} \otimes H_\cV^{t, s'} \;.
\ee
The counit and the antipode map are
\be
\epsilon(H_i) = 1 \;,\qquad \epsilon\bigl( H_\cV^{s,s'} \bigr) = \sqrt2 \, \delta^{s,s'} \;,\qquad S(H_i) = H_i \;,\qquad S\bigl( H_\cV^{s,s'} \bigr) = H_\cV^{s',s} \;.
\ee
In the mathematical literature, the algebra $H_8$ is usually presented (see, \eg, Sec.~4.2 of \cite{Sage:2012}) as generated by the elements $1,x,y,z$ where $1$ is the identity while
\be
x^2 = y^2 = 1 \;,\quad z^2 = \tfrac12 \bigl( 1+x+y-xy \bigr) \;,\quad xy=yx \;,\quad xz = zy \;,\quad yz = zx \;.
\ee
The basis elements are thus $\{ 1,x,y,z, xy, xz, yz, xyz \}$. The Hopf algebra structure is
\bea
\epsilon(x) &= 1 \;,\qquad& S(x) &= x \;,\qquad& \Delta(x) &= x \otimes x \;, \\
\epsilon(y) &= 1 \;,\qquad& S(y) &= y \;,\qquad& \Delta(y) &= y \otimes y \;, \\
\epsilon(z) &= 1 \;,\qquad& S(z) &= z \;,\qquad& \Delta(z) &= \tfrac12 \bigl( z \otimes z + z \otimes xz + yz \otimes z - yz \otimes xz \bigr) \;.
\eea
An isomorphism is given by
\be
1 = H_1 \;,\quad x = H_\eta \;,\quad y = H_{\eta'} \;,\quad z = \frac{1-i}{\sqrt8} \, H_\cV^{0,0} + \frac12 \Bigl( H_\cV^{1,0} + H_\cV^{0,1} \Bigr) + \frac{1+i}{\sqrt8} \, H_\cV^{1,1} \,.
\ee

The algebra has four 1-dimensional and one 2-dimensional irreducible representations. The four 1d representations $\cF_1^\pm$ and $\cF_\eta^\pm$ are
\bea
&\cF_1^\pm \colon \qquad& &H_i \mapsto 1 \;,& H_\cV^{s,s'} &\,\mapsto\, \pm \sqrt2 \; \delta^{s,s'} \;, \\
&\cF_\eta^\pm \colon \qquad& &H_1 \,,\, H_{\eta\eta'} \mapsto 1 \;,\quad H_\eta \,,\, H_{\eta'} \mapsto -1 \;,\quad& H_\cV^{s,s'} &\,\mapsto\, \pm \sqrt2 \; \delta^{s,s'} \, e^{i\pi s s'} \;.
\eea
The 2d representation is given by
\be
\cF_2 \colon \qquad H_{\eta^m \eta^{\prime n}} \,\mapsto\, \biggl( \begin{matrix} (-1)^m & 0 \\ 0 & (-1)^n \end{matrix} \biggr) \;,\qquad H_\cV^{s,s'} \,\mapsto\, \delta^{s,1-s'} \, \sqrt2 \; \biggl( \begin{matrix} 0 & (i)^{s'} \\ (-i)^{s'} & 0 \end{matrix} \biggr) \;.
\ee
The projector, or minimal central idempotent (MCI), to the trivial representation is
\be
P_\epsilon = \frac18 \Bigl( H_1 + H_\eta + H_{\eta'} + H_{\eta\eta'} + \sqrt2\; H_\cV^{0,0} + \sqrt2\; H_\cV^{1,1} \Bigr) \;.
\ee
This is the only MCI on which the counit $\epsilon$ takes value 1. Applying the formula (\ref{symmetrizer Hopf explicit}) for the symmetrizer with respect to a Hopf algebra, one reproduces the result (\ref{symmetrizer of H8}) obtained using the general formula.

\subsubsection{Example: the \btps{(\text{Ising})^2}{(Ising)²} CFT}
\label{sec: example Ising2}

Consider the theory given by two decoupled copies of the Ising CFT (which has central charge $c=1$ and is equivalent to an orbifold of the free compact boson at radius $\sqrt2$ times the self-dual radius). We study a subsymmetry of this theory, given by the $\bZ_2 \times \bZ_2$ Tambara--Yamagami fusion category for the specific choice of bicharacter and Frobenius--Schur indicator already made after (\ref{tyz2z2 fusion ring}), and equivalent to the $\mathsf{Rep}(H_8)$ fusion category. This category admits a strongly-symmetric boundary condition, with which we construct the strip algebra $H_8$.

\paragraph{Case $\boldsymbol{1 \boxtimes 1 + \lambda \, \varepsilon \boxtimes 1}$.}
We study the state on the real line prepared by inserting the operator $\cO(z) = 1 \boxtimes 1 + \lambda \, \varepsilon(z) \boxtimes 1$ at position $z= x - i \tau$. Since $\varepsilon$ has scaling dimension $\Delta_\varepsilon = 1$, the coefficient $\lambda$ has dimensions of a length. The path-integral that prepares the state is done on a half-plane. We study the asymmetry of the reduced density matrix on an interval $A = \bigl[ -\frac\ell2, \frac\ell2 \bigr]$, using the setup of \cref{sec:strongly-symm-bc}: to prepare the state we cut out two discs of size $\varepsilon$ around the entangling surface centered at $\bigl( \pm \frac\ell2, 0 \bigr)$ and impose the strongly-symmetric boundary condition $|\cB\rangle$ there. We compute the second R\'enyi asymmetry using the symmetrization formula derived above. This produces
\be
\Tr \bigl( \rho_\sS^2 \bigr) = \frac18 \Tr\biggl[ \, \sum_{j \,\in\, \bZ_2 \times \bZ_2} \rho \, H_j \, \rho \, H_j + \sum_{s,s'} \rho \, H_\cV^{s,s'} \rho \, H_\cV^{s,s'} \biggr]
= \frac12 \Tr \bigl(  \rho^2 + \rho \rho' \bigr) \;.
\ee
Here $\rho'$ is obtained by swiping a line $\cL_\cV$ across the whole density matrix, transforming the operator $\cO = 1 \boxtimes 1 + \lambda \, \varepsilon \boxtimes 1$ to $\cO' = 1 \boxtimes 1 - \lambda \, \varepsilon \boxtimes 1$. We thus get:
\be
\label{eq:second-renyi-1+epsilon-intermediate-step}
\Delta S^{(2)}[\rho] = - \log \frac{ \Tr \bigl( \rho_\sS^2 \bigr) }{ \Tr \bigl( \rho^2 \bigr) }
= \log 2 - \log\biggl[ 1 + \frac{ \Tr\bigl( \rho \rho' \bigr) }{ \Tr \bigl( \rho^2 \bigr) } \biggr] \;.
\ee
We can express $\Tr(\rho \rho') / \Tr(\rho^2)$ as a ratio of partition functions on a Riemann surface obtained by gluing two copies of the complex plane with the entangling discs cut out. In the $\varepsilon \to 0$ limit, each entangling disc is replaced by a branch point and the partition functions are multiplied by the $g$-factor $g_\cB = \langle \cB | 0 \rangle$ (we already discussed such a procedure in \cref{non-symmetric bc}). The $g$-factors at numerator and denominator are equal and cancel out. The same procedure is discussed in the paragraph above \eqref{eq:asymmetry-group-after-shrink}. In turn, the ratio of partition functions is equal to a ratio of correlation functions on a Riemann surface $\cM_2$ with two branch points. We denote a point on $\cM_2$ by $z_j$ where $z \in \bC$ denotes a point on the complex plane while $j \in \{1,2\}$ labels the sheet. We get:
\be
\label{eq:ratio-correlators-epsilon}
\frac{\Tr(\rho \rho') }{ \Tr( \rho^2) } = \frac{
\Bigl\langle 1+ 2 \lambda^2 \Bigl( \varepsilon(z_1) \, \varepsilon(\bar z_1) - \varepsilon(z_1) \, \varepsilon(z_2) - \varepsilon(z_1) \, \varepsilon(\bar z_2) \Bigr) + \lambda^4 \, \varepsilon(z_1) \, \varepsilon(\bar z_1) \, \varepsilon(z_2) \, \varepsilon(\bar z_2) \Bigr\rangle_{\!\cM_2} }
{ \Bigl\langle 1 + 2 \lambda^2 \Bigl( \varepsilon(z_1) \, \varepsilon(\bar z_1) + \varepsilon(z_1) \, \varepsilon(z_2) + \varepsilon(z_1) \, \varepsilon(\bar z_2) \Bigr) + \lambda^4 \, \varepsilon(z_1) \, \varepsilon(\bar z_1) \, \varepsilon(z_2) \, \varepsilon(\bar z_2) \Bigr\rangle_{\!\cM_2} } .
\ee
We used the symmetry that cyclically exchanges the sheets, and that the 3-point function of $\varepsilon$ vanishes. We can also write
\be
\label{second trace in Ising2}
\frac{\Tr(\rho_\sS^2) }{ \Tr( \rho^2) } = 1 - \frac{
2\lambda^2 \, \Bigl\langle \varepsilon(z_1) \varepsilon(z_2) + \varepsilon(z_1) \varepsilon(\bar z_2) \Bigr\rangle_{\!\cM_2} }
{ \Bigl\langle 1 + 2 \lambda^2 \Bigl( \varepsilon(z_1) \varepsilon(\bar z_1) + \varepsilon(z_1) \varepsilon(z_2) + \varepsilon(z_1) \varepsilon(\bar z_2) \Bigr) + \lambda^4 \, \varepsilon(z_1) \varepsilon(\bar z_1) \varepsilon(z_2) \varepsilon(\bar z_2) \Bigr\rangle_{\!\cM_2} } .
\ee
Correlation functions on $\cM_2$ are obtained from the ones on the plane by a conformal transformation:
\be
\label{eq:conformal-map-two-replicas}
z_1 \,\mapsto\, \sqrt{ \frac{ z - \ell/2}{ z + \ell/2} } \;\equiv\, g(z) \;,\qquad\qquad z_2 \,\mapsto\, -g(z) \;,
\ee
where the square root is taken with its branch cut along the negative real axis. The 2-point functions are then:
\be
\bigl\langle \varepsilon(z_1) \, \varepsilon(\bar z_{1,2}) \bigl\rangle_{\!\cM_2} = \frac{ \bigl\lvert g'(z) \bigr\rvert{}^{2 \Delta_\varepsilon} }{ \bigl\lvert g(z) \mp g(\bar z) \bigr\rvert{}^{2\Delta_\varepsilon} } \;,\qquad\qquad
\bigl\langle \varepsilon(z_1) \, \varepsilon(z_2) \bigr\rangle_{\!\cM_2} = \frac{ \bigl\lvert g'(z) \bigr\rvert{}^{2\Delta_\varepsilon} }{ \bigl\lvert 2g(z) \bigr\rvert{}^{2\Delta_\varepsilon} } \;.
\ee
The 4-point function on $\cM_2$ is related to the one on the complex plane as
\be
\bigl\langle \varepsilon(z_1) \, \varepsilon(\bar z_1) \, \varepsilon(z_2) \, \varepsilon(\bar z_2) \bigr\rangle_{\!\cM_2} = \bigl\lvert g'(z) \bigr\rvert{}^{4 \Delta_\varepsilon} \, \bigl\langle \varepsilon \bigl( g(z) \bigr) \, \varepsilon \bigl( g(\bar z) \bigr) \, \varepsilon \bigl( -g(z) \bigr) \, \varepsilon \bigl( -g(\bar z) \bigr) \bigr \rangle_\bC \;,
\ee
and the latter is given by \cite{Mattis:1986mj}:
\be
\bigl\langle \varepsilon(w_1) \, \varepsilon(w_2) \, \varepsilon(w_3) \, \varepsilon(w_4) \bigr\rangle_\bC = \biggl\lvert \frac{ 1 - \chi + \chi^2 }{ w_{14}\, w_{23} \, \chi} \biggr\rvert^2 \;,
\ee
where $w_{ij} = w_i - w_j$ and $\chi = (w_{12} \, w_{34} ) / ( w_{13} \, w_{24})$ is the conformal cross-ratio.

\begin{figure}[t]
    \centering
    \includegraphics[width=0.45\textwidth]{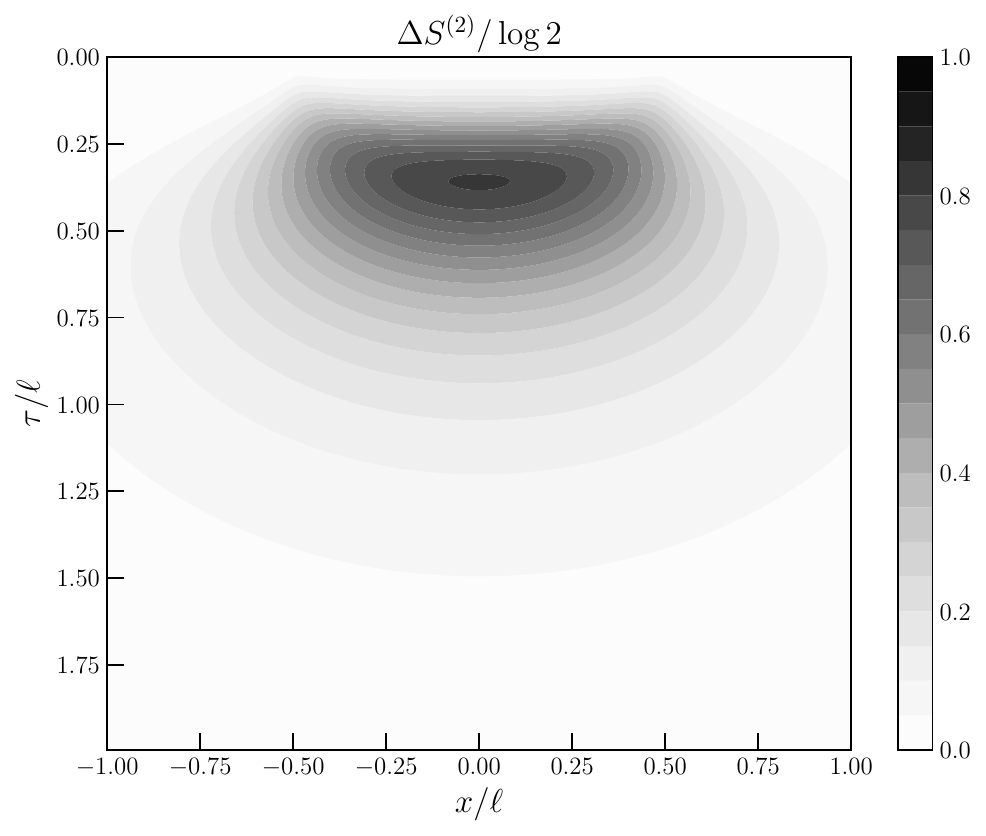}\qquad
    \includegraphics[width=0.45\textwidth]{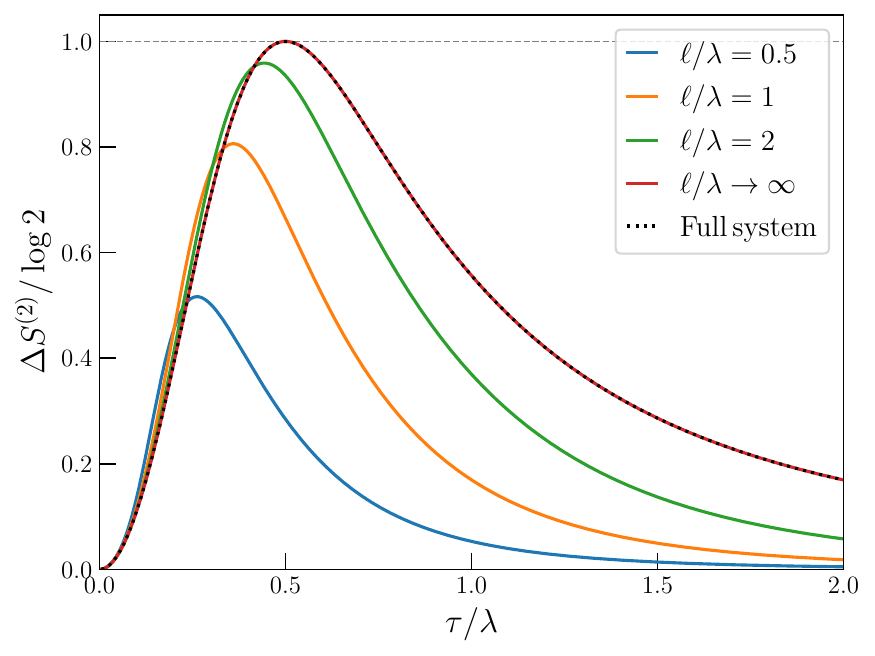} \\
    \includegraphics[width=0.45\linewidth]{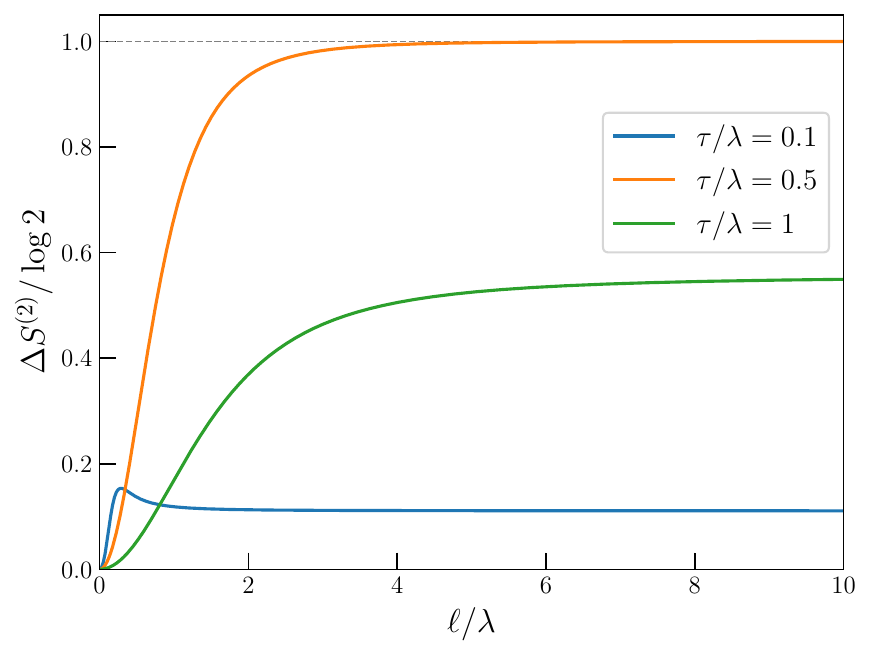}\hspace{7mm}
    \includegraphics[width=0.45\linewidth]{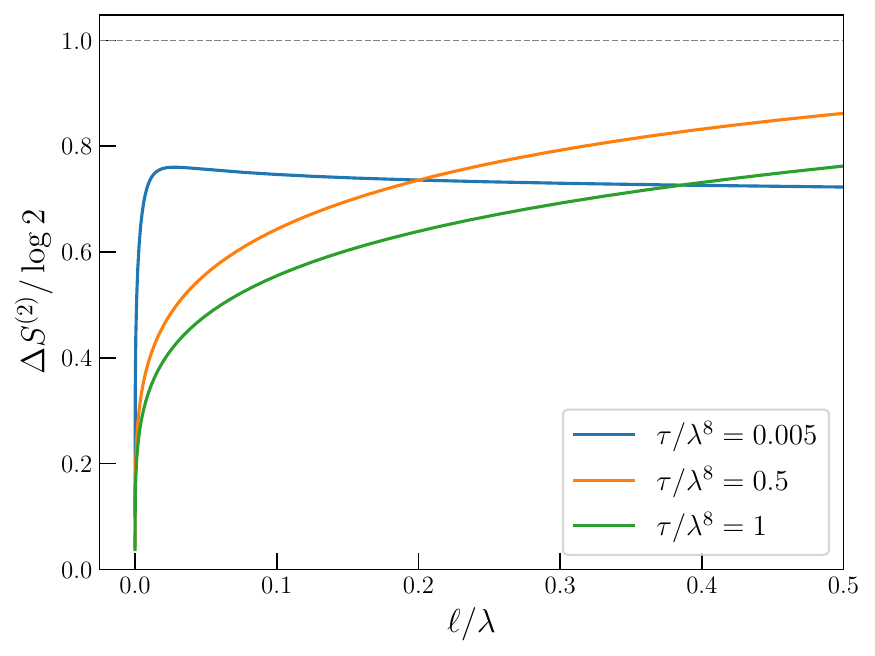}
    \caption{\label{fig:asymmetry_1+epsilon}%
      Top left: density plot of the second R\'enyi asymmetry of the state $\bigl( 1 \boxtimes 1 + \lambda \, \varepsilon(z) \boxtimes 1 \bigr) |0\rangle$ for $\lambda = \ell = 1$ and with $z=x-i\tau$.
      Top right: second R\'enyi asymmetry when the field is inserted at $z=-i\tau$ as a function of $\tau$, for different values of $\ell$ (in units of $\lambda$).
      Bottom left: second R\'enyi asymmetry as a function of $\ell$, for different positions of the field insertion at $z = -i\tau$.
      Bottom right: second R\'enyi asymmetry of the state $\bigl( 1 + \lambda \, \sigma(-i\tau) \bigr)|0\rangle$ in the Ising CFT with respect to the ordinary $\bZ_2$ spin-flip symmetry (that acts as the strip algebra of the $\mathsf{Vec}_{\bZ_2}$ fusion category on the interval Hilbert space with symmetric boundary conditions).}
\end{figure}

We report a few plots of the resulting second R\'enyi asymmetry in \cref{fig:asymmetry_1+epsilon}. Notice that the asymmetry takes values in the interval $[0, \log 2]$. The asymmetry vanishes in both limits $\tau \to \infty$ and $\tau \to 0$: indeed the state is dominated by $1 \boxtimes 1 \, |0\rangle$ or by $\varepsilon \boxtimes 1 \, |0\rangle$, respectively.

\paragraph{Large subsystem limit.} When $\ell \to \infty$, the branch cut extends along the real line and correlation functions factorize. The 2-point functions behave as
\be
\bigl\langle \varepsilon(z_1) \, \varepsilon(\bar z_1) \bigl\rangle_{\!\cM_2} \sim \bigl\langle \varepsilon(z_1) \, \varepsilon(z_2) \bigl\rangle_{\!\cM_2} \sim \frac1{\ell^2} \;,\qquad\qquad \bigl\langle \varepsilon(z_1) \, \varepsilon(\bar z_2) \bigl\rangle_{\!\cM_2} \sim \frac1{(2\tau)^2} \;,
\ee
while the 4-point function factorizes through the identity channel:
\be
\bigl\langle \varepsilon(z_1) \, \varepsilon(\bar z_1) \, \varepsilon(z_2) \, \varepsilon(\bar z_2) \bigr\rangle_{\!\cM_2} \sim \frac1{(2\tau)^4} \;.
\ee
This gives the asymptotic behavior of the second R\'enyi asymmetry:
\be
\label{large subsystem limit in Ising2}
\Delta S^{(2)} = \log \frac{ \bigl[ 1 + \bigl( \frac\lambda{2\tau} \bigr)^2 \bigr]^2 }{ 1 + \bigl( \frac\lambda{2\tau} \bigr)^4 } + \cO\bigl( \lambda^2 / \ell^2 \bigr) \;.
\ee
This expression matches with the R\'enyi asymmetry \eqref{eq:renyi-asymm-1+eps-full} with respect to the fusion algebra of the $\bZ_2 \times \bZ_2$ TY symmetry acting on the circle Hilbert space of the full system.

\paragraph{Non-monotonicity of R\'enyi asymmetry.} It is interesting to notice that R\'enyi asymmetries are not always monotonic in the size of the subsystem. Indeed, when the field is inserted at $z = -i\tau$ and $\tau < \lambda$, the second R\'enyi asymmetry is not monotonic in $\ell$, as shown in the bottom left panel of \cref{fig:asymmetry_1+epsilon}. This raises the question of whether such a counterintuitive behavior is a distinctive feature of asymmetries associated with noninvertible symmetries. The answer is negative: a simpler example involving a group-like symmetry also exhibits non-monotonic behavior. Specifically, we consider the state $\bigl( 1 + \lambda \, \sigma(z) \bigr)|0\rangle$ in the Ising CFT (a single copy) and compute its second R\'enyi asymmetry with respect to the $\bZ_2$ spin-flip symmetry over a subsystem, imposing symmetric conformal boundary conditions at the entangling surface --- thus realizing a strip algebra isomorphic to $\bC[\bZ_2]$. The computation closely parallels the one described above: the second R\'enyi asymmetry is given by \eqref{eq:second-renyi-1+epsilon-intermediate-step} and \eqref{eq:ratio-correlators-epsilon}, with $\varepsilon$ replaced by $\sigma$ and $\lambda$ assigned length dimension $\Delta_\sigma = \frac18$. The relevant 4-point function of $\sigma$ is provided in \eqref{eq:four-pt-function-sigma-mu}. In the bottom right panel of \cref{fig:asymmetry_1+epsilon} we plot the second R\'enyi asymmetry as a function of the subsystem size $\ell$, for a field insertion at $z=-i\tau$, showing a non-monotonic behavior for small values of $\tau$.

\begin{figure}[t]
\centering$\ds
       \rho'  \;\propto\;\;
       \densmatsetup{%
         \draw[dotted] (-2,-1.5) rectangle (2,1.5);
         \densmatvac{}{}
         \coordinate (n) at (0, 0.8);
         \coordinate (s) at (0, -0.8);
         \draw[\colortopoline, thick] (n) -- node[midway, right] {\small$\cL_\eta$} (0, 1.5);
         \draw[\colortopoline, thick] (s) -- node[midway, right] {\small$\cL_\eta$} (0, -1.5);
         \fill (n) circle (1.5pt) node[below] {\small$\mu \boxtimes 1$};
         \fill (s) circle (1.5pt) node[above] {\small$\mu \boxtimes 1$};
       }
$
\caption{\label{fig: rho' with mu}%
Path-integral representation of $\rho'$, described above \eqref{eq:asymmetry-1-box-sigma-subsystem}.}
\end{figure}

\paragraph{Case $\boldsymbol{\sigma \boxtimes 1}$.}
Consider now the state created by inserting the operator $\cO(z) = \sigma(z) \boxtimes 1$ with dimension $\Delta_\sigma = \frac18$ at $z = x - i\tau$. As before, let $\rho$ be the reduced density matrix on $A = \bigl[ - \frac\ell2, \frac\ell2 \bigr]$ obtained using the strongly-symmetric boundary condition. The symmetrized matrix $\rho_\sS$ is obtained applying (\ref{symmetrizer of H8}). The invertible lines can freely pass through $\rho_\sS$, while the noninvertible one changes the untwisted operator $\sigma$ to the twisted (or disorder) operator $\mu$ (with a line $\cL_\eta$ attached). Therefore $\rho_\sS = \frac12 \bigl( \rho + \rho' \bigr)$, where $\rho'$ is the reduced density matrix of the state created by $\cO' = \mu \boxtimes 1$. Such a density matrix contains a line $\cL_\eta$ that goes through infinity, as depicted in \cref{fig: rho' with mu}.

\begin{figure}[t]
\centering
        \includegraphics[width=0.47\textwidth]{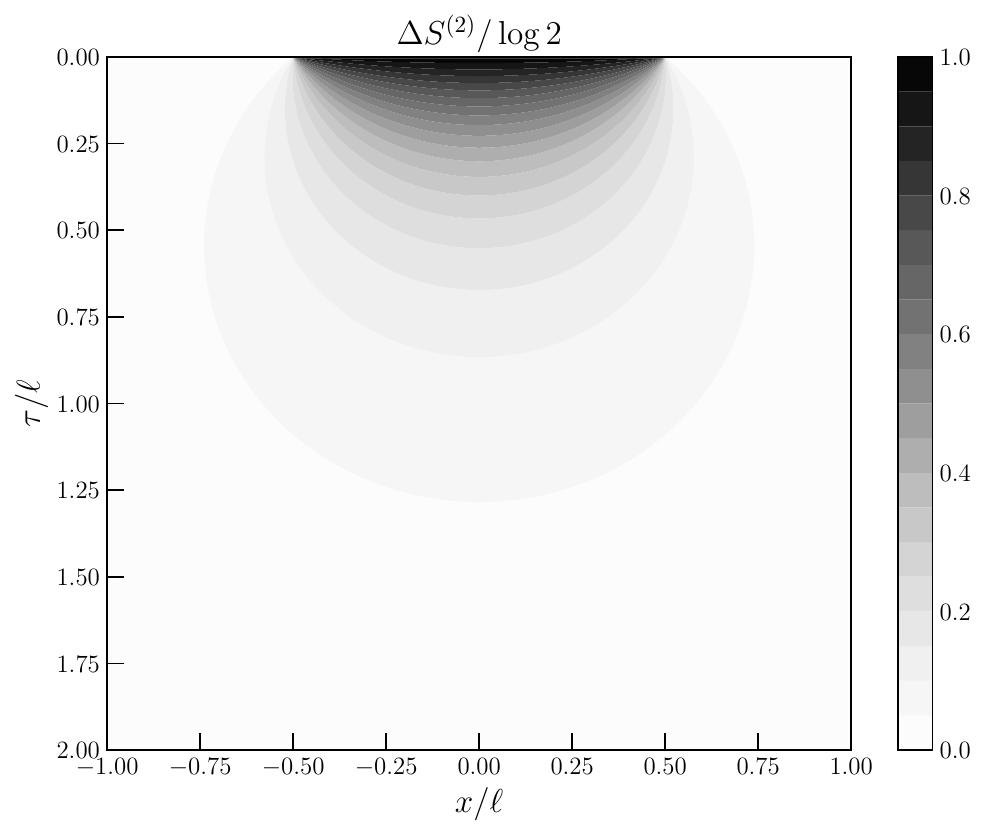}
        \hspace{\stretch{1}}
        \includegraphics[width=0.47\textwidth]{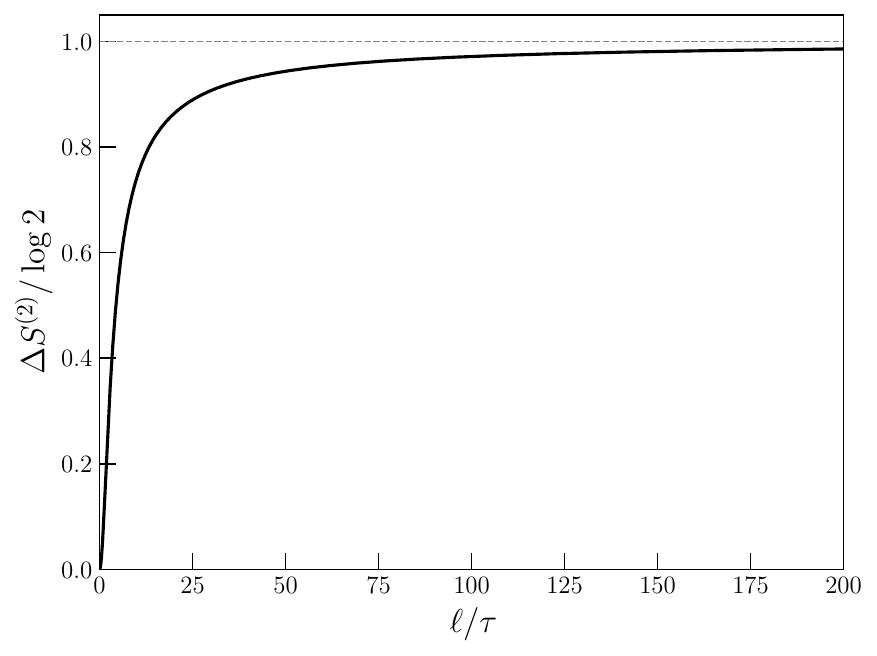}
\caption{\label{fig: asymmetry for sigma}%
Second R\'enyi asymmetry of the state $\bigl( \sigma(z) \boxtimes 1 \bigr)|0\rangle$ in two copies of the Ising CFT.
Left: density plot of the R\'enyi asymmetry, as a function of the Euclidean insertion point $z = x - i\tau$.
Right: R\'enyi asymmetry as a function of the subsystem size $\ell$ (measured in units of $\tau$) for a field insertion at $z = -i\tau$.}
\end{figure}

The second R\'enyi asymmetry takes the form
\be
\label{eq:asymmetry-1-box-sigma-subsystem}
\Delta S^{(2)}[\rho] = \log 2 - \log \biggl[ 1 +  \frac{ \bigl\langle \sigma(z_1) \, \sigma(\bar z_1) \, \mu(z_2) \, \mu(\bar z_2) \bigr\rangle_{\!\cM_2}  }{ \bigl\langle \sigma(z_1) \, \sigma(\bar z_1) \, \sigma(z_2) \, \sigma(\bar z_2) \bigr\rangle_{\!\cM_2} } \biggr] \,,
\ee
where we have used the equality of 4-point functions $\langle\sigma^4 \rangle = \langle \mu^4 \rangle$ as a consequence of Kramers--Wannier duality.
On the plane one finds the following 4-point functions \cite{DiFrancesco:1997nk}:
\begin{align}
\bigl\langle \sigma(w_1) \, \sigma(w_2) \, \sigma(w_3) \, \sigma(w_4) \bigr\rangle_\bC &= \frac{ \bigl( 1 + |\chi| + |1-\chi| \bigr)^{1/2} }{\sqrt2 \; \bigl\lvert w_{14} \, w_{23} \, \chi \bigr\rvert^{1/4}} = \frac{ \bigl\lvert 1 + \sqrt\chi \bigr\rvert + \bigl\lvert 1 - \sqrt\chi \bigr\rvert }{ 2 \bigl\lvert w_{14} \, w_{23} \, \chi \bigr\rvert^{1/4} } \;, \nn\\
\bigl\langle \sigma(w_1) \, \sigma(w_2) \, \mu(w_3) \, \mu(w_4) \bigr\rangle_\bC &= \frac{ \bigl( 1 - |\chi| + |1-\chi| \bigr)^{1/2} }{\sqrt2 \; \bigl\lvert w_{14} \, w_{23} \, \chi \bigr\rvert^{1/4}} \;,
\label{eq:four-pt-function-sigma-mu}
\end{align}
where $\chi$ is the cross-ratio. The fields $\sigma$ and $\mu$, because of duality, have the same conformal dimension, therefore the conformal factors in the map between $\cM_2$ and $\bC$ simplify. This leads to the expression:
\be
\Delta S^{(2)} = \log 2 - \log\biggl[ 1 + \biggl( \frac{ 1 - |\chi| + |1-\chi| }{ 1 + |\chi| + |1-\chi| } \biggr)^{\! 1/2} \, \biggr] \;,
\ee
where $w_1 = g(z)$, $w_2 = g(\bar z)$, $w_3 = -g(z)$, $w_4 = -g(\bar z)$.
At finite subsystem size $\ell$, the asymmetry vanishes in the limit $\tau \to \infty$ in which the field is inserted in the far Euclidean past. This is because symmetry breaking spreads over $\bR$ and thus the subsystem only catches an infinitesimal fraction thereof. On the other hand, in the limit $\tau \to 0$ in which the field is inserted close to the real axis, the asymmetry is equal to $\log(2)$ if $x \in A$ and it vanishes if $x \not\in A$. Finally, in the limit $\ell \to \infty$ that the subsystem size goes to infinity, the asymmetry is equal to $\log(2)$ irrespective of the insertion point. Plots of the R\'enyi asymmetry are in \cref{fig: asymmetry for sigma}.

\subsection{Ising symmetry --- interval Hilbert space}
\label{sec:Ising-symm-interval-Hilbert-space}

As already mentioned in \cref{sec: fusion alg of Ising}, the Ising fusion category admits only one indecomposable module category (that can describe boundary conditions): the regular one. In order to study the asymmetry on a subsystem, then, one has to employ the general procedure of \cref{non-symmetric bc}.
In the regular module the $\tilde F$-symbols are equal to the $F$-symbols, and we indicate the boundary conditions as $\{+, - , f\}$. Recall that the Ising symmetry has a multiplicity-free fusion ring, so no junction labels are necessary. Hence, we use the notation
\be
    H_{r,m}^{s,n}(a) \;\,\equiv\;\;\;
    \strip{r}{m}{a}{s}{n}{}{}
\ee
for basis elements of the strip algebra, which has dimension 34.
The formula (\ref{eq:strip-algebra-product-general}) for the strip-algebra product simplifies considerably, given the fact that every $F$-symbol is a symmetric and involutive matrix, and that every line is self-dual. We will not report it here, though.

The symmetrizer is given by a long expression. Let us write it sector by sector, in terms of the boundary conditions $(r,m)$ for the reduced density matrix $\rho$.
\begin{itemize}[nosep]

\item For a density matrix in the $(+,+)$ sector we have:
\be
\label{eq:ising-regular-symmetrizer-++}
\text{\rule[-0.9em]{0pt}{0em}\small$\ds \rho_\sS = \frac13 \, \Bigl(
H_{++}^{++}(1) \,\rho\, H_{++}^{++}(1) \,+\, H_{++}^{--}(\eta) \,\rho\, H_{--}^{++}(\eta) \,+\, \frac{1}{\sqrt{2}} \, H_{++}^{ff}(\mathcal{N}) \,\rho\, H_{ff}^{++}(\mathcal{N}) \Bigr) \,. $}
\ee

\item In the $(-,-)$ sector:
\be
\text{\rule[-0.9em]{0pt}{0em}\small$\ds \rho_\sS = \frac13 \, \Bigl(
H_{--}^{--}(1) \,\rho\, H_{--}^{--}(1) \,+\, H_{--}^{++}(\eta) \,\rho\, H_{++}^{--}(\eta) \,+\, \frac1{\sqrt{2}} \, H_{--}^{ff}(\mathcal{N}) \,\rho\, H_{ff}^{--}(\mathcal{N})
\Bigr) \,. $}
\ee

\item In the $(+,-)$ sector:
\be
\text{\rule[-0.9em]{0pt}{0em}\small$\ds \rho_\sS = \frac13 \, \Bigl(
H_{+-}^{+-}(1) \,\rho\, H_{+-}^{+-}(1) \,+\, H_{+-}^{-+}(\eta) \,\rho\, H_{-+}^{+-}(\eta) \,+\, \frac{1}{\sqrt{2}} \, H_{+-}^{ff}(\mathcal{N}) \,\rho\, H_{ff}^{+-}(\mathcal{N})
\Bigr) \,. $}
\ee

\item In the $(f,+)$ sector:
\be
\text{\small$\ds \hspace{-0.5em} \rho_\sS = \frac14 \, \Bigl(
H_{f+}^{f+}(1) \,\rho\, H_{f+}^{f+}(1) \,+\, H_{f+}^{f-}(\mathcal{\eta}) \,\rho\, H_{f-}^{f+}(\eta) \,+\, H_{f+}^{+f}(\mathcal{N}) \,\rho\, H_{+f}^{f+}(\mathcal{N}) \,+\, H_{f+}^{-f}(\mathcal{N}) \,\rho\, H_{-f}^{f+}(\mathcal{N})
\Bigr) \,. $}
\ee

\item In the $(f,-)$ sector:
\be
\text{\small$\ds \hspace{-0.5em} \rho_\sS = \frac14 \, \Bigl(
H_{f-}^{f-}(1) \,\rho\, H_{f-}^{f-}(1) \,+\, H_{f-}^{f+}(\mathcal{\eta}) \,\rho\, H_{f+}^{f-}(\eta) \,+\, H_{f-}^{-f}(\mathcal{N}) \,\rho\, H_{-f}^{f-}(\mathcal{N}) \,+\, H_{f-}^{+f}(\mathcal{N}) \,\rho\, H_{+f}^{f-}(\mathcal{N})
\Bigr) \,. $}
\ee

\item In the $(f,f)$ sector:
\be
\label{eq:ising-regular-symmetrizer-ff}
\text{\rule[-0.9em]{0pt}{0em}\small$\ds \rho_\sS =
\frac16 \, \Bigl( H_{ff}^{ff}(1) \,\rho\, H_{ff}^{ff}(1) \,+\, H_{ff}^{ff}(\eta) \,\rho\, H_{ff}^{ff}(\eta) \Bigr) +
\frac1{3\sqrt2} \sum_{s,\, s' \,\in\, \{+,-\}} H_{ff}^{ss'}(\mathcal{N}) \,\rho\, H_{ss'}^{ff}(\mathcal{N}) \;. $}
\ee

\end{itemize}
The sectors not reported are easily obtained by exchanging left and right boundary conditions.

We now proceed to compute the entanglement and R\'enyi asymmetries of the vacuum (\ie, without operator insertions).
The procedure is analogous to the group-like case in \cref{sec:group-regular-strip-algebras}.

Consider first a generic density matrix $\rho$ in the $(+,+)$ sector, that we indicate as $\rho_{++}$ (as in \cref{fig: group symmetry} left), to which we apply the symmetrizer (\ref{eq:ising-regular-symmetrizer-++}). Using the strip algebra products $\bigl[ H_{++}^{++}(1) \bigr]{}^2 = H_{--}^{++}(\eta) H_{++}^{--}(\eta) = \frac1{\sqrt2} H_{ff}^{++}(\cN) H_{++}^{ff}(\cN) = H_{++}^{++}(1)$ and orthogonality between different boundary conditions, we find $\Tr\bigl[ (\rho_{++})_\sS^n \bigr] = 3^{1-n} \Tr(\rho_{++}^n)$. Therefore the asymmetry of a state belonging to this sector is always
\be
\Delta S^{(n)}[\rho_{++}] = \Delta S[\rho_{++}] = \log(3) \;.
\ee
Exactly the same result is found for $\rho_{--}$ and $\rho_{+-}$. This result is valid for any state, and in particular also for the vacuum.
By the same argument one finds $\Tr\bigl[ (\rho_{f+})_\sS^n \bigr] = 4^{1-n} \Tr(\rho_{f+}^n)$ and thus, for a generic density matrix $\rho_{f+}$:
\be
\Delta S^{(n)}[\rho_{f+}] = \Delta S[\rho_{f+}] = \log(4) \;.
\ee
The same result is found for $\rho_{f-}$.
Consider then the reduced density matrix of the vacuum in the $(f,f)$ sector, that we indicate as $\rho_{ff}$, and whose symmetrizer is (\ref{eq:ising-regular-symmetrizer-ff}). When computing $\Tr\bigl[ (\rho_{ff})_\sS^n \bigr]$, to treat the first term in (\ref{eq:ising-regular-symmetrizer-ff}) one uses the products $H_{ff}^{ff}(a) H_{ff}^{ff}(b) = H_{ff}^{ff}(ab)$ for $a,b \in \{1,\eta\} \cong \bZ_2$ and the fact that $H_{ff}^{ff}(a)$ commutes with $\rho_{ff}$. Since the number of lines $\cL_\eta$ is always even, they annihilate among themselves. For the second term one uses $H_{ss'}^{ff}(\cN) H_{ff}^{ss'}(\cN) = \frac1{\sqrt2}\bigl( H_{ff}^{ff}(1) + (ss') H_{ff}^{ff}(\eta) \bigr)$ with $s,s' = \pm1$. When expanding $\Tr\bigl[ (\rho_{ff})_\sS^n \bigr]$, those terms with an odd number of lines $\cL_\eta$ cancel out.
Eventually $\Tr\bigl[ (\rho_{ff})_\sS^n \bigr] = 3^{1-n} \Tr(\rho_{ff}^n)$ and hence
\be
\Delta S^{(n)}[\rho_{ff}] = \Delta S[\rho_{ff}] = \log(3) \;.
\ee
In all cases, the asymmetries of the vacuum in a subsystem with fixed simple boundary conditions attain a nonzero value. This is expected since each simple boundary condition breaks the symmetry.

\begin{figure}[t]
\centering$\ds
{\def\densmatgapw{0.75}
 \def\densmatradx{0.25}
 \def\densmatrady{0.25}
       \tilde \rho_{rm} \equiv\, \densmatsetup{ \densmatvac{r}{m}} \,,\quad
       \tilde \rho_{ff}^{\,\eta} \equiv\,
       \densmatsetup{%
         \draw[\colortopoline] (-\densmatgapw/2 -\densmatradx, \densmatrady)
         arc to [radius=1.5, small, clockwise]
         node[midway, shift={(0,2mm)}] {\small$\eta$} (\densmatgapw/2 + \densmatradx, \densmatrady);
         \densmatvac{f}{f}
       } \,,\quad
       Z_{rm} \equiv Z \Biggl[ \densmatsetup{ \densmatnorm{r}{m} } \Biggr] \,,\quad
       Z_{rm}^\eta \equiv Z \Biggl[ \densmatsetup{%
         \draw[\colortopoline] (-\densmatgapw/2 -\densmatradx, \densmatrady)
         arc to [radius=1.5, small, clockwise]
         node[midway, shift={(0,2mm)}] {\small$\eta$} (\densmatgapw/2 + \densmatradx, \densmatrady);
         \densmatnorm{f}{f}
       } \Biggr]
}$
\caption{\label{fig:ising-regular}%
Path-integral definitions of various density matrices that enter in the symmetrization with respect to the strip algebra of the Ising fusion category with regular boundary conditions.}
\end{figure}

On the other hand, let us apply the averaging procedure of \cref{non-symmetric bc} to the vacuum state and consider the matrix (recall that $\tilde d_m = d_m$ in the regular module):
\be
\label{averaged reduced density matrix}
\rho = \frac1D \, \sum\nolimits_{rm} d_r \, d_m \, \rho_{rm} \;.
\ee
Here $\rho_{rm}$ is the reduced density matrix of the vacuum with boundary conditions $(r,m)$, as in \cref{fig: group symmetry} left but without insertions, the sum is over $r,m \in \{+,-,f\}$, and $D = \bigl( \sum d_m\bigr){}^2$. It is convenient to write $\rho_{rm} = Z_{rm}^{-1} \, \tilde \rho_{rm}$ where, as in \cref{non-symmetric bc}, $\tilde \rho_{rm}$ is the unnormalized density matrix while $Z_{rm}$ is its trace. They are depicted in \cref{fig:ising-regular}. Since $\tilde \rho_{rm}$ come from the vacuum, the lines can move through them and the effect of the symmetrizer can be written as$\,$%
\footnote{Using the tube algebra action (\ref{tube on TQFT general}) on boundary conditions, one can prove various relations among the vacuum traces, for $n \in \bZ_{>0}$: $\Tr(\tilde\rho_{++}^{\,n}) = \Tr(\tilde\rho_{--}^{\,n})$, $\Tr(\tilde\rho_{+-}^{\,n}) = \Tr(\tilde\rho_{-+}^{\,n})$, $\Tr(\tilde\rho_{ff}^{\,n}) = \Tr(\tilde\rho_{++}^{\,n} + \tilde\rho_{+-}^{\,n})$, $\Tr( \tilde\rho_{ff}^{\,n-1} \tilde \rho_{ff}^{\,\eta}) = \Tr(\tilde\rho_{++}^{\,n} - \tilde\rho_{+-}^{\,n})$, as well as $\Tr(\tilde\rho_{f+}^n) = \Tr(\tilde\rho_{f-}^n) = \Tr(\tilde\rho_{+f}^n) = \Tr(\tilde\rho_{-f}^n)$. Besides, one finds the operator relations $\tilde \rho_{ff}^{\,\eta} \tilde \rho_{ff}^{\,\eta} = \tilde\rho_{ff} \tilde\rho_{ff}$ and $[\tilde \rho_{ff}, \tilde\rho_{ff}^{\,\eta} ] = 0$. One can then verify that $\sS$ is trace-preserving.}
\bea
( \tilde\rho_{++})_\sS = (\tilde \rho_{--})_\sS &= \frac13 \Bigl( \tilde\rho_{++} + \tilde\rho_{--} + \frac12 \, \tilde\rho_{ff} + \frac12 \, \tilde\rho_{ff}^{\,\eta} \Bigr) \;, \\
( \tilde\rho_{+-})_\sS = (\tilde \rho_{-+})_\sS &= \frac13 \Bigl( \tilde\rho_{+-} + \tilde\rho_{-+} + \frac12 \, \tilde\rho_{ff} - \frac12 \, \tilde\rho_{ff}^{\,\eta} \Bigr) \;, \\
(\tilde\rho_{f+})_\sS = (\tilde\rho_{f-})_\sS = (\tilde\rho_{+f})_\sS = (\tilde\rho_{-f})_\sS &= \frac14 \Bigl( \tilde\rho_{f+} + \tilde\rho_{f-} + \tilde\rho_{+f} + \tilde\rho_{-f} \Bigr) \;, \\
(\tilde\rho_{ff})_\sS &= \frac13 \Bigl( \tilde\rho_{++} + \tilde\rho_{--} + \tilde\rho_{+-} + \tilde\rho_{-+} + \tilde\rho_{ff} \Bigr) \;.
\eea
Here $\tilde\rho_{ff}^{\,\eta}$ is a density matrix that contains a line $\cL_\eta$ connecting the two boundaries of type $f$, as described in \cref{fig:ising-regular}. To proceed, we use that $\lim_{\varepsilon\to0} Z_{rm} \propto d_r d_m$, proven in \cref{non-symmetric bc}. In the $\varepsilon \to 0$ limit, then, one immediately obtains the weak identity $\rho_\sS = \rho$, valid when inserted into traces. We thus explicitly confirm that, in the limit,  the asymmetries of the reduced vacuum density matrix vanish: $\Delta S^{(n)}[\rho] = \Delta S[\rho] = 0$.

\begin{figure}[t]
\centering
	\includegraphics[width=0.45\textwidth]{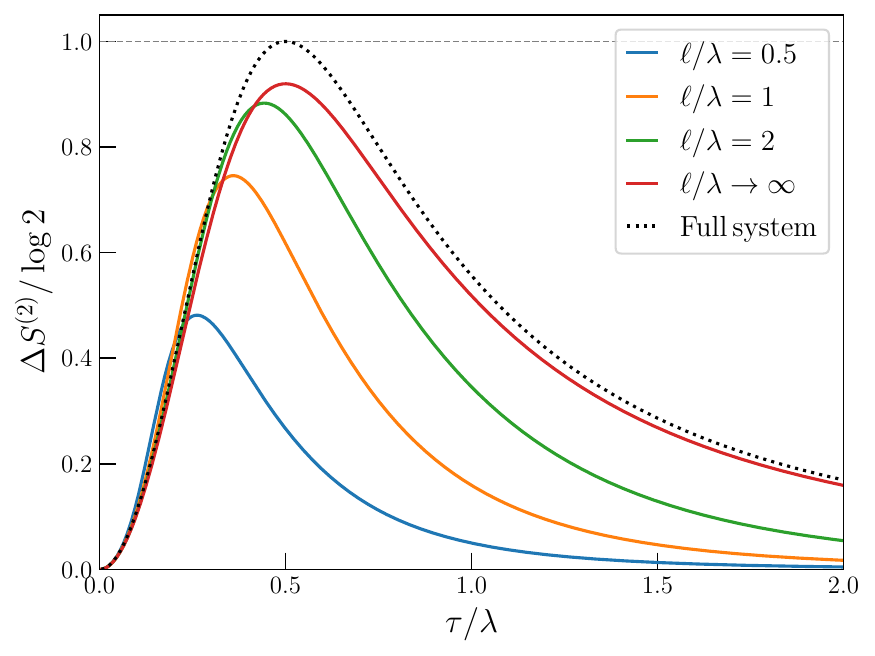}
	\includegraphics[width=0.45\textwidth]{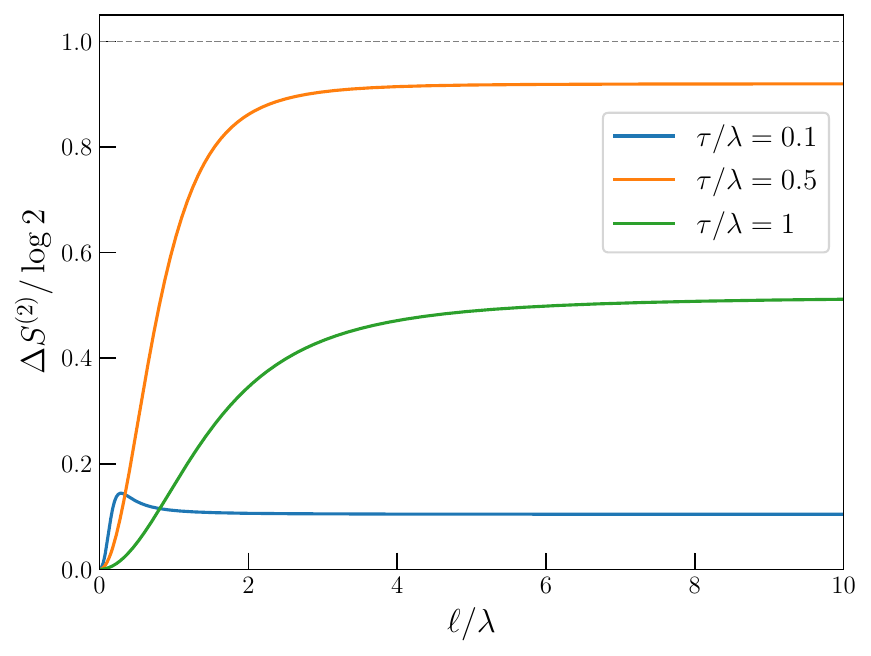}
\caption{\label{fig:asymmetry-composite-bc}%
Second R\'enyi asymmetry of the state $\bigl(1 + \lambda \, \varepsilon(z) \bigr) \ket{0}$ in the Ising CFT, with ``averaged'' boundary conditions at the entangling surface.
Left: asymmetry when the field is inserted at $z = -i \tau$, as a function of $\tau$, for different values of $\ell$ (in units of $\lambda$). We compare it with the asymmetry of the full system (dotted line) computed with respect to the fusion algebra.
Right: asymmetry as a function of $\ell$, for different positions of the field insertion at $z = -i \tau$.}
\end{figure}

\paragraph{Case $\boldsymbol{1 + \lambda \, \varepsilon}$.}
We now wish to compute the asymmetry of an excited state $\cO(z) |0\rangle$ obtained by the operator insertion $\cO(z) = 1 + \lambda \, \varepsilon(z)$ at $z = -i\tau$, along with its mirror image $\cO^\dag(\bar z)$. The reduced density matrix is as in (\ref{averaged reduced density matrix}), but now $\rho_{rm}$ contain the insertions. When we swipe a line $\cL_\cN$ through them, this time $\cO \mapsto \cO' = 1 - \lambda\, \varepsilon$ and we call $\rho'_{rm}$ the reduced density matrices with that insertion. The action of the symmetrizer can be written as
\begin{align}
( \tilde\rho_{++})_\sS &= (\tilde \rho_{--})_\sS = \frac{ \tilde\rho_{++} + \tilde\rho_{--} }3 + \frac{ \tilde\rho_{ff}^{\,\prime} + \tilde\rho_{ff}^{\,\eta \,\prime} }6 , \quad &
(\tilde\rho_{f+})_\sS &= (\tilde\rho_{f-})_\sS = \frac{ \tilde\rho_{f+} + \tilde\rho_{f-} + \tilde\rho_{+f}^{\,\prime} + \tilde\rho_{-f}^{\,\prime} }4 , \nn\\
( \tilde\rho_{+-})_\sS &= (\tilde \rho_{-+})_\sS = \frac{ \tilde\rho_{+-} + \tilde\rho_{-+} }3 + \frac{ \tilde\rho_{ff}^{\,\prime} - \tilde\rho_{ff}^{\,\eta \,\prime} }6 , &
(\tilde\rho_{+f})_\sS &= (\tilde\rho_{-f})_\sS = \frac{ \tilde\rho_{f+}^{\,\prime} + \tilde\rho_{f-}^{\,\prime} + \tilde\rho_{+f} + \tilde\rho_{-f} }4 , \nn\\
(\tilde\rho_{ff})_\sS &= \frac{ \tilde\rho_{++}^{\,\prime} + \tilde\rho_{--}^{\,\prime} + \tilde\rho_{+-}^{\,\prime} + \tilde\rho_{-+}^{\,\prime} + \tilde\rho_{ff} }3 .
\end{align}
This allows one to compute traces of powers of $\rho_\sS$. For example, in the $\varepsilon \to 0$ limit, one can write the compact expression
\be
\lim_{\varepsilon \to 0} \; \frac{\Tr(\rho_\sS^2)}{\Tr(\rho^2)} = \frac13 \, \frac{ \Tr\bigl[ (3 - \sqrt2) \, \tilde\rho^{\,2} + \sqrt2 \, \tilde\rho \, \tilde\rho^{\,\prime} \bigr] }{ \Tr (\tilde\rho^{\,2}) } \;.
\ee
The notation on the r.h.s.\ should merely be interpreted as an instruction to construct a branched Riemann surface, obtained from the gluing of unnormalized density matrices followed by the $\varepsilon\to0$ limit. The expression can be rewritten in terms of correlation functions on a two-sheeted Riemann surface $\cM_2$:
\be
\label{second trace Ising}
\frac{\Tr(\rho_\sS^2) }{ \Tr( \rho^2) } = 1 - \frac{
\frac{4\sqrt2}3 \, \lambda^2 \, \Bigl\langle \varepsilon(z_1) \varepsilon(z_2) + \varepsilon(z_1) \varepsilon(\bar z_2) \Bigr\rangle_{\!\cM_2} }
{ \Bigl\langle 1 + 2 \lambda^2 \Bigl( \varepsilon(z_1) \varepsilon(\bar z_1) + \varepsilon(z_1) \varepsilon(z_2) + \varepsilon(z_1) \varepsilon(\bar z_2) \Bigr) + \lambda^4 \, \varepsilon(z_1) \varepsilon(\bar z_1) \varepsilon(z_2) \varepsilon(\bar z_2) \Bigr\rangle_{\!\cM_2} }
\ee
where we used the same notation as in \cref{sec: example Ising2} and cyclic permutation symmetry of the sheets. This can be used to obtain the second R\'enyi asymmetry, of which we present some plots in \cref{fig:asymmetry-composite-bc} (more generally, the $n$-th R\'enyi asymmetry can be written in terms of correlation functions up to $2n$ points). Comparing (\ref{second trace Ising}) with (\ref{second trace in Ising2}), the two expressions are very similar. Indeed, the second R\'enyi asymmetry $\Delta S^{(2)}[\rho] = - \log \bigl[ \Tr(\rho_\sS^2) / \Tr(\rho^2) \bigr]$ of $(1+\lambda\,\varepsilon)|0\rangle$ in the Ising CFT is very similar to (in fact, just slightly smaller than) the one of $(1 + \lambda\, \varepsilon \boxtimes 1 )|0\rangle$ in two copies of the Ising CFT, plotted in \cref{fig:asymmetry_1+epsilon}.

In the large subsystem limit $\ell \to \infty$, correlation functions factorize as discussed around (\ref{large subsystem limit in Ising2}). For the Ising CFT, where $\Delta_\varepsilon = 1$, we obtain:
\be
\label{Ising asymm large volume}
\lim_{\ell \to \infty} \; \Delta S^{(2)} = - \log \Biggl[ 1 - \frac{ \frac{4\sqrt2}3 \, \bigl( \frac{\lambda}{2\tau} \bigr)^2 }{ \bigl( 1 + \bigl( \frac{\lambda}{2\tau} \bigr)^2 \bigr)^2 } \Biggr] \;.
\ee
This, as a function of $\tau/\lambda$, yields a very similar curve to the second R\'enyi asymmetry of the full system (\ref{eq:renyi-asymm-1+eps-full}) computed with respect to the fusion algebra, just slightly smaller (the maximum of (\ref{Ising asymm large volume}) is $\log\bigl( \frac3{3-\sqrt2} \bigr)$, smaller than $\log(2)$) . We compare them in \cref{fig:asymmetry-composite-bc}. So in this case the large subsystem limit of the asymmetry computed with the averaging procedure is not identical to the asymmetry of the full system.

\subsection{Fibonacci fusion category}
\label{sec: Fibonacci}

The Fibonacci fusion category has 2 simple objects: $\cL_1$, $\cL_W$. Their fusion rule is$\,$%
\footnote{The Fibonacci fusion category is also a specific instance ($N=1$) of the Haagerup--Izumi categories, which we study in \cref{sec: Haagerup}.}
\be
\label{Fibonacci fusion algebra}
\cL_W \times \cL_W = \cL_1 + \cL_W
\ee
and it is commutative. The quantum dimension of $\cL_W$ is
\be
d_W = \frac{1 + \sqrt{5}}{2} \equiv \varphi
\ee
which is the golden ratio (\ie, the positive solution of the equation $\varphi^2 = \varphi +1$). Both lines are self-dual and $\varkappa_W = 1$.
The only nontrivial $F$-symbol, in terms of $a,b \in \{ 1,W \}$, is$\,$%
\footnote{There is a unique fusion category whose fusion ring is \eqref{Fibonacci fusion algebra} and that has unitary $F$-symbols, see \eg \cite{Bartsch:2025drc}. This category occurs for example as a subsymmetry of the tricritical Ising model \cite{Chang:2018iay}.}
\be
\label{F-symbols Fibonacci}
\bigl[ F_{WWW}^W \bigr]_{ab} = \mat{ \varphi^{-1} & \varphi^{-1/2} \\ \varphi^{-1/2} & -\varphi^{-1} } \;,
\ee
while all other $F$-symbols allowed by fusion are equal to 1. The symmetrizer for the fusion algebra is determined by the inverse bilinear form $\tilde K$
and takes the form:
\be
\label{Fibonacci symmetrizer}
\rho_\sS = \frac15 \, \Bigl( 3 \, \rho - \cL_W \, \rho - \rho \, \cL_W + 2 \, \cL_W \, \rho \, \cL_W \Bigr) \;.
\ee
The fusion algebra has two one-dimensional representations $R_{\pm}$, where $R_+(\cL_W) = \varphi$ while $R_-(\cL_W) = - \varphi^{-1}$.
The corresponding projectors are given by
\be
P_+ = 5^{-\frac12} \, \bigl( \varphi^{-1} \cL_1 + \cL_W \bigr) \;,\qquad\qquad P_- = 5^{-\frac12} \, \bigl( \varphi \, \cL_1 - \cL_W \bigr) \;.
\ee

\paragraph{Spontaneous symmetry breaking.} Like in the case of Ising, the only indecomposable module category over the Fibonacci fusion category is the regular one. This means that in a Fibonacci-symmetric gapped phase, the symmetry is always completely spontaneously broken. The clustering vacua form a 2-dimensional NIM-rep spanned by \( \ket{1} \) and \( \ket{W} \), and
\be
\cL_W \ket{1} = \ket{W} \;, \qquad\qquad \cL_W \ket{W} = \ket{1} + \ket{W} \;,
\ee
where $\langle 1 | 1 \rangle = \langle W | W \rangle = 1$. Consider the pure state $\rho = |1 \rangle \langle 1 |$. Using the symmetrizer \eqref{Fibonacci symmetrizer}, in the basis above we find
\be
\rho =  |1 \rangle \langle 1 | \qquad\Rightarrow\qquad \rho_\sS = \frac15 \mat{ 3 & -1 \\ -1 & 2} \;.
\ee
Defining
\be
\phi_+ = \frac{\sqrt5 + 1}{2 \sqrt5} = \frac{\varphi}{\sqrt5} \;,\qquad\qquad \phi_- = \frac{\sqrt5 - 1}{2\sqrt5} = \frac{\varphi^{-1}}{\sqrt5} \;,
\ee
the eigenvalues of $\rho_\sS$ are $\{\phi_+ , \phi_-\}$. Then the R\'enyi asymmetries are $\Delta S^{(n)}= \frac1{1-n} \log \bigl( \phi_+^n + \phi_-^n \bigr)$ while the entanglement asymmetry is
\be
\Delta S = - \phi_+ \log \phi_+ - \phi_- \log \phi_- = \frac12 \log 5 - \frac1{\sqrt5} \operatorname{arccoth}(\sqrt5) \simeq 0.5895 \;.
\ee
This value is smaller than $\log(2)$, the asymmetry for a spontaneously broken group of order 2. For the other vacuum:
\be
\rho = |W \rangle \langle W| \qquad\Rightarrow\qquad \rho_\sS = \frac15 \mat{ 2 & 1 \\ 1 & 3}
\ee
which gives the same asymmetries (because $\rho_\sS$ has the same eigenvalues).

\begin{figure}[t]
\centering$\ds
\begin{aligned}
& \tube{1}{W}{}{1} \,\times \tube{1}{W}{}{1} = \tube{1}{1}{}{1} \,+ \tube{1}{W}{}{1} \,,\quad \tube{W}{W}{}{1} \,\times \tube{1}{W}{}{W} = \varphi^{\frac12} \tube{1}{1}{}{1} \,-\, \varphi^{-\frac12} \tube{1}{W}{}{1} \,, \\
& \tube{1}{W}{}{1} \,\times \tube{W}{W}{}{1} = - \varphi^{-1} \tube{W}{W}{}{1} \,,\qquad \tube{1}{W}{}{W} \,\times \tube{1}{W}{}{1} = - \varphi^{-1} \tube{1}{W}{}{W} \,, \\
& \tube{1}{W}{}{W} \,\times \tube{W}{W}{}{1} = \varphi^{-\frac12} \tube{W}{1}{}{W} \,+\, \varphi^{-\frac12} \tube{W}{W}{1}{W} \,+\, \varphi^{-2} \tube{W}{W}{W}{W} \,, \\
& \tube{W}{W}{}{1} \,\times \tube{W}{W}{a}{W} = (1, \varphi^{-\frac32})_a \tube{W}{W}{}{1} \,,\qquad \tube{W}{W}{a}{W} \,\times \tube{1}{W}{}{W} = (1, \varphi^{-\frac32})_a \tube{1}{W}{}{W} \,, \\
& \tube{W}{W}{a}{W} \,\times \tube{W}{W}{b}{W} = \left( \begin{matrix} \varphi^{-1} & \varphi^{-\frac12} \\ \varphi^{-\frac12} & - \varphi^{-1} \end{matrix} \!\right)_{\!\!\!ab} \hspace{-0.5em} \tube{W}{1}{}{W} \,+\, \mat{ 0 & 0 \\ 0 & 1 }_{\!\!\!ab} \hspace{-0.5em} \tube{W}{W}{1}{W} \,+\, \mat{ \varphi^{-\frac12} & - \varphi^{-1} \\ - \varphi^{-1} & - \varphi^{-\frac52} }_{\!\!\!ab} \hspace{-0.5em} \tube{W}{W}{W}{W} \,.
\end{aligned}
$
\caption{\label{fig: tube of Fibonacci}%
Nontrivial products in the tube algebra of the Fibonacci fusion category.}
\end{figure}

\paragraph{Tube algebra.} This has been discussed, \eg, in \cite{Bartsch:2025drc}. It has dimension 7. The elements
\be
\tube{1}{1}{}{1} \qquad\text{and}\qquad \tube{W}{1}{}{W}
\ee
behave as the identity in the untwisted and twisted sector, respectively. The nonvanishing products are in \cref{fig: tube of Fibonacci}. The algebra has three 1-dimensional and one 2-dimensional irreducible representations, that we now describe.
\begin{itemize}

\item A 1-dimensional representation $\cF_1$ with space $V = \bC_1$:
\be
\cF_1 \biggl(\!\! \tube{1}{1}{}{1} \biggr) = 1 \;,\qquad\qquad \cF_1 \biggl(\!\! \tube{1}{W}{}{1} \biggr) = \varphi \;,
\ee
while all other ones are mapped to zero. It represents genuine local operators invariant under the symmetry (up to multiplication by the quantum dimension of $\cL_W$).

\item Two 1-dimensional representations $\cF_W$ and $\cF_W^*$ with space $V = \bC_{W}$, and
\be
\cF_W \biggl(\!\! \tube{W}{1}{}{W} \biggr) = 1 \;,\qquad \cF_W \biggl(\!\! \tube{W}{W}{1}{W} \biggr) = e^{\frac{4\pi i}5} \;,\qquad \cF_W \biggl(\!\! \tube{W}{W}{W}{W} \biggr) = e^{- \frac{3\pi i}5} \varphi^{\frac12} \;,
\ee
while all other ones are zero. It represents $W$-twisted operators, that remain $W$-twisted under the action of $\cL_W$. Notice that there are two actions of $\cL_W$ on these operators.

\item One 2-dimensional representation $\cF_2$ with space $V = \bC_{1}\oplus \bC_{W}$ and
\begin{align}
\label{eq:fib-tube-rep-2d}
\cF_2 \biggl(\!\! \tube{1}{1}{}{1} \biggr) &= \biggl( \begin{matrix} 1 & 0 \\ 0 & 0 \end{matrix} \biggr) \,,\qquad
\cF_2 \biggl(\!\! \tube{1}{W}{}{1} \biggr) = \biggl( \begin{matrix} - \varphi^{-1} & 0 \\ 0 & 0 \end{matrix} \biggr) \,,\\
\cF_2 \biggl(\!\! \tube{W}{1}{}{W} \biggr) &= \biggl( \begin{matrix} 0 & 0 \\ 0 & 1 \end{matrix} \biggr) \,,\qquad
\cF_2 \biggl(\!\! \tube{W}{W}{1}{W} \biggr) = \biggl( \begin{matrix} 0 & 0 \\ 0 & 1 \end{matrix} \biggr) \,,\qquad
\cF_2 \biggl(\!\! \tube{W}{W}{W}{W} \biggr) = \biggl( \begin{matrix} 0 & 0 \\ 0 & \varphi^{-\frac32} \end{matrix} \biggr) \,, \nn\\
\cF_2 \biggl(\!\! \tube{1}{W}{}{W} \biggr) &= \biggl( \begin{matrix} 0 & 0 \\ x & 0 \end{matrix} \biggr) \,,\qquad
\cF_2 \biggl(\!\! \tube{W}{W}{}{1} \biggr) = \biggl( \begin{matrix} 0 & x \\ 0 & 0 \end{matrix} \biggr) \,,\qquad\quad
x = 5^{\frac14} \varphi^{-\frac14} \,. \nn
\end{align}
It represents a doublet of an untwisted local operator $\cO_1$ and a $W$-twisted operator $\cO_W$ that get mixed under the action of $\cL_W$.
\end{itemize}

\paragraph{Full and reduced strip algebras.} The only indecomposable module category is the regular one.
The corresponding strip algebra has dimension 13, and a basis is given by the elements:
\begin{equation*}
\rule{0pt}{2.3em}
    \begin{tikzpicture}[line width=0.6]
        \draw [densely dashed, gray] (-0.4, -0.4) -- (-0.4, 0.4);
        \draw [densely dashed, gray] (0.4, -0.4) -- (0.4, 0.4);
        \draw [densely dashed, gray] (-0.4, 0) -- (0.4, 0);
    \end{tikzpicture} \;,\;
    \begin{tikzpicture}[line width=0.6]
        \draw (-0.4, -0.4) -- (-0.4, 0.4);
        \draw [densely dashed, gray] (0.4, -0.4) -- (0.4, 0.4);
        \draw [densely dashed, gray] (-0.4, 0) -- (0.4, 0);
    \end{tikzpicture} \;,\;
    \begin{tikzpicture}[line width=0.6]
        \draw [densely dashed, gray] (-0.4, -0.4) -- (-0.4, 0.4);
        \draw (0.4, -0.4) -- (0.4, 0.4);
        \draw [densely dashed, gray] (-0.4, 0) -- (0.4, 0);
    \end{tikzpicture} \;,\;
    \begin{tikzpicture}[line width=0.6]
        \draw (-0.4, -0.4) -- (-0.4, 0.4);
        \draw (0.4, -0.4) -- (0.4, 0.4);
        \draw [densely dashed, gray] (-0.4, 0) -- (0.4, 0);
    \end{tikzpicture} \;,\;
    \begin{tikzpicture}[line width=0.6]
        \draw [densely dashed, gray] (-0.4, -0.4) -- (-0.4, 0);
        \draw [densely dashed, gray] (0.4, -0.4) -- (0.4, 0);
        \draw (-0.4, 0) -- (0.4, 0);
        \draw (-0.4, 0) -- (-0.4, 0.4);
        \draw (0.4, 0) -- (0.4, 0.4);
    \end{tikzpicture} \;,\;
    \begin{tikzpicture}[line width=0.6]
        \draw (-0.4, -0.4) -- (-0.4, 0);
        \draw [densely dashed, gray] (0.4, -0.4) -- (0.4, 0);
        \draw (-0.4, 0) -- (0.4, 0);
        \draw [densely dashed, gray] (-0.4, 0) -- (-0.4, 0.4);
        \draw (0.4, 0) -- (0.4, 0.4);
    \end{tikzpicture} \;,\;
    \begin{tikzpicture}[line width=0.6]
        \draw (-0.4, -0.4) -- (-0.4, 0);
        \draw [densely dashed, gray] (0.4, -0.4) -- (0.4, 0);
        \draw (-0.4, 0) -- (0.4, 0);
        \draw (-0.4, 0) -- (-0.4, 0.4);
        \draw (0.4, 0) -- (0.4, 0.4);
    \end{tikzpicture} \;,\;
    \begin{tikzpicture}[line width=0.6]
        \draw [densely dashed, gray] (-0.4, -0.4) -- (-0.4, 0);
        \draw (0.4, -0.4) -- (0.4, 0);
        \draw (-0.4, 0) -- (0.4, 0);
        \draw (-0.4, 0) -- (-0.4, 0.4);
        \draw [densely dashed, gray] (0.4, 0) -- (0.4, 0.4);
    \end{tikzpicture} \;,\;
    \begin{tikzpicture}[line width=0.6]
        \draw [densely dashed, gray] (-0.4, -0.4) -- (-0.4, 0);
        \draw (0.4, -0.4) -- (0.4, 0);
        \draw (-0.4, 0) -- (0.4, 0);
        \draw (-0.4, 0) -- (-0.4, 0.4);
        \draw (0.4, 0) -- (0.4, 0.4);
    \end{tikzpicture} \;,\;
    \begin{tikzpicture}[line width=0.6]
        \draw (-0.4, -0.4) -- (-0.4, 0);
        \draw (0.4, -0.4) -- (0.4, 0);
        \draw (-0.4, 0) -- (0.4, 0);
        \draw [densely dashed, gray] (-0.4, 0) -- (-0.4, 0.4);
        \draw [densely dashed, gray] (0.4, 0) -- (0.4, 0.4);
    \end{tikzpicture} \;,\;
    \begin{tikzpicture}[line width=0.6]
        \draw (-0.4, -0.4) -- (-0.4, 0);
        \draw (0.4, -0.4) -- (0.4, 0);
        \draw (-0.4, 0) -- (0.4, 0);
        \draw (-0.4, 0) -- (-0.4, 0.4);
        \draw [densely dashed, gray] (0.4, 0) -- (0.4, 0.4);
    \end{tikzpicture} \;,\;
    \begin{tikzpicture}[line width=0.6]
        \draw (-0.4, -0.4) -- (-0.4, 0);
        \draw (0.4, -0.4) -- (0.4, 0);
        \draw (-0.4, 0) -- (0.4, 0);
        \draw [densely dashed, gray] (-0.4, 0) -- (-0.4, 0.4);
        \draw (0.4, 0) -- (0.4, 0.4);
    \end{tikzpicture} \;,\;
    \begin{tikzpicture}[line width=0.6]
        \draw (-0.4, -0.4) -- (-0.4, 0);
        \draw (0.4, -0.4) -- (0.4, 0);
        \draw (-0.4, 0) -- (0.4, 0);
        \draw (-0.4, 0) -- (-0.4, 0.4);
        \draw (0.4, 0) -- (0.4, 0.4);
    \end{tikzpicture} \;,\;
\end{equation*}
where dashed lines are $\cL_1$ while solid lines are $\cL_W$. One easily verifies that the boundary $\hat\cB = \cL_W$ provides a weakly-symmetric boundary condition \cite{Choi:2023xjw}, and the corresponding reduced strip algebra has dimension 2. In terms of the generators
\be
\label{eq:fib-reduced-strip}
    \cS_1 \;=\;\;
    \begin{tikzpicture}[baseline=\mathaxis, line width=0.6]
        \draw (-0.5, -0.5) -- (-0.5, 0.5);
        \draw (0.5, -0.5) -- (0.5, 0.5);
        \draw [densely dashed, gray] (-0.5, 0) -- (0.5, 0);
    \end{tikzpicture}
    \;\;,\qquad\qquad
    \cS_W \;=\;\;
    \begin{tikzpicture}[baseline=\mathaxis, line width=0.6]
        \draw (-0.5, -0.5) -- (-0.5, 0.5);
        \draw (0.5, -0.5) -- (0.5, 0.5);
        \draw (-0.5, 0) -- (0.5, 0);
    \end{tikzpicture}
    \;\;,
\ee
the reduced strip algebra is commutative, $\cS_1$ is its identity, and the nontrivial product is
\be
\cS_W \times \cS_W = \cS_1 + \varphi^{-\frac32} \, \cS_W \;.
\ee
This algebra is very similar to the original fusion algebra (\ref{Fibonacci fusion algebra}), and it is semisimple. The symmetrizer with respect to the reduced strip algebra, computed using (\ref{symmetrizer}), takes the form
\be
\rho_\sS = \frac12 \, \biggl( \rho \,+\, \frac1{4\varphi^3+1} \, \bigl( 1 - 2 \varphi^{3/2} \cS_W \bigr) \, \rho \, \bigl( 1 - 2 \varphi^{3/2} \cS_W \bigr) \biggr) \;.
\ee

For the asymmetry of the vacuum, it is clear that the density matrix $\rho = \tilde \rho_{WW} / Z_{WW}$ (using the notation of \cref{fig:ising-regular}) commutes with the action of the reduced strip algebra \eqref{eq:fib-reduced-strip} (because there is a single boundary condition involved).
Hence its asymmetry is zero.

\begin{figure}[t]
\centering$\ds
    \def\tuberadius{0.65}
    \def\twistlength{1.1}
        \begin{aligned}
            \begin{tikzpicture}[baseline=\mathaxis, line width=0.6]
                \draw [\colortopoline] (0,0) circle [radius=\tuberadius];
                \draw [\colortopoline, densely dashed] (0,0) -- (0, \twistlength);
                \draw [fill] (0,0) node[below] {\small$\cO_1$} circle [radius = 0.04];
                \draw [\colortopojun, fill] (0,\tuberadius) circle [radius = 0.04];
            \end{tikzpicture}
            &\;=\; - \varphi^{-1} \;\;
            \begin{tikzpicture}[baseline=\mathaxis, line width=0.6]
                \draw [fill] (0, 0) node[right] {\small$\cO_1$} circle [radius = 0.04];
            \end{tikzpicture}
            \\[1ex]
            \begin{tikzpicture}[baseline=\mathaxis, line width=0.6]
                \draw [\colortopoline] (0,0) circle [radius=\tuberadius];
                \draw [\colortopoline] (0,0) -- (0,\tuberadius);
                \draw [\colortopoline, densely dashed] (0,\tuberadius) -- (0,\twistlength);
                \draw [fill] (0,0) node[below] {\small$\cO_W$} circle [radius = 0.04];
                \draw [\colortopojun, fill] (0,\tuberadius) circle [radius = 0.04];
            \end{tikzpicture}
            &\;=\; 5^{1/4} \varphi^{-1/4} \;\;
            \begin{tikzpicture}[baseline=\mathaxis, line width=0.6]
                \draw [fill] (0, 0) node[right] {\small$\cO_1$} circle [radius = 0.04];
            \end{tikzpicture}
        \end{aligned}
        \qquad\qquad
        \begin{tikzpicture}[baseline=\mathaxis, line width=0.6]
            \draw[fill] (0,0) node[right] {\small$\cO_1$} circle [radius = 0.04];
            \draw[\colortopoline] (-0.7,0.3) -- (0.7,0.3);
        \end{tikzpicture}
        \;=\; - \varphi^{-2} \;
        \begin{tikzpicture}[baseline=\mathaxis, line width=0.6]
            \draw[fill] (0,0) node[right] {\small$\cO_1$} circle [radius = 0.04];
            \draw[\colortopoline] (-0.7,-0.3) -- (0.7,-0.3);
        \end{tikzpicture}
        \;+\, 5^{1/4} \varphi^{-3/4} \;
        \begin{tikzpicture}[baseline=\mathaxis, line width=0.6]
            \draw[\colortopoline] (-0.7,-0.3) -- (0.7,-0.3);
            \draw[\colortopoline] (0,0) -- (0,-0.3);
            \draw[fill] (0,0) node[right] {\small$\cO_W$} circle [radius = 0.04];
        \end{tikzpicture}
$
\caption{\label{fig:fibonacci-lasso}%
Left: lasso action of the Fibonacci tube algebra on the doublet of operators $(\cO_1, \cO_W)$, extracted from (\ref{eq:fib-tube-rep-2d}).
Right: swiping action of the line $\cL_W$ across $\cO_1$. The coefficients have been fixed by separately completing the diagram from above and from below to form a circle, and by using the lasso actions on the left.}
\end{figure}

We then compute the asymmetry of an excited state created by the insertion of an untwisted operator. We choose an operator transforming as $\cO_1$ in the two-dimensional tube-algebra representation $\cF_2$ in \eqref{eq:fib-tube-rep-2d} (because the one-dimensional representations are either trivial or contain twisted-sector operators only).
As above, we denote by $\cO_1$ and $\cO_W$ the two operators spanning the 2d irreducible representation, and we let $\rho$ be the reduced density matrix for the state $\cO_1(z) |0\rangle$. In order to compute the second R\'enyi asymmetry, we need to understand the swiping action of the line $\cL_W$ across the operator $\cO_1$. By looking at \eqref{eq:fib-tube-rep-2d}, it is clear that this will be a linear combination of $\cO_1$ and $\cO_W$. The result is reported in \cref{fig:fibonacci-lasso}. When swiping $\cL_W$ through $\rho$, which contains two instances of $\cO_1$, it is also useful to perform an $F$-move described by the $F$-symbols (\ref{F-symbols Fibonacci}).
Consider now the limit $\varepsilon \to 0$ of small radius for the discs at the entangling surfaces. The traces of products of density matrices can be written in terms of partition functions on a replicated manifold, with the weakly-symmetric boundary condition at the entangling surfaces, and possibly with a line $\cL_W$ connecting the two boundaries or connecting $\cO_W$ to one of the boundaries. As discussed around \eqref{eq:Z-as-amplitude-cardy}, those partition functions can be analyzed by mapping the geometry to a cylinder and expressing them as amplitudes of Cardy states evolved in the closed channel. If the Cardy states are untwisted, the limit $\varepsilon \to 0$ simply produces two multiplicative $g$-factors, as in \eqref{eq:shrinking-discs-Z}. If the Cardy states are twisted, instead, the ground state in the closed channel necessarily has higher energy than the untwisted vacuum, and the amplitude is suppressed by a positive power of $\varepsilon$. It follows that, in the $\varepsilon \to 0$ limit, partition functions involving a residual line $\cL_W$ connected to one or both boundaries do not contribute.

For instance, in the computation of the second moment $\Tr(\rho_\sS^2)$, we can write
\be
\rho \, \bigl( 1 - 2 \varphi^{3/2} \cS_W \bigr) \, \rho \, \bigl( 1 - 2 \varphi^{3/2} \cS_W \bigr) = (1 + 4 \varphi^{-1}) \, \rho + 4 \cdot 5^{1/2} \varphi  \, \rho' + \ldots \;,
\ee
where $\rho'$ is as $\rho$ but with $\cO_W$ in place of $\cO_1$ and with a line $\cL_W$ connecting the two operators, similarly to \cref{fig: rho' with mu}, while the dots represent terms that are suppressed in the $\varepsilon\to0$ limit once the trace is taken. In the end, up to an overall normalization which cancels in the ratio, we find:
\bea
\Tr \bigl( \rho_\sS^2 \bigr)
&\,\propto\, \bigl( 11\sqrt5 \, - 24 \bigr) \braket{\cO_1^\dag \cO_1 \cO_1^\dag \cO_1}_{\cM_2} + \bigl( 25 - 11 \sqrt5 \bigr) \braket{ \cO_1^\dag \cO_1 \cO_W^\dag \cO_W }_{\cM_2} \;, \\
\Tr \bigl( \rho^2 \bigr)
&\,\propto\, \braket{\cO_1^\dag \cO_1 \cO_1^\dag \cO_1}_{\cM_2} \;,
\eea
where the correlators are computed on the 2-sheeted Riemann surface $\cM_2$. Hence:
\be
\label{eq:strip_fib_asymm}
\Delta S^{(2)} = - \log\left[ \bigl( 11\sqrt5 \, - 24 \bigr) + \bigl( 25 - 11 \sqrt{5} \bigr) \, \frac{ \braket{ \cO_1^\dag \cO_1 \cO_W^\dag \cO_W }_{\cM_2} }{ \braket{\cO_1^\dag \cO_1 \cO_1^\dag \cO_1}_{\cM_2} } \right] \;.
\ee

In the limit of a large subsystem, the correlation four-point functions factorize into two-point functions, see also the discussion around \eqref{eq:half-space-fact}. While the correlator at denominator $\braket{\cO_1^\dag \cO_1 \cO_1^\dag \cO_1}_{\cM_2}$ has a conformal block decomposition dominated by the identity channel and thus $\braket{\cO_1^\dag \cO_1 \cO_1^\dag \cO_1}_{\cM_2} \simeq \braket{ \cO_1^\dag \cO_1}_\bC^2$, in the numerator the OPE $\cO_1^\dag \cO_W$ does not contain the identity channel and therefore, keeping into account that $\cO_1$ and $\cO_W$ have the same conformal dimension, the correlator $\braket{ \cO_1^\dag \cO_1 \cO_W^\dag \cO_W }_{\cM_2}$ is further suppressed. We conclude that, in the large subsystem limit, the second R\'enyi asymmetry becomes
\be
\Delta S^{(2)} = -\log \bigl( 11\sqrt5\, -24 \bigr) \simeq 0.516 \;.
\ee
It is interesting to notice that this value is smaller than the second R\'enyi asymmetry of the full system computed with respect to the tube algebra. Indeed, by symmetrizing with respect to the tube algebra, one immediately gets $\Delta S^{(n)} = \Delta S = \log(2)$ because $\cO_1$ belongs to an irreducible representation of dimension 2, see the discussion after \eqref{eq:symm-tube-ising-2d}.%
\footnote{On the other hand, the asymmetry with respect to the fusion algebra vanishes: $\Delta S^{(n)} = \Delta S = 0$. To see this, we should restrict the 2d representation $\cF_2$ of the tube algebra to its untwisted sector, corresponding to the fusion algebra, and determine how $\cO_1$ decomposes in irreducible fusion-algebra representations. From \eqref{eq:fib-tube-rep-2d}, $\cO_1$ lands in the single representation $R_-$. Since this is one-dimensional, the asymmetry vanishes.}

\subsection{Haagerup fusion ring}
\label{sec: Haagerup}

The simplest fusion category whose product rules are both noninvertible and non-Abelian is the Haagerup fusion category \cite{Haagerup:1994, Aseada:1998, Grossman:2011}, that we indicate as $H_{(3)}$. There exists a generalization to Haagerup--Izumi fusion categories \cite{Izumi:2001mi, Grossman:2011, Grossman:2015dza, Huang:2020lox} that we indicate as $H_{(N)}$, with $N > 3$.
In this section we will only analyze the fusion algebra, hence we do not report the $F$-symbols of the category.%
\footnote{The $F$-symbols of $H_{(3)}$ can be found in \cite{Osborne:2019, Huang:2020lox}. The $F$-symbols in \cite{Huang:2020lox} are in a particularly convenient gauge, which makes the expressions extremely compact. Numerical evidence for the existence of a CFT with Haagerup symmetry was given in \cite{Huang:2021nvb}. The $F$-symbols of $H_{(N)}$ with $3 \leq N \leq 15$ odd can be found in \cite{Huang:2020lox}.}
The fusion ring of $H_{(N)}$ is generated by the objects $\{ \unit, a, \varrho \}$ subject to the conditions
\be
a^N = \unit \;,\qquad\qquad a \times \varrho = \varrho \times a^{-1} \;,\qquad\qquad \varrho^2 = \unit + \sum\nolimits_{j=0}^{N-1} a^j \times \varrho \;.
\ee
We see that there is a $\bZ_N$ subsymmetry and that, for $N\geq 3$, the element $\varrho$ does not commute with $a$ and thus the multiplication is non-Abelian.
Conjugation is given by
\be
\bar a = a^{-1} \;,\qquad\qquad \bar\varrho = \varrho \;,\qquad\qquad \wb{a^j \times \varrho} = a^j \times \varrho \;.
\ee
Notice that $H_{(1)}$ coincides with the Fibonacci (commutative) fusion ring that we already discussed in \cref{sec: Fibonacci}, while $H_{(2)}$ is the fusion ring of the $SO(3)_6$ WZW model.

The fusion algebra is $2N$-dimensional and a basis is given by the elements
\be
\{ \unit, a , \ldots, a^{N-1}, \varrho, a\varrho, \ldots, a^{N-1}\varrho \} \;.
\ee
The symmetrizer with respect to it can be written as:
\be
\rho_\sS = \frac1{2N} \sum_\cL \cL \, \rho \, \cL^\dag + \frac{1}{2(N^2+4)} \Bigl( A \, \rho \, A - B \, \rho \, B \Bigr) - \frac1{N(N^2+4)} \Bigl( A \, \rho \, B + B \, \rho \, A \Bigr)
\ee
where in the first term we summed over all simple lines, while
\be
A = \unit + a + \ldots + a^{N-1} = A^\dag \;,\qquad\qquad B = \varrho + a\varrho + \ldots + a^{N-1} \varrho = B^\dag \;.
\ee

Let us list the irreducible representations of the fusion algebra.
For $N$ odd, there are two 1-dimensional representations $R_+$, $R_-$ and $\frac{N-1}2$ 2-dimensional representations $R_j$. For $N$ even, there are four 1-dimensional representations $R_+$, $R_-$, $R_{\tilde+}$, $R_{\tilde-}$ and $\frac{N-2}2$ 2-dimensional representations $R_j$.
The 1-dimensional representations are:
\bea
R_\pm \colon \qquad a &= 1 \;,\qquad& \varrho &= \tfrac12 \bigl( N \pm \sqrt{N^2 + 4} \bigr) \;, \\
R_{\tilde\pm} \colon \qquad a &= -1 \;,\qquad& \varrho &= \pm 1 \;. \\
\eea
The 2-dimensional representations $R_j$ with $j=1, \ldots, \frac{N-1}2$ (for $N$ odd) or $j=1, \ldots, \frac{N-2}2$ (for $N$ even) are:
\be
R_j\colon \qquad a = \mat{ \zeta^j & 0 \\ 0 & \zeta^{-j} } \;,\qquad \varrho = \mat{ 0 & 1 \\ 1 & 0 } \;,\qquad \zeta = e^{\frac{2\pi i}N} \;.
\ee

\paragraph{Spontaneous symmetry breaking.} For the Haagerup fusion ring with $N=3$, let us consider a TQFT describing the full SSB of the symmetry, thus giving 6 vacua. These transform in the regular representation.
The inverse bilinear form in the fusion algebra is given by
\be
\tilde K =
    \frac{1}{39}
    \begin{pmatrix}
      8 & \frac{3}{2} & \frac{3}{2} & -1 & -1 & -1 \\
      \frac{3}{2} & \frac{3}{2} & 8 & -1 & -1 & -1 \\
      \frac{3}{2} & 8 & \frac{3}{2} & -1 & -1 & -1 \\
      -1 & -1 & -1 & 5 & -\frac{3}{2} & -\frac{3}{2} \\
      -1 & -1 & -1 & -\frac{3}{2} & 5 & -\frac{3}{2} \\
      -1 & -1 & -1 & -\frac{3}{2} & -\frac{3}{2} & 5
    \end{pmatrix} \;.
\ee
We take a pure density matrix $\rho = |b \rangle \langle b|$.
The symmetrized density matrix $\rho_\sS$ has eigenvalues
\be
\lambda_i \,=\, \biggl\{\, \frac{\sqrt{13} + 3}{6\sqrt{13}} ,\; \frac16 ,\; \frac16 ,\; \frac16 ,\; \frac16 ,\; \frac{\sqrt{13} - 3}{6\sqrt{13}} \,\biggr\}
\ee
for any of the six vacua. This leads to the entanglement asymmetry $\Delta S \simeq 1.6568$. This value is smaller than $\log(6)$, the asymmetry for a spontaneously broken group of order 6.

\subsection[The \tps{\frac12 E_6}{1/2 E₆} fusion ring]{The \btps{\frac12 E_6}{1/2 E₆} fusion ring}

Finally, we briefly discuss the example of the $\frac12 E_6$ fusion ring, where a multiplicity greater than 1 appears.
Since we limit ourselves to the study of the fusion ring, we will not be concerned with its categorifications, which do in fact exist (see \cite{Hagge:2007} and Sec.~6.4 of \cite{Chang:2018iay}). One of the corresponding fusion categories appears as a subsymmetry of the non-diagonal $SU(2)_{10}$ WZW model of $E_6$ type, or of the $(A_{10}, E_6)$ minimal model. Interestingly, the ring does not admit any braiding despite being Abelian.

The ring has three elements $\{ \cL_\unit, \cL_X, \cL_Y \}$ with Abelian product rules
\be
\cL_X \times \cL_X = \cL_\unit \;,\qquad \cL_X \times \cL_Y = \cL_Y \times \cL_X = \cL_Y \;,\qquad \cL_Y \times \cL_Y = \cL_\unit + \cL_X + 2 \cL_Y \;.
\ee
The non-degenerate inverse bilinear form, that provides the symmetrizer, is
\be
\tilde K = \frac1{12} \mat{ 5 & -1 & -1 \\ -1 & 5 & -1 \\ -1 & -1 & 2} \;.
\ee
With this, we can compute the asymmetry in the case of fully broken symmetry.
We get, for the three vacua, the following eigenvalues of $\rho_{\mathsf{S}}$ and the corresponding asymmetries:
\be
\begin{alignedat}{3}
      \ket{1}, \ket{X}:& \qquad\qquad & & \lambda_i = \biggl\{ \, \frac12 ,\; \frac{3 + \sqrt3 }{12} ,\; \frac{3 - \sqrt3}{12} \,\biggr\} & \qquad\qquad \Delta S &\simeq 0.951 \;, \\
      \ket{Y}:& & & \lambda_i = \biggl\{\, \frac{3+\sqrt{3}}{6} ,\; \frac{3-\sqrt{3}}{6} ,\; 0 \,\biggr\} & \Delta S &\simeq 0.516 \;.
\end{alignedat}
\ee
As in the case of the Ising symmetry, the first two vacua are related by a $\bZ_2$ action and thus share the same asymmetry. The third vacuum, instead, has a different asymmetry.


\section*{Acknowledgments}

We are grateful to Filiberto Ares and Kantaro Ohmori for helpful discussions.
We acknowledge support
by the ERC-COG grant NP-QFT No.~864583 ``Non-perturbative dynamics of quantum fields: from new deconfined phases of matter to quantum black holes'',
by the MUR-FARE grant EmGrav No.~R20E8NR3HX ``The Emergence of Quantum Gravity from Strong Coupling Dynamics'',
by the MUR-PRIN2022 grant No.~2022NY2MXY, and
by the INFN ``Iniziativa Specifica ST\&FI''.
PC acknowledges support
by the ERC-AdG grant MOSE No.~101199196 and PRIN 2022 Project HIGHEST no.~2022SJCKAH\_002.


\appendix
\crefalias{section}{appendix}
\crefalias{subsection}{appendix}
\addtocontents{toc}{\setcounter{tocdepth}{1}}

\section{Strongly separable algebras and the symmetrizer}
\label{app: strongly separable algebras}

Consider a finite-dimensional unital semisimple associative algebra $\cX_a \times \cX_b = \sum_c T_{ab}^c \, \cX_c$ (often we will omit the product sign $\times$) over $\bC$ as in (\ref{algebra product}). Here $\{ \cX_a\}$ is a basis of the algebra, and one defines a trace as $\Tr(\cX_a) = \sum_m T_{am}^m$ then extended by linearity. In the finite-dimensional case, such algebras are automatically strongly separable and the symmetric bilinear form $K_{ab}$ defined in (\ref{bilinear form}) is non-degenerate. We call $\tilde K \equiv K^{-1}$ its inverse. Indeed, the Wedderburn--Artin theorem states that every finite-dimensional unital semisimple associative algebra over $\bC$ is isomorphic to a sum of matrix algebras,
\be
\label{WA isomorphism}
\cA \,\cong\, \bigoplus\nolimits_{\alpha=1}^r \mathrm{Mat}_{N_\alpha \times N_\alpha}(\bC) \;,
\ee
where $r$ is the number of blocks, $N_\alpha$ are the sizes of the matrices, and those numbers are unique up to permutation. Note that $r$ is also the number of irreducible representations, and $N_\alpha$ are their dimensions. Using the isomorphism (\ref{WA isomorphism}), all properties of the algebras can be explicitly checked. Following \cite{Aguiar:2000}, we construct the dual basis $\{\sX^a \}$ with respect to the bilinear form:
\be
\sX^a = \sum\nolimits_b \tilde K^{ab} \cX_b \;.
\ee
It satisfies $\Tr(\cX_a\, \sX^b) = \delta_a^b$. Going to a basis, one proves that for every $x \in \cA$:
\be
\label{trace relation in algebra}
x = \sum\nolimits_a \Tr(x \,\cX_a) \, \sX^a = \sum\nolimits_a \Tr(x \, \sX^a) \, \cX_a \;.
\ee
The symmetric separability idempotent (SSI) can be written as
\be
e = \sum\nolimits_{ab} \tilde K^{ab} \cX_a \otimes \cX_b = \sum\nolimits_a \cX_a \otimes \sX^a = \sum\nolimits_a \sX^a \otimes \cX_a \;.
\ee
It is clearly symmetric in the exchange of the two factors. One proves that $e$ commutes with every element of $\cA$. Using the shorthand notations $\cX_b \, e \equiv (\cX_b \otimes \unit) \, e$ and $e \, \cX_b \equiv e \, (\unit \otimes \cX_b)$:
\begin{multline}
\cX_b \, e = \sum\nolimits_a \cX_b \cX_a \otimes \sX^a = \sum\nolimits_{ac} \Tr\bigl( \cX_b \cX_a \sX^c \bigr) \cX_c \otimes \sX^a \\
= \sum\nolimits_{ac} \cX_c \otimes \Tr\bigl( \sX^c \cX_b \cX_a \bigr) \sX^a = \sum\nolimits_c \cX_c \otimes \sX^c \cX_b = e\, \cX_b \;.
\end{multline}
We can use $K$ to obtain a pivotal relation. Consider the following element of $\cA \otimes \cA$:
\be
\sum\nolimits_b \cX_b \otimes \cX_a \sX^b = \sum\nolimits_b \cX_b \cX_a \otimes \sX^b = \sum\nolimits_{bc} T_{ba}^c \cX_c \otimes \sX^b = \sum\nolimits_{bc} \cX_b \otimes T_{ca}^b \sX^c \;.
\ee
In the first equality we used that $e$ commutes with $\cA$ also ``from the inside'', which follows from the fact that $e$ is symmetric. In the last equality we moved the scalars and renamed the indices. Since $\{\cX_b\}$ are a basis on the first factor of $\cA \otimes \cA$, we conclude that
\be
\label{modified fusion algebra}
\cX_a \times \sX^b = \sum\nolimits_c T_{ca}^b \, \sX^c \;.
\ee
This can be rewritten as the pivotal relation (\ref{pivotal relation}). One can also obtain
\be
\label{inverse fusion algebra}
\sum\nolimits_{ab} T_{ab}^c \, \sX^b \, \sX^a = \sX^c \;.
\ee
Using the trace relation (\ref{trace relation in algebra}), the identity element (\ref{identity element}) can be written as
\be
\cX_\unit = \sum\nolimits_a \Tr( \cX_\unit \cX_a) \, \sX^a = \sum\nolimits_{ab} T_{ab}^b \, \sX^a = \sum\nolimits_b \cX_b \, \sX^b = \sum\nolimits_b \sX^b \, \cX_b \;.
\ee
In the third equality we used (\ref{modified fusion algebra}). Using the notation $xy \equiv \mu(x \otimes y)$ for the product, we can regard the previous relation as the contraction $\mu(e) = \cX_\unit$. Finally, let us show that $e$ is idempotent as an element of $\cA \otimes \cA^\text{op}$ (where $\cA^\text{op}$ is the same as $\cA$ but elements are multiplied in the opposite order). In other words, $e^2 = e$ where $e$ acts on itself from the outside:
\be
e^2 = \sum\nolimits_{ab} \cX_a \cX_b \otimes \sX^b \sX^a = \sum\nolimits_{abc} T_{ab}^c \, \cX_c \otimes \sX^b \sX^a = \sum\nolimits_c \cX_c \otimes \sX^c = e \;.
\ee
In the third equality we moved the scalars to the other side and used (\ref{inverse fusion algebra}). We have thus shown that $e$ is a \emph{symmetric separability idempotent}, that commutes with the algebra and gives the identity upon contraction.

\paragraph{Relation with special Frobenius algebras.}
The strong separability condition allows one to make $\cA$ into a special Frobenius algebra $(\cA, \mu , u, \Delta, \epsilon)$. The product $\mu\colon \cA \otimes \cA \to \cA$ is the one already defined. The unital map $u\colon \bC \to \cA$ is given by $u(1) = \cX_\unit$ and linearity. They satisfy $\mu \bigl( x \otimes u(1) \bigr) = \mu \bigl( u(1) \otimes x \bigr) = x$. The coproduct $\Delta \colon \cA \to \cA \otimes \cA$ is defined as
\be
\Delta(x) = x \, e = e \, x \;.
\ee
It satisfies $\mu \circ \Delta = \id_\cA$ because $\mu \bigl( \Delta(x) \bigr) = \mu(xe) = \mu \bigl( \sum_a x \cX_a \otimes \sX^a \bigr)= \sum_a x \cX_a \sX^a = x \cX_\unit$. Note that $\Delta \circ \mu \neq \id_{\cA \otimes \cA}$. The Frobenius relation
\be
(\id_\cA \otimes \mu) \circ (\Delta \otimes \id_\cA) = (\mu \otimes \text{id}_\cA) \circ (\id_\cA \otimes \Delta) = \Delta \circ \mu
\ee
simply follows from $xey = xye$. One can also define the counital map $\epsilon\colon \cA \to \bC$ using the trace, $\epsilon(x) = \Tr(x)$. It satisfies $(\id_\cA \otimes \epsilon) \circ \Delta = \id_\cA = (\epsilon \otimes \id_\cA ) \circ \Delta$. The first equation is
\be
(\id_\cA \otimes \epsilon) \circ \Delta(x) = (\id_\cA \otimes \epsilon) \, \Bigl( {\ts \sum_a } \, x \cX_a \otimes \sX^a \Bigr) = \sum\nolimits_a x \, \cX_a \, \Tr(\sX^a) = x \, \cX_\unit = x \;.
\ee
The second equation is similar. One finds $\Delta \bigl( u(1) \bigr) = e$ and $\epsilon \bigl( \mu(e) \bigr) = \dim\cA$.

\paragraph{\btps{*}{}-structure.}
A $*$-structure on an algebra $\cA$ over the complex field $\bC$ is an anti-linear involution $\dagger\colon \cA \to \cA$ (\ie, such that $X^{\dag\dag} = X$ and $(\alpha X + \beta Y)^\dag = \alpha^* X^\dag + \beta^* Y^\dag$ for $\alpha,\beta \in \bC$) which is also an anti-homomorphism: $(XY)^\dag = Y^\dag X^\dag$.

If $\cA$ is a finite-dimensional $C^*$-algebra, a version of the Wedderburn--Artin theorem (see \cite{Murphy:1990} Th.~6.3.8) states that the isomorphism to a sum of matrix algebras can be chosen such that the star operation is mapped to the standard matrix adjoint operation. Consider for simplicity the case of a single block of size $N$: $\cA \cong \mathrm{Mat}_{N \times N}(\bC)$, which means that $\cA$ is simple (the following arguments extend to the general case of multiple blocks with minimal modifications). Then we can represent the index $a$ as $a \equiv ij$ with $i,j=1, \dots, N$. Let us work in a special basis in which $\cX_a \equiv \cX_{ij}$ is mapped by the $*$-algebra isomorphism to the matrix with a single $1$ at position $(i,j)$ and all other entries vanishing. In this basis the product is
\be
\label{use of basis Xij}
\cX_a \times \cX_b \equiv \cX_{ij} \times \cX_{kl} = \delta_{jk} \, \cX_{il} \;,
\ee
in other words the structure constants are
\be
T_{ab}^c \equiv T_{ij, kl}^{mn} = \delta_i^m \delta_{jk} \delta_l^n \;.
\ee
The symmetric bilinear form is$\,$%
\footnote{In the general case (\ref{WA isomorphism}), $\cA$ is isomorphic to a sum of matrix algebras. The basis elements can be labeled as $\cX_{\alpha ij}$. Then $K_{ab}$ is as in (\ref{matrix K in special basis}) in each block, namely $K_{\alpha ij, \beta kl} = N_\alpha \, \delta_{\alpha\beta} \, \delta_{il} \delta_{jk}$. All the following formulas apply block by block.}
\be
\label{matrix K in special basis}
K_{ab} \equiv K_{ij, kl} = \sum\nolimits_{mnrs} T_{ij, mn}^{rs} \, T_{kl, rs}^{mn} = N \, \delta_{il} \delta_{jk} \;.
\ee
This is indeed invertible, and the inverse is
\be
\tilde K^{ab} \equiv \tilde K^{ij, kl} = N^{-1} \, \delta^{il} \delta^{jk} \;.
\ee
In this basis the symmetric separability idempotent takes the simple form
\be
e = \sum\nolimits_{ijkl} \tilde K^{ij, kl} \, \cX_{ij} \otimes \cX_{kl} = N^{-1} \sum\nolimits_{ij} \cX_{ij} \otimes \cX_{ji} \;.
\ee
The star operation takes a simple form as well:
\be
\cX_a^\dag \equiv \cX_{ij}^\dag = \cX_{ji} \;.
\ee
We see that $e$ is Hermitian: $e^\dag = e$. Since Hermiticity does not depend on the basis (it is invariant under isomorphism), we conclude that $e$ is Hermitian in general:
\be
e^\dag = \sum\nolimits_{ab} (\tilde K^{ab})^* \, \cX_a^\dag \otimes \cX_b^\dag = e \;.
\ee
In the special basis the SSI can also be written as follows:
\be
e = N^{-1} \sum\nolimits_{ij} \cX_{ij}^\dag \otimes \cX_{ij} \;.
\ee
By a rescaling of the basis, $\cX_{\underline a} \equiv N^{-\frac12} \cX_{ij}$, it reads $e = \sum_{\underline a} \cX_{\underline a}^\dag \otimes \cX_{\underline a}$. This very simple form (which is reached also in the general case of multiple blocks) is not invariant under change of basis, however it is useful that such an expression exists.
The special basis provides a Kraus representation of the symmetrizer. This implies that our symmetrizer is in fact a (trace preserving) quantum channel, and in particular is completely positive \cite{Nielsen:2012yss}.

\paragraph{Consequences for the symmetrizer.}
The symmetrizer can be defined in terms of the symmetric separability idempotent $e$ as the linear map $\rho_\sS = e(\rho) = \sum_{ab} \tilde K^{ab} \, \cX_a \rho \cX_b$. Idempotency of $e$ implies that $(\rho_\sS)_\sS = \rho_\sS$ and thus the symmetrizer is a projector. The property $\mu(e) = \cX_\unit$ and symmetry of $e$ imply that $\Tr(\rho_\sS) = \Tr(\rho)$ and that, if $\rho$ commutes with $\cA$, then $\rho_\sS = \rho$. Since $e$ commutes with $\cA$ then $\rho_\sS$ commutes with $\cA$. In other words, the image of the projector is precisely given by the matrices that commute with the action of the algebra, and not a subspace thereof.

Since $e = e^\dag$, when $\rho = \rho^\dag$ is Hermitian one finds that also $\rho_\sS$ is Hermitian. Finally, assume that the density matrix $\rho$ is nonnegative, namely that $\langle \psi | \rho | \psi\rangle \geq 0$ for every vector $|\psi\rangle$ in the Hilbert space. Then
\be
\langle \psi | \rho_\sS | \psi \rangle = \sum\nolimits_{ab} \tilde K^{ab} \langle \psi | \cX_a \rho \cX_b | \psi \rangle \stackrel{\text{special basis}}{=} \sum\nolimits_{\underline a} \langle \psi | \cX_{\underline a}^\dag \rho \cX_{\underline a} | \psi \rangle \geq 0 \;.
\ee
In the second equality we used the special basis adapted to the $*$-structure. In that basis we find a sum of nonnegative numbers, which is nonnegative. This proves that also $\rho_\sS$ is nonnegative.

\subsection[\tps{C^*}{C*}-structure of fusion algebras]{\btps{C^*}{C*}-structure of fusion algebras}
\label{app: C* algebras}

A $C^*$-algebra $\cA$ is a Banach algebra%
\footnote{Meaning that it is associative, it has a norm such that $\lVert xy \rVert \leq \lVert x \rVert \cdot \lVert y \rVert$, and it is complete with respect to the metric induced by that norm.}
over $\bC$, with a $*$-structure $\dagger\colon \cA \to \cA$ and with a norm such that
\be
\label{eq:c-star-condition}
\lVert x^\dag x \rVert = \lVert x \rVert^2 \qquad \text{for all } x \in \mathcal{A} \;.
\ee
This condition is called the $C^*$ relation and it is very restrictive, indeed, if it exists, such a norm is uniquely fixed by the algebraic structure as
\be
\lVert x \rVert^2 = \lVert x^\dag x \rVert = \sup \bigl\{ \lvert \lambda \rvert \bigm| (x^\dag x - \lambda \unit) \text{ is not invertible} \bigr\} \;.
\ee
One can prove that $C^*$-algebras are semisimple.

Let us show that finite-dimensional fusion algebras (summarized in \cref{sec: fusion alg}) are $C^*$-algebras. We show that
\be
x^\dag x = 0 \qquad \Rightarrow \qquad x = 0 \;.
\ee
For finite-dimensional unital $*$-algebras, this implies that they are $C^*$ (see, \eg, Prop.~2.1 in \cite{Muger:2000} or Th.~2.4.4 in \cite{PenneysNotes}). Decompose $x = \sum_a x_a \cL_a$. Then
\be
x^\dag x = \sum\nolimits_{abcd} x_a^* x_b \, N_{ac}^1 \, N_{cb}^d \, \cL_d \;.
\ee
Assume that $x^\dag x = 0$, then $\sum_{abc} x_a^* x_b \, N_{ac}^1 \, N_{cb}^d = 0$ for all $d$. Taking $d=1$ we get $\sum_a |x_a|^2 = 0$, therefore $x=0$. We conclude that finite-dimensional fusion algebras are $C^*$-algebras, and thus they are also semisimple.

\subsection{Explicit form of \btps{\rho_\sS}{ρₛ}}
\label{sec:schur}

Let us consider an algebra $\cA$, with the same hypotheses as above, which we identify with a matrix algebra by virtue of \eqref{WA isomorphism}. We pick a basis of $\cA$ given by $\{ \cX_{\alpha,ij} \}$ as above (\ref{use of basis Xij}), where
\be
\cX_{\alpha,ij} = \left(
        \begin{array}{c|c|c}
          0 & 0 \tikzmark{up} & 0 \\ \hline
          0 & \raisebox{0.15em}{$e_{ij}^\alpha$} & 0 \tikzmark{right} \\ \hline
          0 & 0 & 0
        \end{array}
\right) \qquad\qquad\qquad\qquad\qquad
\begin{aligned} \alpha &= 1, \ldots, r \\ i,j &= 1, \ldots, N_\alpha \end{aligned}
\tikz[overlay, remember picture]{
    \draw[->] ([yshift = 1.2mm, xshift = 10mm]right) node[right, shift={(0, 0.03)}] {\scriptsize$\alpha$-th block} -- ([yshift = 1.2mm, xshift = 7mm]right);
}
\ee
and $e_{ij}^\alpha$ is the $N_\alpha \times N_\alpha$ matrix that has a single 1 in position $(i,j)$ and 0 elsewhere. The $\beta$-th representation of $\cA$ merely selects the $\beta$-th diagonal block from the matrix above. The most generic representation $\fR \colon \cA \to \End(V)$ is a direct sum of irreducible representations:
\be
\label{eq:hilb-irrep-decomp}
V = \bigoplus\nolimits_{\alpha=1}^r \, n_\alpha \, V_\alpha \;,
\ee
where $\dim(V_\alpha) = N_\alpha$, and thus the matrix that represents $\cX_{\alpha,ij}$ is of the form:
\be
\label{eq:rho-a-blocks}
\fR(\cX_{\alpha,ij}) = \left(
        \begin{array}{c|c|c}
          0 & 0 \tikzmark{up2} & 0 \\ \hline
          0 & \rule[-1.7em]{0pt}{3.8em}
              \begingroup
              \def\arraystretch{0.75}
              \def\arraycolsep{0mm}
              \begin{array}{ccc}
                \smash{e_{ij}^\alpha} &  & \smash{0} \\
                                      & \raisebox{-1mm}{$\smash{\ddots}$} &   \\
                \smash{0}          &  & \smash{e_{ij}^\alpha}
              \end{array}
              \endgroup
          & 0 \tikzmark{right2} \\ \hline
          0 & 0 & 0
        \end{array}
\right)
\tikz[overlay,remember picture] {
  \draw[->] ([yshift = 1mm, xshift = 10mm]right2) node[right, shift={(0, 0.03)}] {\scriptsize$\alpha$-th block} -- ([yshift = 1mm, xshift = 7mm]right2);
}
\ee
where $e^\alpha_{ij}$ is repeated $n_\alpha$ times. We now assume that the symmetrizer, which in principle can be any element of $\End \bigl( \End(V) \bigr)$, is written in the following form:%
\footnote{This assumption is referred to as ``universality'' in \cref{sec: asymmetry}.}
\be
\label{eq:symm-with-lines}
M_\sS = \sum\nolimits_{\alpha,\, i,\, j,\, \alpha'\!,\, i'\!,\, j'} \; c_{(\alpha,ij)(\alpha',i'j')} \;\; \fR(\cX_{\alpha,ij}) \; M \; \fR(\cX_{\alpha',i'j'}) \;,
\end{equation}
where $M \in \End(V)$ is any matrix, that we also decompose in blocks $M^{\alpha \alpha'}$ according to \eqref{eq:hilb-irrep-decomp}.
Thanks to property 2.\ and to Schur's lemma,%
\footnote{\label{fn:schur}%
Schur's lemma \cite{Lang:2002} can be stated (under our working hypotheses) as the fact that the module-endomorphisms of an irreducible $\cA$-module form a division algebra. But there is only one such algebra over $\bC$: the complex numbers themselves. Since the identity is clearly a module-endomorphism, we conclude that all such endomorphisms must be proportional to the identity. In more concrete terms, matrices that commute with all elements of an irreducible representation of $\cA$ must be proportional to the identity matrix. This result generalizes the well-known analogous result for groups and group algebras.}
we can conclude that, as a matrix, $M_\sS$ has vanishing off-diagonal blocks, $(M_\sS)^{\alpha \alpha'} = 0$ for $\alpha \neq \alpha'$, while each diagonal block is of the form
\be
(M_\sS)^{\alpha \alpha} =
    \def\arraystretch{0.75}
    \def\arraycolsep{0mm}
    \begin{pmatrix}
      \lambda^{\alpha\alpha}_{11} \Id_{N_\alpha} & \cdots & \lambda^{\alpha\alpha}_{1 n_\alpha} \Id_{N_\alpha} \\
      \vdots & \ddots & \vdots \\
      \lambda^{\alpha\alpha}_{n_\alpha 1} \Id_{N_\alpha} & \cdots & \lambda^{\alpha\alpha}_{n_\alpha n_\alpha} \Id_{N_\alpha}
    \end{pmatrix}
\,.
\ee
In principle each coefficient $\lambda^{\alpha\alpha}_{ab}$ is completely generic. However, using the additional assumption \eqref{eq:symm-with-lines}, we have
\be
M_\sS = \sum_{\alpha,\, i,\, j,\, i'\!,\, j'} \, c_{(\alpha,ij)(\alpha,i'j')} \; \left(
      \begin{array}{c|c|c}
        0 & 0 \tikzmark{up3} & 0 \\ \hline
        0 & \rule[-2.0em]{0pt}{4.5em}
            \begingroup
            \def\arraystretch{0.75}
            \def\arraycolsep{0.2em}
            \begin{array}{ccc}
              e_{ij}^{\alpha} m^{\alpha\alpha}_{11} e_{i'j'}^{\alpha} & \cdots & e_{ij}^{\alpha} m^{\alpha\alpha}_{1n_{\alpha}} e_{i'j'}^{\alpha} \\
              \vdots & \ddots & \vdots \\
              e_{ij}^{\alpha} m^{\alpha\alpha}_{n_{\alpha}1} e_{i'j'}^{\alpha} & \cdots & e_{ij}^{\alpha} m^{\alpha\alpha}_{n_{\alpha}n_{\alpha}} e_{i'j'}^{\alpha} \\
            \end{array}
            \endgroup
        & 0 \tikzmark{right3} \\ \hline
        0 & 0 & 0
      \end{array}
\right)
\tikz[overlay, remember picture]{
      \draw[->] ([yshift = 1mm, xshift = 10mm]right3) node[right, shift={(0, 0.03)}] {\scriptsize$\alpha$-th block} -- ([yshift = 1mm, xshift = 7mm]right3);
}
\qquad\qquad
\ee
where $m^{\alpha \alpha}_{ab}$ with $a, b \in \{ 1, \dots, n_\alpha \}$ are sub-blocks of $M^{\alpha \alpha}$. The symmetrization in each of the sub-blocks $m^{\alpha \alpha}_{ab}$ depends only on the entries of $M$ of that same sub-block. Let us now focus on such a sub-block. Its symmetrized version is
\be
\sum\nolimits_{i,\, j,\, i'\!,\, j'} c_{(\alpha,ij)(\alpha,i'j')} \, e_{ij}^\alpha \, m^{\alpha \alpha}_{ab} \, e_{i'j'}^\alpha = \sum\nolimits_{i,\, j,\, i'\!,\, j'} c_{(\alpha,ij)(\alpha,i'j')} \, (m^{\alpha\alpha}_{ab})_{ji'} \, e_{ij'}^\alpha \;.
\ee
This must be proportional to the identity, so it must be $c_{(\alpha,ij)(\alpha,i'j')} = \delta_{ij'} \, \tilde c^{\alpha}_{i'j}$. Consider now a matrix $M$ that is nonzero only in the block $m^{\alpha\alpha}_{aa}$. Imposing conservation of the trace gives $\sum_j (m^{\alpha\alpha}_{aa})_{jj} = \Tr(M) \overset{!}{=} \Tr(M_\sS) = N_\alpha \sum_{ji'} \tilde c_{i'j}^\alpha (m^{\alpha\alpha}_{aa})_{ji'}$. We conclude it must be $\tilde c^{\alpha}_{i'j} = \delta_{i'j}/N_\alpha$ and $c_{(\alpha,ij)(\alpha,i'j')} = \delta_{ij'} \delta_{i'j} / N_\alpha$. This means that the sub-block $(a,b)$ of the block $(\alpha, \alpha)$ is
\be
\label{eq:blocks-trace-identity}
\sum\nolimits_{i,\, j,\, i'\!,\, j'} c_{(\alpha,ij)(\alpha,i'j')} \, e_{ij}^\alpha \, m^{\alpha\alpha}_{ab} \, e_{i'j'}^\alpha = \frac1{N_\alpha} \Tr(m^{\alpha\alpha}_{ab}) \, \Id_{ab}^\alpha \;.
\ee
Notice that conservation of the trace has also fixed those blocks with $a \neq b$.

Said in words, the matrix $M_\sS$ is proportional to the identity in blocks that connect isomorphic irreducible representations, and is zero elsewhere. In each block, the proportionality constant is the trace of that block, divided by its dimension%
\footnote{Implicit in this statement is a choice of basis such that all copies of a given representation use the same matrices, as was the case in \eqref{eq:rho-a-blocks}. This choice implies that also the off-diagonal blocks are proportional to the identity (see \cref{fn:schur}). Indeed, such blocks are nonzero module-endomorphisms between isomorphic representations and thus, as matrices, commute with all elements of the corresponding irreducible module of $\cA$.}
(see also \cref{fig: action on matrices}).

\section{Hopf algebras, modules, and tensor products}
\label{app: Hopf algebras}

We review here the definition of Hopf algebras, following \cite{Bohm:1998iu, Cordova:2024iti, Schweigert}. An associative algebra $\cA$ over $\bC$ has an associative product $\mu\colon \cA \otimes \cA \to \cA$ and a unit $u\colon \bC \to \cA$ such that
\be
\mu \circ \bigl( u(1) \otimes \id \bigr) = \id = \mu \circ \bigl( \id \otimes u(1) \bigr) \;.
\ee
We will henceforth denote $u(1) \equiv \unit$. Both $\mu$ and $u$ are linear, and $u$ is unique when it exists. A coassociative coalgebra $\cA$ over $\bC$, instead, has a coassociative coproduct $\Delta\colon \cA \to \cA \otimes \cA$, such that $( \Delta \otimes \id) \circ \Delta = (\id \otimes \Delta) \circ \Delta \equiv \Delta^2$, and a counit $\epsilon\colon \cA \to \bC$ such that
\be
\label{eq:counit-property}
( \epsilon \otimes \id ) \circ \Delta = \id = (\id \otimes \epsilon) \circ \Delta \;.
\ee
Both $\Delta$ and $\epsilon$ are linear, and $\epsilon$ is unique.
It is convenient to use Sweedler's notation: it is always possible to write (in a non-unique way) coproducts as sums: $\Delta(x) = \sum_i x_{(1)}^i \otimes x_{(2)}^i$. This is concisely written as $\Delta(x) = x_{(1)} \otimes x_{(2)}$ with the sum kept implicit.

A weak bialgebra $\cA$ has both the algebra and coalgebra structures, which must be compatible in the following sense. The coproduct is an algebra homomorphism: $\Delta(xy) = \Delta(x) \, \Delta(y)$, where on the \rhs\ pairwise multiplication is used. The counit is weakly multiplicative:
\be
\label{weak multiplicativity counit}
\epsilon(xyz) = \epsilon\bigl( x \, y_{(1)} \bigr) \, \epsilon\bigl( y_{(2)} \, z \bigr) = \epsilon\bigl( x \, y_{(2)} \bigr) \, \epsilon\bigl( y_{(1)}\, z \bigr) \;,
\ee
where associativity is assumed. The unit is weakly comultiplicative, that can be written as
\be
\Delta^2 (\unit) = \bigl( \Delta(\unit) \otimes \unit \bigr) \bigl( \unit \otimes \Delta(\unit) \bigr) = \bigl( \unit \otimes \Delta(\unit) \bigr) \bigl( \Delta(\unit) \otimes \unit \bigr) \;,
\ee
where coassociativity is assumed in the first expression, while in the second and third expressions pairwise multiplication is used.
A \emph{weak Hopf algebra} is a weak bialgebra in which there exists an antipode map $S\colon \cA \to \cA$ satisfying the following three properties:
\be
\label{antipode in weak Hopf}
x_{(1)} \, S\bigl( x_{(2)} \bigr) = \epsilon\bigl( \unit_{(1)} \, x \bigr) \, \unit_{(2)} \;,\qquad
S\bigl( x_{(1)} \bigr) \, x_{(2)} = \unit_{(1)} \, \epsilon\bigl( x \, \unit_{(2)} \bigr) \;,\qquad
S\bigl( x_{(1)} \bigr) \, x_{(2)} \, S\bigl( x_{(3)} \bigr) = S(x) \;.
\ee
In the last equation we used Sweedler's notation for $\Delta^2 (x)$. It follows \cite{Bohm:1998iu} that the antipode map is unique, and that it is an algebra anti-homomorphism satisfying:
\bea
\label{props of S}
S(xy) &= S(y) \, S(x) \;,\qquad& S(\unit) &= \unit \;, \\
S(x)_{(1)} \otimes S(x)_{(2)} &= S\bigl( x_{(2)} ) \otimes S\bigl( x_{(1)} \bigr) \;,\qquad& \epsilon \circ S &= \epsilon \;.
\eea
In finite-dimensional weak Hopf algebras, $S$ is always invertible.
Weak $C^*$-Hopf algebras in which the antipode map is involutive, $S^2 = \id$, are called weak Kac algebras \cite{Nikshych:1998, Bohm:1999}.

A bialgebra (sometimes called strong bialgebra) is a weak bialgebra in which the counit is multiplicative and the unit is comultiplicative:
\be
\label{eq:mult-counit}
\epsilon(xy) = \epsilon(x) \, \epsilon(y) \;,\qquad\qquad\qquad \Delta(\unit) = \unit \otimes \unit \;.
\ee
These stronger properties imply the corresponding weaker ones. A \emph{Hopf algebra} (sometimes called a strong Hopf algebra) is a weak Hopf algebra in which one requires, equivalently, multiplicativity of the counit, comultiplicativity of the unit, or the hexagon diagram:
\be
\label{eq:hopf-hexagon}
x_{(1)} \, S\bigl( x_{(2)} \bigr) = S\bigl( x_{(1)} \bigr) \, x_{(2)} = \epsilon(x) \, \unit \;.
\ee
If one of these three properties holds for a weak Hopf algebra, then the other two follow automatically. This means that a Hopf algebra is in particular a bialgebra. Also in this case the antipode $S$ is unique, it satisfies (\ref{props of S}), and it is always invertible in the finite-dimensional case \cite{Larson:1969}. Besides,%
\footnote{In finite-dimensional semisimple weak Hopf algebras over $\bC$, $S$ is not necessarily involutive.}
\be
S^2 = \id
\ee
in finite-dimensional semisimple Hopf algebras over $\bC$ \cite{Larson:1988a, Larson:1988b}.

\paragraph{Group algebra.} The group algebra $\bC[G]$ is a typical example of Hopf algebra, where
\bea
\mu(g_1 \otimes g_2) = (g_1 g_2) \;,\quad\; u(1) = \unit_G \;,\quad\; \Delta(g) = g \otimes g \;,\quad\; \epsilon(g) = 1 \;,\quad\; S(g) = g^{-1} \;.
\eea
One easily verifies that all properties are satisfied.

\paragraph{Integrals.} Following \cite{Bohm:1998iu}, in a weak Hopf algebra one introduces the left and right projectors
\bea
\Pi^\mathrm{L}(x) &= \epsilon\bigl( \unit_{(1)} \, x \bigr) \, \unit_{(2)} = x_{(1)} \, S\bigl( x_{(2)} \bigr) \;, \\
\Pi^\mathrm{R}(x) &= \unit_{(1)} \, \epsilon\bigl( x \, \unit_{(2)} \bigr) = S\bigl( x_{(1)} \bigr) \, x_{(2)} \;,
\eea
acting on $x \in \cA$. These are projectors because of (\ref{weak multiplicativity counit}), and the second equality is as in (\ref{antipode in weak Hopf}). They are related by the antipode map:%
\footnote{Let us prove the first relation. First, notice that $\Pi^\mathrm{L}(\unit) = \Pi^\mathrm{R}(\unit) = \unit$ and $\epsilon \circ \Pi^\mathrm{L} = \epsilon \circ \Pi^\mathrm{R} = \epsilon$. Then $\Pi^\mathrm{L} \circ \Pi^\mathrm{R}(x) = \epsilon\bigl( \unit_{(1)} S(x_{(1)}) x_{(2)} \bigr) \unit_{(2)} = \epsilon\bigl( \unit_{(1)} S(x_{(1)})_{(2)} \bigr) \epsilon\bigl( S(x_{(1)})_{(1)} x_{(2)} \bigr) \unit_{(2)} = \epsilon\bigl( \unit_{(1)} S(x_{(1)}) \bigr) \epsilon\bigl( S( x_{(2)}) x_{(3)} \bigr) \unit_{(2)}$. Use $\epsilon \bigl( S(x_{(2)}) x_{(3)} \bigr) = \epsilon \circ \Pi^\mathrm{R}(x_{(2)}) = \epsilon(x_{(2)})$, so that $\Pi^\mathrm{L} \circ \Pi^\mathrm{R}(x) = \epsilon\bigl( \unit_{(1)} S(x) \bigr) \unit_{(2)} = \Pi^\mathrm{L} \circ S(x)$.}
\be
\Pi^\mathrm{L} \circ S = \Pi^\mathrm{L} \circ \Pi^\mathrm{R} = S \circ  \Pi^\mathrm{R} \;,\qquad\qquad
\Pi^\mathrm{R} \circ S = \Pi^\mathrm{R} \circ \Pi^\mathrm{L} = S \circ  \Pi^\mathrm{L} \;.
\ee
Then one defines left and right normalized integrals $\ell$ and $r$, respectively, such that:
\be
\label{left right integrals}
x\, \ell = \Pi^\mathrm{L}(x) \, \ell \;,\qquad \Pi^\mathrm{L}(\ell) = \unit \;, \qquad\text{or}\qquad r \, x = r \, \Pi^\mathrm{R}(x) \;,\qquad \Pi^\mathrm{R}(r) = \unit \;,
\ee
for all $x \in \cA$. Note that the normalized integrals $\ell,r$ are themselves projectors. Besides, $S(\ell)$ is a normalized right integral, and $S(r)$ is a normalized left integral (proven by applying $S$ to \eqref{left right integrals} and using that $S$ is invertible). An element that is simultaneously a left and right normalized integral is called a Haar integral $h$. When it exists, the Haar integral is unique. It follows that $S(h) = h$.

\paragraph{(Weak) Hopf algebra modules.} A left-module, or representation, consists of a vector space $V$ and a homomorphism $r\colon \cA \to \End(V)$ that acts from the left and respects the algebra structure. For a general algebra there is no concept of conjugate representation or of tensor product of representations. Those concepts become available in a (weak) Hopf algebra \cite{Bohm:1999}.

Let $\cM = (V, r)$ be a left-module for a weak Hopf algebra $\cA$ over $\bC$. There is a canonical right action on the dual space $V^\vee$, induced by the left action on $V$  simply as $\langle f \, r(x), v \rangle \equiv \langle f, r(x) \, v \rangle$ for all $f \in V^\vee$, $v \in V$ and $x \in \cA$. In order to construct a left action, instead, one uses the antipode map $S$:
\be
r^*(x) \, f \equiv f \, r\bigl( S(x) \bigr) \;.
\ee
The fact that $S(xy) = S(y) \, S(x)$ guarantees that $r^*$ is an algebra homomorphism. Thus $\cM^* = (V^\vee, r^*)$ is a left-module that we call the conjugate representation to $\cM$.

Let $\cM_1$, $\cM_2$ be two left-modules. The tensor product representation $\cM_1 \otimes \cM_2$ on $V_1 \otimes V_2$ is constructed using the coproduct:
\be
r_{12} \equiv (r_1 \otimes r_2) \circ \Delta \qquad\text{that is}\qquad r_{12} (x) \, v_1 \otimes v_2 \,\equiv\, \sum\nolimits_i \Bigl[ r_1\bigl( x_{(1)}^i \bigr) \, v_1 \Bigr] \otimes \Bigl[ r_2 \bigl( x_{(2)}^i \bigr) \, v_2 \Bigr] \,.
\ee
This defines an algebra homomorphism $r_{12}\colon \cA \to \End(V_1 \otimes V_2)$ in the case of Hopf algebras. In strictly weak Hopf algebras, since $\Delta(\unit) \neq \unit \otimes \unit$, there are vectors annihilated by $r_{12}(\unit)$. To remedy, notice that $\Delta(\unit)$ is a projector, therefore one constructs the representation on the subspace of $V_1 \otimes V_2$ where $\Delta(\unit)$ acts as the identity.

In the case of Hopf algebras one can also define the so-called ``trivial'' one-dimensional representation $r_\epsilon\colon \cA \to \bC$ using the counit:
\be
r_\epsilon(x) = \epsilon(x) \;.
\ee
This is an algebra homomorphism thanks to multiplicativity of the counit (\ref{eq:mult-counit}). The trivial representation multiplies trivially with all other representations (\ie, it is the identity) under tensor product, since $\bigl( (\epsilon \otimes r) \circ \Delta \bigr) (x) = \bigl( ( r \otimes \epsilon) \circ \Delta \bigr) (x) = r(x)$ using (\ref{eq:counit-property}), and besides it is self-conjugate since $\epsilon \circ S = \epsilon$. In strictly weak Hopf algebras, a representation that multiplies trivially with all the other ones exists as well, however it is not one-dimensional.

\subsection{Strip algebras with strongly-symmetric boundaries}
\label{app: symm strip Hopf}

It was shown in \cite{Kitaev:2011dxc, Cordova:2024iti} that strip algebras are weak Hopf algebras. Here we show that a strip algebra constructed using strongly-symmetric boundary conditions is actually a Hopf algebra. We only need to show that $\Delta(\unit) = \unit \otimes \unit$.

A general formula to construct the coproduct $\Delta$ in strip algebras was provided in \cite{Cordova:2024iti} Sec.~2.1. The identity element $\unit$ (that we sometimes called $\cX_\unit$) of the strip algebra, and its coproduct computed using that formula, are:
\be
\unit = \sum_{r,m} \;\; \strip{r}{m}{1}{r}{m}{}{} \;\;,\qquad
\Delta(\unit) = \sum_{r,n,m} \;\;\; \strip{r}{n}{1}{r}{n}{}{} \;\; \otimes \;\; \strip{n}{m}{1}{n}{m}{}{} \;.
\ee
For strongly-symmetric boundary conditions (and only in that case) the module has only one simple object, $r,m,n = \bdot\,$, and therefore $\Delta(\unit) = \unit \otimes \unit$.

\subsection{Symmetrizer in Hopf and weak Kac algebras}
\label{app: symmetrizer in Hopf}

In semisimple Hopf algebras, the symmetrizer is constructed using the object
\be
\bP = (\id \otimes S) \circ \Delta(P_\epsilon) \; \in \cA \otimes \cA^\text{op} \;,
\ee
that we called $e$ in (\ref{symmetrizer Hopf abstract}). This is known to provide a separability idempotent with respect to the canonical Frobenius structure of a Hopf algebra, see \cite{Schweigert} Remark~3.2.31. This implies that $\bP$ satisfies the properties 1.~--~3.\ listed in \cref{sec: asymmetry}. Let us explicitly check 1., 3.\ and 4. Idempotency is
\bea
\bP \times \bP &= P_{\epsilon \, (1)'} \, P_{\epsilon \, (1)} \otimes S\bigl( P_{\epsilon \, (2)} \bigr) \, S\bigl( P_{\epsilon \, (2)'} \bigr) = P_{\epsilon \, (1)'} \, P_{\epsilon \, (1)} \otimes S\bigl( P_{\epsilon \, (2)'} \, P_{\epsilon \, (2)} \bigr) \\
&= (\id \otimes S) \bigl( \Delta(P_\epsilon) \, \Delta(P_\epsilon) \bigr) = (\id\otimes S) \circ \Delta \bigl( P_\epsilon^2 \bigr) = \bP \;.
\eea
The preservation of symmetry-invariant matrices is $\mu(\bP) = P_{\epsilon\, (1)} \, S\bigl( P_{\epsilon\, (2)} \bigr) = \unit$, which immediately follows from (\ref{eq:hopf-hexagon}) and $\epsilon(P_\epsilon) = 1$. The property of being trace-preserving follows from
\be
S\bigl( P_{\epsilon\, (2)} \bigr) \, P_{\epsilon \, (1)} = S\Bigl( S\bigl( P_{\epsilon\, (1)} \bigr) \, P_{\epsilon \, (2)} \Bigr) = S\bigl( \epsilon(P_\epsilon) \, \unit \bigr) = \unit \;,
\ee
where we used $S^2 = \id$ repeatedly.

In semisimple weak Hopf algebras where $S^2 = \id$, the symmetrizer is constructed using
\be
\bP = (\id \otimes S) \circ \Delta(h) \;,
\ee
where $h$ is the Haar integral. By Th.~3.13 of \cite{Bohm:1998iu}, the properties 1.~--~3.\ are satisfied. In particular, the preservation of symmetry-invariant matrices follows from
\be
\mu(\bP) = h_{(1)} \, S\bigl( h_{(2)} \bigr) = \Pi^\mathrm{L}(h) = \unit \;.
\ee
The property of being trace-preserving instead follows from
\be
S\bigl( h_{(2)} \bigr) \, h_{(1)} = S \Bigl( S\bigl( h_{(1)} \bigr) \, h_{(2)} \Bigr) = S\bigl( \Pi^\mathrm{R}(h) \bigr) = S(\unit) = \unit \;,
\ee
where we used $S^2 = \id$.

\section{\btps{F}{F}-symbols and the tube algebra}
\label{app: F-symbols}

\begin{figure}[t]
\centering\begin{tabular}{ccc}
\begin{tikzpicture}[line width=0.6]
        \draw (-0.7, -1.2) node[below] {\small$a$} -- (0,0) node[shift={(-0.4, -0.2)}] {\small$e$} -- (0.7, -1.2) node[below] {\small$c$};
        \draw (0, -1.2) node[below] {\small$b$} -- (-0.35, -0.6) node[shift={(0.23, 0)}] {\scriptsize$\mu$};
        \draw (0,0) node[shift={(0.2, 0)}] {\scriptsize$\nu$} -- (0, 0.4) node[above] {\small$d$};
        \node at (2.9, -0.6) {$\ds{} = \sum_{f, \rho, \sigma} \; \bigl[ F_{abc}^d \bigr]_{(e; \mu\nu)(f; \rho\sigma)}$};
\begin{scope}[shift={(5.9,0)}]
        \draw (-0.7, -1.2) node[below] {\small$a$} -- (0,0) node[shift={(0.4, -0.2)}] {\small$f$} -- (0.7, -1.2) node[below] {\small$c$};
        \draw (0, -1.2) node[below] {\small$b$} -- (0.35, -0.6) node[shift={(-0.2, 0)}] {\scriptsize$\rho$};
        \draw (0,0) node[shift={(-0.15, 0.05)}] {\scriptsize$\sigma$} -- (0, 0.4) node[above] {\small$d$};
\end{scope}
\end{tikzpicture}
&\hspace{0.1em}&
\begin{tikzpicture}[line width=0.6]
        \draw (-0.7, -1.2) node[below] {\small$a$} -- (0,0) node[shift={(0.4, -0.2)}] {\small$f$} -- (0.7, -1.2) node[below] {\small$c$};
        \draw (0, -1.2) node[below] {\small$b$} -- (0.35, -0.6) node[shift={(-0.2, 0)}] {\scriptsize$\rho$};
        \draw (0,0) node[shift={(-0.15, 0.05)}] {\scriptsize$\sigma$} -- (0, 0.4) node[above] {\small$d$};
        \node at (2.9, -0.6) {$\ds{} = \sum_{e, \mu, \nu} \; \bigl[ F_{abc}^d \bigr]^*_{(e; \mu\nu)(f; \rho\sigma)}$};
\begin{scope}[shift={(5.9,0)}]
        \draw (-0.7, -1.2) node[below] {\small$a$} -- (0,0) node[shift={(-0.4, -0.2)}] {\small$e$} -- (0.7, -1.2) node[below] {\small$c$};
        \draw (0, -1.2) node[below] {\small$b$} -- (-0.35, -0.6) node[shift={(0.23, 0)}] {\scriptsize$\mu$};
        \draw (0,0) node[shift={(0.2, 0)}] {\scriptsize$\nu$} -- (0, 0.4) node[above] {\small$d$};
\end{scope}
\end{tikzpicture}
\\
\begin{tikzpicture}[line width=0.6]
        \draw (-0.7, 1.2) node[above] {\small$a$} -- (0,0) node[shift={(-0.4, 0.2)}] {\small$e$} -- (0.7, 1.2) node[above] {\small$c$};
        \draw (0, 1.2) node[above] {\small$b$} -- (-0.35, 0.6) node[shift={(0.23, 0)}] {\scriptsize$\mu$};
        \draw (0,0) node[shift={(0.2, 0)}] {\scriptsize$\nu$} -- (0, -0.4) node[below] {\small$d$};
        \node at (2.9, 0.6) {$\ds{} = \sum_{f, \rho, \sigma} \; \bigl[ F_{abc}^d \bigr]^*_{(e; \mu\nu)(f; \rho\sigma)}$};
\begin{scope}[shift={(5.9,0)}]
        \draw (-0.7, 1.2) node[above] {\small$a$} -- (0,0) node[shift={(0.4, 0.2)}] {\small$f$} -- (0.7, 1.2) node[above] {\small$c$};
        \draw (0, 1.2) node[above] {\small$b$} -- (0.35, 0.6) node[shift={(-0.2, 0)}] {\scriptsize$\rho$};
        \draw (0,0) node[shift={(-0.15, -0.05)}] {\scriptsize$\sigma$} -- (0, -0.4) node[below] {\small$d$};
\end{scope}
\end{tikzpicture}
&&
\begin{tikzpicture}[line width=0.6]
        \draw (-0.7, 1.2) node[above] {\small$a$} -- (0,0) node[shift={(0.4, 0.2)}] {\small$f$} -- (0.7, 1.2) node[above] {\small$c$};
        \draw (0, 1.2) node[above] {\small$b$} -- (0.35, 0.6) node[shift={(-0.2, 0)}] {\scriptsize$\rho$};
        \draw (0,0) node[shift={(-0.15, -0.05)}] {\scriptsize$\sigma$} -- (0, -0.4) node[below] {\small$d$};
        \node at (2.9, 0.6) {$\ds{} = \sum_{e, \mu, \nu} \; \bigl[ F_{abc}^d \bigr]_{(e; \mu\nu)(f; \rho\sigma)}$};
\begin{scope}[shift={(5.9,0)}]
        \draw (-0.7, 1.2) node[above] {\small$a$} -- (0,0) node[shift={(-0.4, 0.2)}] {\small$e$} -- (0.7, 1.2) node[above] {\small$c$};
        \draw (0, 1.2) node[above] {\small$b$} -- (-0.35, 0.6) node[shift={(0.23, 0)}] {\scriptsize$\mu$};
        \draw (0,0) node[shift={(0.2, 0)}] {\scriptsize$\nu$} -- (0, -0.4) node[below] {\small$d$};
\end{scope}
\end{tikzpicture}
\end{tabular}
\caption{\label{fig: F-moves}%
Different ways of writing the $F$-move. All lines are oriented from bottom to top. We take as definition the one in the upper-left corner. The other ones are obtained using that $F_{abc}^d$ is a unitary matrix, as well as manipulating the associator diagrams
\raisebox{-0.5em}{\begin{tikzpicture}[line width=0.6]
        \protect\draw (0, -0.3) -- (0, -0.2) -- (-0.2, 0) -- (0, 0.2) -- (0, 0.3);
        \protect\draw (0, -0.2) -- (0.2, 0) -- (0, 0.2);
        \protect\draw (0.1, -0.1) -- (-0.1, 0.1);
\end{tikzpicture}}
and
\raisebox{-0.5em}{\begin{tikzpicture}[line width=0.6]
        \protect\draw (0, -0.3) -- (0, -0.2) -- (-0.2, 0) -- (0, 0.2) -- (0, 0.3);
        \protect\draw (0, -0.2) -- (0.2, 0) -- (0, 0.2);
        \protect\draw (-0.1, -0.1) -- (0.1, 0.1);
\end{tikzpicture}}.}
\end{figure}

\begin{figure}[t]
\centering\begin{tabular}{c}
\begin{tikzpicture}[line width=0.6]
        \draw (-0.5, -0.8) node[left] {\small$a$} -- (-0.3, 0.2) -- (-0.5, 0.8) node[left] {\small$c$};
        \draw (0.5, -0.8) node[right] {\small$b$} -- (0.3, -0.2) -- (0.5, 0.8) node[right] {\small$d$};
        \draw (0.3, -0.2) node[shift={(0.17,0)}] {\scriptsize$\mu$} -- (-0.3, 0.2) node[shift={(-0.17,0)}] {\scriptsize$\nu$} node[pos=0.5, shift={(80: 0.2)}] {\small$e$};
        \node at (4.1, 0) {$\ds{} =\;\; \sum_{f, \rho, \sigma} \, \sqrt{ \frac{d_e d_f}{d_b d_c}} \; \bigl[ F_{aed}^f \bigr]_{(c; \nu\sigma)(b; \mu\rho)}$};
\begin{scope}[shift={(8.1, 0)}]
        \draw (-0.5, -0.8) node[left] {\small$a$} -- (0, -0.4) -- (0, 0.4) node[pos=0.5, right] {\small$f$} -- (-0.5, 0.8) node[left] {\small$c$};
        \draw (0.5, -0.8) node[right] {\small$b$} -- (0, -0.4) node[shift={(-0.2, 0.05)}] {\scriptsize$\rho$} -- (0, 0.4) node[shift={(-0.2, -0.05)}] {\scriptsize$\sigma$}-- (0.5, 0.8) node[right] {\small$d$};
\end{scope}
\end{tikzpicture}
\\
\begin{tikzpicture}[line width=0.6]
        \draw (-0.5, -0.8) node[left] {\small$a$} -- (-0.3, -0.2) -- (-0.5, 0.8) node[left] {\small$c$};
        \draw (0.5, -0.8) node[right] {\small$b$} -- (0.3, 0.2) -- (0.5, 0.8) node[right] {\small$d$};
        \draw (-0.3, -0.2) node[shift={(-0.17,0)}] {\scriptsize$\mu$} -- (0.3, 0.2) node[shift={(0.17,0)}] {\scriptsize$\nu$} node[pos=0.5, shift={(100: 0.2)}] {\small$e$};
        \node at (4.1, 0) {$\ds{} =\;\; \sum_{f, \rho, \sigma} \, \sqrt{ \frac{d_e d_f}{d_a d_d}} \; \bigl[ F_{ceb}^f \bigr]^*_{(a; \mu\rho)(d; \nu\sigma)}$};
\begin{scope}[shift={(8.1, 0)}]
        \draw (-0.5, -0.8) node[left] {\small$a$} -- (0, -0.4) -- (0, 0.4) node[pos=0.5, right] {\small$f$} -- (-0.5, 0.8) node[left] {\small$c$};
        \draw (0.5, -0.8) node[right] {\small$b$} -- (0, -0.4) node[shift={(-0.2, 0.05)}] {\scriptsize$\rho$} -- (0, 0.4) node[shift={(-0.2, -0.05)}] {\scriptsize$\sigma$}-- (0.5, 0.8) node[right] {\small$d$};
\end{scope}
\end{tikzpicture}
\end{tabular}
\caption{\label{fig: F-moves bis}%
Other $F$-moves, derived from the ones in \cref{fig: F-moves} by manipulating the associator diagrams. All lines run from bottom to top.}
\end{figure}

In the definition of the monoidal structure of $\scrC$ we follow the normalizations of \cite{Barkeshli:2014cna}. The $F$-symbols $F_{abc}^d$ are unitary isomorphisms
\be
    \bigoplus_{e} \Hom(\cL_{a} \otimes \cL_{b}, \cL_{e}) \otimes \Hom(\cL_{e} \otimes \cL_{c}, \cL_{d})
    \,\xrightarrow{\sim}\,
    \bigoplus_{f} \Hom(\cL_{a} \otimes \cL_{f}, \cL_{d}) \otimes \Hom(\cL_{b} \otimes \cL_{c}, \cL_{f})
\ee
represented by a unitary square matrix of size
\be
N_{abc}^d = \sum\nolimits_e N_{ab}^e \, N_{ec}^d = \sum\nolimits_f N_{af}^d \, N_{bc}^f \;.
\ee
We use the definition in (\ref{def F-symbols}). The vector spaces of morphisms are normalized such that the relations in (\ref{fusion and q dim}) hold. Those relations, together with unitarity, allow one to fix various moves that we summarize in \cref{fig: F-moves,fig: F-moves bis}.

The $F$-symbols satisfy the pentagon equation:
\be
\label{pentagon equation}
\sum_\delta [F_{fcd}^e]_{(g; \beta \gamma)(l; \nu\delta)} \, [F_{abl}^e]_{(f; \alpha\delta)(k; \mu\lambda)} = \sum_{h,\psi, \sigma, \rho} [F_{abc}^g]_{(f; \alpha\beta)(h; \psi\sigma)} \, [F_{ahd}^e]_{(g; \sigma\gamma)(k; \rho\lambda)} \, [F_{bcd}^k]_{(h; \psi\rho)(l; \nu\mu)} \,.
\ee
Unitarity can be used to move the first $F$-symbol and the sum over $(h; \psi\sigma)$ (or the last $F$-symbol and the sum over $(h;\psi\rho)$) from the right to the left, obtaining in this way many other equivalent relations.

The basis elements of the tube algebra are defined in (\ref{tube algebra basis}), and the structure coefficients $M$ of the algebra are introduced in (\ref{def product tube algebra}). To compute them we proceed as follows. First we fuse $b$ with $a$ into $c$ using (\ref{fusion and q dim}), then we section the diagram horizontally intro three parts:
\bea
\text{(\ref{def product tube algebra})} = \sum_{c,\gamma} \sqrt{ \frac{d_c}{d_a d_b}} \quad \begin{tikzpicture}[line width = 0.6, baseline = {(0,0)}]
        \draw [gray!80!white, dashed] (-1.3, -0.4) -- (1.3, -0.4);
        \draw [gray!80!white, dashed] (-1.3, 0.4) -- (1.3, 0.4);
        \draw [|-|] (-0.9, -1.3) node[left] {\small$c$} -- (-0.5, -1) node[shift={(-60: 0.17)}] {\scriptsize$\gamma$} -- (0.5, -0.6) node[pos=0.5, above] {\small$a$} node[shift={(-45: 0.17)}] {\scriptsize$\mu$} -- (0.5, -0.2) node[pos=0.5, right] {\small$d$} node[shift={(45: 0.17)}] {\scriptsize$\nu$} -- (-0.5, 0.2) node[pos=0.5, above] {\small$h$} node[shift={(-135: 0.17)}] {\scriptsize$\rho$} -- (-0.5, 0.6) node[pos=0.5, left] {\small$e$} node[shift={(135: 0.17)}] {\scriptsize$\sigma$} -- (0.5, 1) node[pos=0.5, above] {\small$b$} node[shift={(120: 0.17)}] {\scriptsize$\gamma$} -- (0.9, 1.3) node[right] {\small$c$};
        \draw (-0.5, -1) -- (-0.5, 1.3) node[pos=0.26, left] {\small$b$} node[left] {\small$k$};
        \draw (0.5, -1.3) node[right] {\small$g$} -- (0.5, 1) node[pos=0.74, right] {\small$a$};
\end{tikzpicture}
\eea
and finally we apply the $F$-moves in \cref{fig: F-moves bis}. The resulting expression for $M$ is in (\ref{product tube algebra}).

\paragraph{Case of an invertible symmetry.} In the special case of an invertible symmetry group $G$, possibly with 't~Hooft anomaly $\omega$, the fusion category $\scrC$ is $\mathsf{Vec}_G^\omega$. The anomaly is parametrized by $\omega \in H^3\bigl( G, U(1) \bigr)$. The only nonvanishing $F$-symbols are
\be
\bigl[ F_{abc}^{(abc)} \bigr]_{(ab)(bc)} \equiv F_{abc} = \omega(a,b,c) \;.
\ee
On the left we used parenthesis $(gh)$ to indicate the group multiplication, while in the middle we introduced a lighter notation for the $F$-symbols since they only depend on three group elements. The pentagon equation (\ref{pentagon equation}) is nonvanishing only if $e = (abcd)$, $f=(ab)$, $g=(abc)$, $h=(bc)$, $k=(bcd)$, $l=(cd)$, and it reads $F_{(ab)cd} \, F_{ab(cd)} = F_{abc} \, F_{a(bc)d} \, F_{bcd}$. This is precisely the cocycle condition $d\omega = 1$.

The elements of the tube algebra are specified by two labels: the vertical line incoming from below, and the horizontal line. Referring to (\ref{tube algebra basis}), we specify $g$ and $a$, while $d = ag$ and $h = aga^{-1}$. The nonvanishing structure coefficients of the tube algebra are then specified by three labels:
\be
M_{g, a, ag, aga^{-1}, b, baga^{-1}, baga^{-1}b^{-1}}^{ba, bag} \,\equiv\, M_{g,a,b} \;.
\ee
Formula (\ref{product tube algebra reduced}) gives:
\be
M_{g,a,b} = \bigl[ F_{baga^{-1}b^{-1},b,a} \bigr]^* \bigl[ F_{b,aga^{-1},a} \bigr] \bigl[ F_{b,a,g} \bigr]^* = \frac{ \omega(b, aga^{-1}, a) }{ \omega(baga^{-1}b^{-1}, b, a) \, \omega(b,a,g) } \;.
\ee
Associativity of the tube algebra is guaranteed by the following identity:
\be
M_{g,a,b} \, M_{g, ba, c} = M_{g,a,cb} \, M_{aga^{-1}, b, c} \;.
\ee
When expanded in terms of $\omega$, this identity becomes equivalent to
\be
\frac{d\omega(c,b,a,g) \; d\omega(c, baga^{-1}b^{-1}, b, a) }{ d\omega(c,b,aga^{-1}, a) \; d\omega(cbaga^{-1}b^{-1}c^{-1}, c, b, a) } = 1 \;,
\ee
which is indeed a true equation because $\omega$ is a cocycle.

\paragraph{Associativity in the noninvertible case.} In the general case, the tube algebra product is given by (\ref{product tube algebra}). Using an argument similar to the previous one, but with lengthier algebra, and in particular applying the pentagon equation (\ref{pentagon equation}) four times, one can show that the product is associative. Let us sketch the proof in the case without multiplicities, \ie, for $N_{ab}^c \in \{0,1\}$. Associativity is the equation
\bea
\sum_{ds} M_{g,a,m,h,b,n,k}^{d,s} \, M_{g,d,s,k,c,p,l}^{f,t} = \sum_{er} M_{g,a,m,h,e,r,l}^{f,t} \, M_{h,b,n,k,c,p,l}^{e,r}
\qquad\text{for}\qquad
\raisebox{-3em}{\begin{tikzpicture}
        \draw (0, -1.2) node[shift={(0.26, 0.15)}] {\small$g$} -- (0, 1.2) node[shift={(0.26, -0.15)}] {\small$l$} node[pos=0.36, right] {\small$h$} node[pos=0.65, right] {\small$k$};
        \draw (-1, -0.7) node[left] {\small$a$} -- ++(2, 0) node[pos=0.4, shift={(0, 0.15)}] {\scriptsize$m$};
        \draw (-1, 0) node[left] {\small$b$} -- (1, 0) node[pos=0.42, shift={(0, 0.15)}] {\scriptsize$n$};
        \draw (-1, 0.7) node[left] {\small$c$} -- ++(2, 0) node[pos=0.42, shift={(0, 0.15)}] {\scriptsize$p$};
\end{tikzpicture}} \;.
\eea
Substituting the expression for $M$ in (\ref{product tube algebra reduced}), the quantum dimensions are the same on the two sides and can be dropped. One is left with the relation
\begin{multline}
\label{associativity to prove}
\sum\nolimits_{ds} \bigl[ F_{kba}^s \bigr]_{nd}^* \, \bigl[ F_{bha}^s \bigr]_{nm} \, \bigl[ F_{bag}^s \bigr]_{dm}^* \,\cdot\, \bigl[ F_{lcd}^t \bigr]_{pf}^* \, \bigl[ F_{ckd}^t \bigr]_{ps} \, \bigl[ F_{cdg}^t \bigr]_{fs}^* \\
= \sum\nolimits_{er} \bigl[ F_{lea}^t \bigr]_{rf}^* \, \bigl[ F_{eha}^t \bigr]_{rm} \, \bigl[ F_{eag}^t \bigr]_{fm}^* \,\cdot\, \bigl[ F_{lcb}^r \bigr]_{pe}^* \, \bigl[ F_{ckb}^r \bigr]_{pn} \, \bigl[ F_{cbh}^r \bigr]_{en}^* \,.
\end{multline}
This can be proven using the following set of four pentagon equations, obtained from (\ref{pentagon equation}) with various indices and possibly applying unitarity once or twice:
\bea
\bigl[ F_{ckd}^t \bigr]_{ps} \, \bigl[ F_{kba}^s \bigr]_{nd}^* &= \sum\nolimits_r \bigl[ F_{pba}^t \bigr]_{rd}^* \, \bigl[ F_{ckb}^r \bigr]_{pn} \, \bigl[ F_{cna}^t \bigr]_{rs} \;, \\
\sum\nolimits_e \bigl[ F_{cba}^f \bigr]_{ed}^* \, \bigl[ F_{eag}^t \bigr]_{fm} \, \bigl[ F_{cbm}^t \bigr]_{es} &= \bigl[ F_{cdg}^t \bigr]_{fs} \, \bigl[ F_{bag}^s \bigr]_{dm} \;, \\
\bigl[ F_{cbh}^r \bigr]_{en}^* \, \bigl[ F_{eha}^t \bigr]_{rm} &= \sum\nolimits_s \bigl[ F_{cna}^t \bigr]_{rs} \, \bigl[ F_{bha}^s \bigr]_{nm} \, \bigl[ F_{cbm}^t \bigr]_{es}^* \;, \\
\sum\nolimits_d \bigl[ F_{pba}^t \bigr]_{rd} \, \bigl[ F_{lcd}^t \bigr]_{pf} \, \bigl[ F_{cba}^f \bigr]_{ed}^* &= \bigl[ F_{lcb}^r \bigr]_{pe} \, \bigl[ F_{lea}^t \bigr]_{rf} \;.
\eea
The first one and the complex conjugate of the second one are used on the \lhs\ of (\ref{associativity to prove}), while the third one and the complex conjugate of the fourth one are used on the \rhs\@ On each side one ends up with a quadruple sum $\sum_{dser}$ of the same string, proving associativity.

\section{\btps{\tilde F}{\~F}-symbols and the strip algebra}
\label{app: strip algebra}

The $\tilde F$-symbols of a left-module category $\cM$, as defined in (\ref{def F tilde symbols}), satisfy a pentagon equation that also involves the $F$-symbols of the underlying fusion category $\scrC$:
\be
\label{left-module pentagon}
\sum_\rho \bigl[ \tilde F_{dcm}^n \bigr]_{(e; \beta \mu)(p; \nu \rho)} \bigl[ \tilde F_{abp}^n \bigr]_{(d; \alpha \rho)(q; \sigma \tau)}
= \! \sum_{f, \gamma, \delta, \omega} \! \bigl[ F_{abc}^e \bigr]_{(d; \alpha \beta)(f; \gamma \delta)} \bigl[ \tilde F_{afm}^n \bigr]_{(e; \delta \mu)(q; \omega \tau)} \bigl[ \tilde F_{bcm}^q \bigr]_{(f; \gamma \omega)(p; \nu\sigma)} \;.
\ee
Unitarity of the $F$ and $\tilde F$ symbols allows one to move the first or third term on the \lhs\ and the sum to the \rhs\@ Similarly, the symbols ${}^R\!\tilde F$ satisfy a right-module pentagon equation:
\begin{multline}
\label{right-module pentagon}
\sum\nolimits_\omega \bigl[ {}^R\!\tilde F_{\bar q bc}^{\bar m} \bigr]_{(\bar p; \sigma\nu)(f; \gamma\omega)} \, \bigl[ {}^R\!\tilde F_{\bar n af}^{\bar m} \bigr]_{(\bar q; \tau\omega)(e; \delta\mu)} = \\
= \sum\nolimits_{d \alpha \beta \rho} \bigl[ {}^R\!\tilde F_{\bar n ab}^{\bar p} \bigr]_{(\bar q; \tau\sigma)(d; \alpha\rho)} \, \bigl[ {}^R\!\tilde F_{\bar n dc}^{\bar m} \bigr]_{(\bar p; \rho\nu)(e; \beta\mu)} \, \bigl[ F_{abc}^e \bigr]_{(d; \alpha\beta)(f; \gamma\delta)}\;.
\end{multline}

\paragraph{Associativity of the strip algebra.} The strip algebra product was given in~\eqref{eq:strip-algebra-product-general}. Let us verify that it gives rise to an associative algebra, working for simplicity in the case that we use the same module on the right and on the left, and that there are no multiplicities. Associativity is the identity:
\be
\sum\nolimits_d Q_{m,r,a,n,s,b,p,t}^d \, Q_{m,r,d,p,t,c,q,u}^f = \sum\nolimits_e Q_{m,r,a,n,s,e,q,u}^f \, Q_{n,s,b,p,t,c,q,u}^e \;.
\ee
Substituting the expression for $Q$, the quantum dimensions cancel out and we are left with
\be
\label{relation for strip associativity}
\sum\nolimits_d \bigl[ \tilde F_{bar}^t \bigr]_{ds} \, \bigl[ \tilde F_{bam}^p \bigr]_{dn}^* \, \bigl[ \tilde F_{cdr}^u \bigr]_{ft} \, \bigl[ \tilde F_{cdm}^q \bigr]_{fp}^* =
\sum\nolimits_e \bigl[ \tilde F_{ear}^u \bigr]_{fs} \, \bigl[ \tilde F_{eam}^q \bigr]_{fn}^* \, \bigl[ \tilde F_{cbs}^u \bigr]_{et} \, \bigl[ \tilde F_{cbn}^q \bigr]_{ep}^* \,.
\ee
This identity can be verified using the module pentagon equation (\ref{left-module pentagon}) twice, in particular in the following two forms:
\bea
\sum\nolimits_e \bigl[ F_{cba}^f \bigr]_{ed}^* \, \bigl[ \tilde F_{ear}^u \bigr]_{fs} \, \bigl[ \tilde F_{cbs}^u \bigr]_{et} &= \bigl[ \tilde F_{cdr}^u \bigr]_{ft} \, \bigr[ \tilde F_{bar}^t \bigr]_{ds} \;, \\
\bigl[ \tilde F_{eam}^q \bigr]_{fn} \, \bigl[ \tilde F_{cbn}^q \bigr]_{ep} &= \sum\nolimits_d \bigl[ F_{cba}^f \bigr]_{ed} \, \bigl[ \tilde F_{cdm}^q \bigr]_{fp} \, \bigr[ \tilde F_{bam}^p \bigr]_{dn} \;.
\eea
The second one is used after complex conjugation. On both sides of (\ref{relation for strip associativity}) we obtain a double sum $\sum_{de}$ with the same string. This proves the identity.

\section{Module categories over \tps{\mathsf{Vec}_{\boldsymbol{G}}^{\boldsymbol{\omega}}}{VecGω}}
\label{app:group-modules}

Consider a symmetry group $G$. We first consider the case that there is no anomaly, $\omega = 1$, therefore the fusion category is $\mathsf{Vec}_G$. Module categories are labeled by conjugacy classes of subgroups $H \subseteq G$ and cocycles $\psi \in H^2\bigl( H, U(1) \bigr)$ (see Example~7.4.10 in \cite{etingof2016tensor}). Let us explain how the modules are constructed.

The simple objects $\cB_m$ correspond to the right-cosets $G/H$. We indicate them as $\mu \in G/H$. There is a projection map $[ \,\cdot\, ] \colon G \to G/H$ so that $[m] = \mu$ if $m$ is a representative of the class $\mu$. The module action is simply
\be
a \times [m] = [a m] \qquad\qquad \text{for } a \in G \;,\quad [m] \in G/H \;.
\ee
This product is well defined, in the sense that it does not depend on the representative $m$ chosen, because the coset is a right-coset. We can then write $a \times \mu = a\mu$.
Notice that $H$ and $k H k^{-1}$ (where $k \in G$) define equivalent module categories. If we denote $[g]_{1} \in G/H$ and $[g]_{k} \in G/(kHk^{-1})$, an isomorphism is given by $\varphi \bigl( [g]_1 \bigr) = [gk^{-1}]_k$.

In order to construct the $\tilde F$-symbols we need a map $h \colon G \times G/H \to H$. First, choose a lift $r \colon G/H \to G$ of $[ \,\cdot\, ]$, that is a map
\be
r(\mu) \qquad\text{such that}\qquad [ r(\mu) ] = \mu \;.
\ee
In general, $a \cdot r(\mu) \neq r (a\mu)$, however $[ a \cdot r(\mu)]= a\mu = [r(a\mu)]$, therefore the two terms in the inequality differ by the action of $H$ from the right. This allows us to define
\be
\label{def map h}
h(a,\mu) \qquad\text{such that}\qquad a \cdot r(\mu) = r(a\mu) \cdot h(a,\mu) \;.
\ee
By considering $ab\cdot r(\mu)$ one obtains the identity
\be
h(ab,\mu) = h(a, b\mu) \cdot h(b,\mu) \;.
\ee
Besides, $h(1, \mu) = 1$ and so $h(a^{-1},\mu) = h(a, a^{-1}\mu)^{-1}$. We then define the $\tilde F$-symbols as
\be
\tilde F_{ab\mu} = \psi\bigl( h(a,b\mu), h(b,\mu) \bigr) \;.
\ee
The indices not indicated are fixed. The left-module pentagon equation reads:
\be
\tilde F_{ab, c, \mu} \; \tilde F_{a,b,c\mu} = F_{a,b,c} \; \tilde F_{a, bc, \mu} \; \tilde F_{b,c,\mu} \;.
\ee
As we are assuming that there are no anomalies, $\omega=1$, we have $F_{a,b,c} = 1$. The remaining terms organize as
\be
\frac{ \psi \bigl[ h(b, c\mu) , h(c,\mu) \bigr] \, \psi \bigl[ h(a, bc\mu) , h(bc,\mu) \bigr] }{ \psi\bigl[ h(ab, c\mu) , h(c,\mu) \bigr] \, \psi\bigl[ h(a, bc\mu) , h(b, c\mu) \bigr] } = d\psi\bigl[ h(a, bc\mu) , h(b, c\mu) , h(c, \mu) \bigr] = 1 \;.
\ee
This equation is satisfied because $\psi$ is a cocycle. Thus the pentagon equation is satisfied.

There are two extreme cases. When $H = \{1\}$ we obtain the regular module, whose simple objects are all the elements of $G$. Such a module is completely fixed by $G$. Indeed $r(\mu) = \mu \in G$, $h(a,\mu) = 1$, and $\psi$ is trivial so that $\tilde F = \omega$ is trivial.
When $H = G$, instead, the module has a single simple object (which is then a strongly-symmetric boundary condition) and the inequivalent consistent $\tilde F$-symbols parametrize the different SPT phases of $G$ --- or equivalently projective representations of $G$.

Consider now the general case that $\omega$ is not necessarily vanishing. This includes the case that $\omega$ vanishes in cohomology, but we have chosen a representative that is not identically 1. We will however use normalized cocycles. The fusion category is $\mathsf{Vec}_G^\omega$. Then module categories are labeled by:
\begin{itemize}
\item the conjugacy class of a subgroup $H \subseteq G$ such that $\omega \big|_H$ is trivial in $H^3\bigl( H, U(1) \bigr)$;
\item a cochain $\psi \in C^2\bigl( H, U(1) \bigr)$ such that (using multiplicative notation):
\be
\label{condition on psi}
d\psi = \omega^{-1} \big|_H \;.
\ee
Such cochains form a torsor over $H^2 \bigl( H, U(1) \bigr)$.
\end{itemize}
As before, the simple objects are classes of $G/H$ and fusion is defined in the natural way. Given a lift $r$ of the projection and the map $h(a,\mu) \in H$ defined in (\ref{def map h}), the $\tilde F$-symbols are
\be
\tilde F_{ab\mu} =  \psi\bigl( h(a,b\mu), h(b,\mu) \bigr) \; \frac{ \omega\bigl( r(ab\mu) , h(a, b\mu) , h(b,\mu) \bigr) \, \omega\bigl( a, b, r(\mu) \bigr) }{ \omega\bigl( a, r(b\mu) , h(b,\mu) \bigr) } \;.
\ee

The only thing to check is that the left-module pentagon equation is satisfied. The condition \eqref{condition on psi} on $\psi$ reads:
\be
\frac{ \psi \bigl[ h(b, c\mu) , h(c,\mu) \bigr] \, \psi \bigl[ h(a, bc\mu) , h(bc,\mu) \bigr] }{ \psi\bigl[ h(ab, c\mu) , h(c,\mu) \bigr] \, \psi\bigl[ h(a, bc\mu) , h(b, c\mu) \bigr] } = d\psi = \omega\bigl[ h(a, bc\mu) , h(b, c\mu) , h(c, \mu) \bigr]^{-1} \;.
\ee
When substituting into the pentagon equation, we obtain
\begin{align}
& \Bigl\{ \omega[a,b,c] \, \omega[a, bc, r(\mu)] \, \omega[ a, r(bc\mu), h_{b, c\mu}] \, \omega[b,c,r(\mu)] \, \omega[ab, r(c\mu), h_{c,\mu}] \times {} \\
& \; \omega[ r(bc\mu), h_{b, c\mu}, h_{c,\mu}] \, \omega[ r(abc\mu), h_{a, bc\mu} , h_{bc, \mu} ] \Bigr\} \Big/ \Bigl\{ \omega[ a,b, r(c\mu)] \, \omega[a, r(bc\mu) , h_{bc, \mu} ] \, \omega[ b, r(c\mu), h_{c,\mu} ] \times {} \nn\\
& \; \omega[ h_{a, bc\mu} , h_{b, c\mu} , h_{c,\mu}] \, \omega[ ab, c, r(\mu) ] \, \omega[ r(abc\mu) , h_{a, bc\mu} , h_{b, c\mu} ] \, \omega[ r(abc\mu) , h_{ab, c\mu} , h_{c,\mu} ] \Bigr\} \stackrel{?}{=} 1 \nn \;,
\end{align}
where we used a more compact notation for $h$. The expression on the left is indeed equal to 1, using that $d\omega[ a, b, c, r(\mu) ]$, $d\omega[ a, b, r(c\mu), h_{c,\mu} ]$, $d\omega[ a, r(bc\mu) , h_{b, c\mu} , h_{c, \mu} ]$, $d\omega[ r(abc\mu) , h_{a, bc\mu} , h_{b, c\mu} , h_{c,\mu} ]$ are all equal to 1, as well as the multiplication properties of $r$ and $h$ given above.


\bibliographystyle{ytphys}
\baselineskip=0.92\baselineskip
\bibliography{Entanglement}

\end{document}